\documentclass[traditabstract]{aa}
\usepackage{amssymb,amsmath}
\usepackage{fixltx2e}
\usepackage{graphicx}
\usepackage{longtable}
\usepackage{txfonts}
\usepackage{natbib}
\bibpunct{(}{)}{;}{a}{}{,} 
\usepackage{setspace}

\graphicspath{{figures/}}
\newcommand{\be}{\begin{equation}}
\newcommand{\ee}{\end{equation}}
\newcommand{\bea}{\begin{eqnarray}}
\newcommand{\eea}{\end{eqnarray}}
\newcommand{\bc}{\begin{center}}
\newcommand{\ec}{\end{center}}

\newcommand{\msun}{{\rm M}_{\odot}}
\newcommand{\rsun}{{\rm R}_{\odot}}
\newcommand{\lsun}{{\rm L}_{\odot}}

\begin{document}

\title{3D simulations of Betelgeuse's bow shock}
\author{S. Mohamed\inst{1,2}, J. Mackey\inst{1}, and N. Langer\inst{1}} 
\institute{Argelander Institut f\"{u}r Astronomie, Auf dem H\"{u}gel 71, Bonn D-53121, Germany
\and
South African Astronomical Observatory, P.O. Box 9, 7935 Observatory, South Africa}
\offprints{shazrene@saao.ac.za}

\abstract
{Betelgeuse, the bright, cool red supergiant in Orion, is moving supersonically 
relative to the local interstellar medium. The star emits a 
powerful stellar wind that collides with this medium, forming   
a cometary structure, a bow shock, pointing in the direction of motion.
We present the first 3D hydrodynamic simulations of the formation and evolution of 
Betelgeuse's bow shock.  The models include realistic low-temperature cooling and 
cover a range of plausible interstellar medium densities of  0.3 - 1.9 cm$^{-3}$ and 
stellar velocities of 28 - 73 km\,s$^{-1}$. We show that the flow dynamics and morphology 
of the bow shock differ substantially because of the growth of Rayleigh-Taylor or 
Kelvin-Helmholtz instabilities. The former dominate the models with slow stellar 
velocities resulting in a clumpy bow shock substructure, whereas the latter produce a 
smoother, more layered substructure in the fast models. If the mass in the bow shock shell is 
low, as seems to be implied by the \textit{AKARI} luminosities ($\sim$3$\times10^{-3}\,\msun$), 
then Betelgeuse's bow shock is very young and is unlikely to have reached a steady state.  
The circular nature of the bow shock shell is consistent with this conclusion. 
Thus, our results suggest that Betelgeuse only entered the red supergiant phase recently.}   
 
\authorrunning{S. Mohamed, J. Mackey, N. Langer}
\titlerunning{3D simulations of Betelgeuse's bow shock}
\maketitle

\section{Introduction}
\label{sec: intro}

Massive stars have a tremendous impact on their circumstellar 
environment. Betelgeuse ($\alpha$ Orionis, HD 39801, HIP 27989), the 
prototype red supergiant (RSG) and the brightest M supergiant in the 
sky (M1-M2 Ia-Iab), is no exception. Although Betelgeuse's initial mass is 
uncertain, it is thought to be $\gtrsim$10\,$\msun$ (see Table.~\ref{tab: bet}). 
Thus, the star is likely in the core helium-burning phase and is approaching its 
final stages of evolution -- ultimately exploding as a 
core-collapse supernova and enriching the interstellar medium (ISM)  
 with radiation, mechanical energy, mass, and heavy elements. 
 Owing to its high luminosity, $\sim$$10^5\lsun$, and powerful stellar wind, 
 $\dot{M}_{\rm wind}$$\sim$3$\times 10^{-6}\, \msun$yr$^{-1}$ (see Table.~\ref{tab: bet}), 
the star is already injecting copious amounts of mass and energy into 
the ISM. 

For stationary stars, the wind-ISM interface is expected to be a 
spherical shell as shown, for example, by the analytic and hydrodynamic  
models of \citet{Wea77} and \citet{Gar96}, respectively. 
Betelgeuse has an average radial (heliocentric) velocity of $V_{\rm rad}$ = +20.7 $\pm$ 
0.4 km\,s$^{-1}$ away from the sun, and the tangential velocities  are $V_{\alpha\cos\delta}$ = 23.7($D$ [pc]/200) 
km\,s$^{-1}$ and $V_\delta$ = 9.1($D$ [pc]/200) km\,s$^{-1}$ \citep[and references therein]{Harp08}. The 
distance $D$ to Betelgeuse is not yet well established, the most recent 
estimate being 197$\pm$ 45 pc \citep[see][for a detailed discussion]{Harp08}. 
For this range of distances, Betelgeuse is moving supersonically 
relative to the local ISM, and the collision of its stellar wind with this medium has 
resulted in the formation of a cometary structure, a bow shock, 
pointing in the direction of motion (at a position angle of 
$69.0^{\circ}$ east of north). Although it is the only known 
RSG runaway, theoretical models predict that up to 30\% of RSGs 
can be runaway stars \citep{Eld11}.   

\begin{figure}
\centering
\includegraphics[scale=.45, angle=0,trim= 0 0 0 0, clip=true]{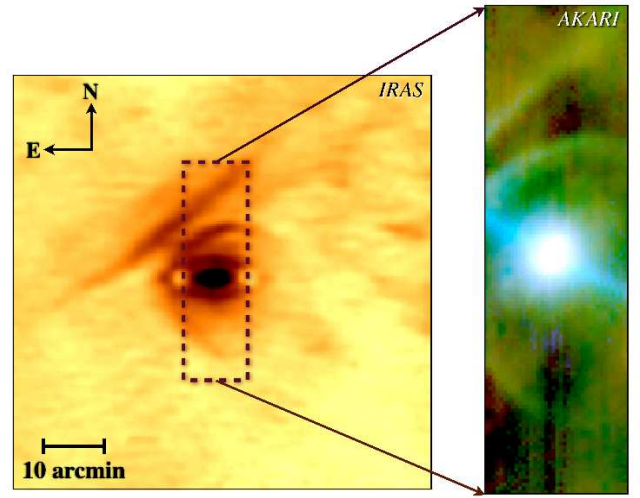}
\caption{Bow shock and `bar' of Betelgeuse. [Left] The 60 $\mu$m  
 {\it IRAS} observation retrieved from the {\it IRAS} Galaxy Atlas (IGA) Image
Server\protect\footnote \protect\citep[for details about IGA, see][]{Cao97}.
[Right] The high-resolution, composite {\it AKARI} image with 65\,$\mu$m, 90$\mu$m, and 140 $\mu$m 
emission in blue, green, and red, respectively \protect\citep{Ueta08} [Credit: {\it AKARI} MLHES team].
\label{fig: iras}}
\end{figure}

Early observations of the circumstellar medium around 
Betelgeuse suggested there is extended emission 
in the far-infrared \citep[e.g.~][]{Sten88,You93a}. However, because 
the star is so bright at these wavelengths, it was only in 1997 
that the bow shock was imaged successfully using {\it IRAS} at 60 and 
100 $\mu$m \citep{Nor97}. 
\footnotetext{http://irsa.ipac.caltech.edu/applications/IRAS/IGA/}
More recently, \cite{Ueta08} have imaged 
the bow shock with {\it AKARI} at higher spatial resolution in 
the 65, 90, 140, and 160 $\mu$m bands. The bow shock is a smooth, 
rather circular arc that has a  thickness of $\sim$1.5 arcmin and an inner 
radius of  $\sim$5 arcmin  (see Fig.~\ref{fig: iras}).  
Also detected by both {\it IRAS} and {\it AKARI} is a mysterious 
`bar' structure, of unknown origin, located just ahead of 
the bow shock arc. 

Many of the observed properties of the bow shock can be 
derived in terms of parameters of the interacting flows, making this 
an invaluable probe of the physical conditions in both the stellar 
wind and the ISM. The characteristic size of the bow shock shell, known 
as the stand-off distance, $R_{\rm SO}$, is determined by the location 
where the ram pressures of the two `flows' are in equilibrium:
\be
\rho_{\rm ISM} v^2_* = \rho_{\rm w} v_{\rm w}^2\,,
\label{eq:  ram}
\ee
where $\rho_{\rm ISM}$ and $\rho_{\rm w}$ are the
density of the ISM and stellar wind, respectively; $v_*$ is 
the velocity of the star with respect to the ISM, and $v_{\rm w}$ is 
the stellar wind velocity. 
Assuming spherical mass loss from the star,
\be
\rho_{\rm w} v_{\rm w} = \dot{M}_{\rm w} / 4 \pi r^2 \,,
\ee
and rearranging the resulting equation yields the well known solution: 
\be
R_{\rm SO} = \sqrt{\frac{\dot{M}_{\rm w} v_{\rm w}}{4 \pi \rho_{\rm ISM} v^2_*}} \,.
\
\ee
Making the additional assumption of momentum conservation and assuming that 
the material mixes and cools instantaneously (so that the dense shell 
has negligible thickness, i.e.~the thin-shell approximation), \cite{Wil96} derived 
an analytic expression for the shape of the bow shock:  
\be
R(\theta) = R_{\rm SO}\, \mathrm{cosec}\, \theta \sqrt{3 (1 - \theta\cot \theta)}\,,
\label{eq: shape}
\ee
where $\theta$ is the polar angle measured from the axis of symmetry.  

Utilising the above analytic models and current estimates of Betelgeuse's wind 
properties and distance (see Table.~\ref{tab: bet}), \cite{Ueta08} derived a space 
velocity of $v_*$ = 40 n$_{\rm H}^{-1/2}$ km\,s$^{-1}$ with respect to the local ISM.  
Estimates of the ISM density, n$_{\rm H}$, range from 0.3 cm$^{-3}$ for a flow emanating from the Orion 
Nebula Complex (ONC) \citep{Fri90} to 1.5 - 1.9 cm$^{-3}$ if the origin of the flow is the Orion 
OB 1 association \citep{Ueta08,Ueta09}. Given this range of ISM densities, Betelgeuse's space velocity with 
respect to the ambient medium is therefore likely to be between 73 km\,s$^{-1}$ and 28 km\,s$^{-1}$. If we assume 
 strong shock conditions, the post-shock temperature corresponding to 
these stellar velocities, is at least 10\,000 K. To date, the bow shock has only 
been detected in the far-infrared, where the bulk of the emission is probably caused by   
dust grains with an uncertain contribution from oxygen and carbon fine-structure lines.

There are several multi-dimensional hydrodynamic models of wind-ISM interactions 
with relevant parameters for Betelgeuse. For example, \cite{Gar96} have investigated the 
interaction of the ISM with a wind ejected from a stationary star evolving from the main sequence to the  
RSG and then Wolf-Rayet (WR) phase. They found that the RSG wind had a significant 
effect on the subsequent WR phase. The same conclusion was reached by \cite{Bri95} who studied 
the circumstellar medium (CSM) resulting from a moving star in 2D. As expected, in this case a 
bow shock arc was formed rather than a spherical shell.  The implications of an RSG 
phase for the WR progenitors of gamma-ray bursts was also discussed in a similar investigation 
by \cite{Van06}. \cite{War07a} and \cite{Vil03} calculated a series of 3D and 2D 
hydrodynamic models, respectively, of the interaction of a stellar wind from an 
asymptotic giant branch (AGB) star with the ISM and the subsequent structure 
formed during the planetary nebula phase. Other models of the CSM around 
fast-moving AGB stars include 3D and axisymmetric models of Mira's bow shock \citep[][respectively]{Rag08,Esq10}. 
More recently, \cite{Van11} have modelled Betelgeuse's bow shock with 2D high-resolution simulations 
that include a simple dust tracking scheme. They show that the flow for grains of various sizes differed 
and can have important consequences for interpreting infrared observations of bow shocks. 

We build on the work above by including a realistic treatment of the thermal physics 
and chemistry in 3D models and consider a broader range of parameters. The numerical 
method and model parameters are described in Sect.~\ref{sec: method}. In Sect.~\ref{sec: adiab}, 
we present an adiabatic model that serves as a code test and a point of reference 
for the simulations that include radiative cooling. The flow characteristics, morphology,  
and shell properties of the latter are described in Sect.~\ref{sec: cool}. 
Finally, the implications of this work, particularly with regard to Betelgeuse's 
bow shock morphology, shell mass and age are discussed in Sect.~\ref{sec: disc}.  

\begin{table}
\begin{center}
\begin{spacing}{1.5}
\caption{Summary of Betelgeuse's parameters\label{tab: bet}.}
\begin{tabular}{|l|ll|}
\hline
Parameter & Value & Refs.\\
\hline
Mass ($M_*$)                         	& 11.6$^{+5.0}_{-3.9}$, 15 - 20 $\msun$    & [1,2]\\
Temperature ($T_*$)             	& 3\,300 K                                 & [2]\\
Radius ($R_*$)                 		& 950 - 1\,200 $\rsun$                     & [2]\\
Luminosity ($L_*$)              	& 0.9 - 1.5 $\times 10^5\, \lsun$          & [2]\\
Wind velocity ($v_{\rm w}$)     	& 17 $\pm$ 1 km\,s$^{-1}$                  & [3]\\
Mass-loss rate ($\dot{M}_{\rm w}$)    	& 2 - 4 $\times 10^{-6}\, \msun$yr$^{-1}$  & [4]\\
ONC ISM density ($n_{\rm H}$)	        & 0.3 cm$^{-3}$                            & [5]\\
ISM density ($n_{\rm H}$)	        & 1.0 - 1.9 cm$^{-3}$                      & [6]\\
Stellar velocity ($v_*$)$^{a}$          & 73 - 28 km\,s$^{-1}$                     & [6]\\
Stand-off distance ($R_{\rm SO}$)       & $(8.5 \pm 1.9) \times 10^{17}$ cm        & [6]\\
Distance ($D$)                  	& $197 \pm 45$ pc            		   & [7]\\
\hline
\end{tabular}
\end{spacing}
\end{center}
\vspace{-0.25in}
\footnotesize{{\bf Notes.} $a$. Stellar velocity with respect to the local ambient ISM. \\
              {\bf References.} (1) \cite{Hil11}.
              (2) \cite{Smi09} and references therein. 
              (3) \cite{Ber79}.
	      (4) \cite{Nor97}.
	      (5) Observed ISM density of the Orion Nebulae Complex (ONC), \cite{Fri90}.
              (6) \cite{Ueta08}, bow shock fitting.
              (7) \cite{Harp08}.}
\end{table}

\section{Numerical method}
\label{sec: method}

\subsection{Smoothed particle hydrodynamics}
Smoothed particle hydrodynamics (SPH) is a Lagrangian method
in which particles behave like discrete fluid elements.
However, each of their fluid properties, e.g.~density, temperature, or 
velocity, is the result of mutually overlapping summations
and interpolations of the same properties of neighbouring
particles. In this way, a continuous fluid is realised; 
i.e.~physical fluid properties that vary smoothly over all
points in space can be defined from interpolations on a finite 
number of particles.

SPH has been utilised to study a wide range of problems, 
not only in astrophysics, but also in other fields,
e.g.~engineering. Thus, the SPH equations are often
adapted and optimised for a specific field or application.
We briefly outline the particular SPH formulation
implemented in the code, which serves as a basis
for this work, {\small GADGET-2} \citep{Spr05,Spr01}, and
refer the reader to detailed reviews of the method
in \cite{Mon92}, \cite{Ros09}, and \cite{Spr10}.

The momentum equation is derived based on an `entropy
formulation' for SPH, which conserves entropy, energy, and
momentum by construction \citep{Spr02}. In this
formulation, a function of the entropy,
$A(s) = P /\rho^ \gamma$, where $\gamma$ is the ratio of
specific heats, $P$ the pressure, and $\rho$ the density,
is evolved instead of the internal energy, $\epsilon$.
To model shocks, viscosity terms that transform kinetic motion of
the gas into internal energy are also included in the momentum equation.
When the fluid is compressed, the viscosity acts as an excess pressure
assigned to the particles in the equation of motion.

{\small GADGET-2} also includes self-gravity and
utilises a hierarchical Barnes and Hut 
oct-tree \citep{BH86} algorithm to calculate the
gravitational accelerations. The simulations carried out
in this work are characterised by a broad dynamic
range in densities. To obtain better and more efficient
spatial and time resolution, particles have adaptive
smoothing lengths and are evolved with individual and
adaptive timesteps.

\begin{table}
\begin{spacing}{1.5}
\caption{Summary of the parameters used for the different computed CSM evolution models.
\label{tab: models}}
\begin{tabular}{|l|lllll|}
\hline
Model & v$_*$ & n$_{\rm H}$ & T$_{\rm ISM}$  & $\bar{h}^1$ & Comment\\
 & (km\,s$^{-1}$) & (cm$^{-3}$)  &  (K) &  (0.01 pc) &\\
\hline
A            	    & 28.8 & 1.9  & 650    & 0.3 - 2   & Cooling\\
B           	    & 32.4 & 1.5  & 1\,000 & 0.2 - 2   & Cooling\\
C                   & 39.7 & 1.0  & 1\,600 & 0.2 - 2   & Cooling\\
D                   & 72.5 & 0.3  & 8\,000 & 0.1 - 2   & Cooling\\
D$_{\rm L}$         & 72.5 & 0.3  & 8\,000 & 0.5 - 4   & Cooling\\
C$_{\rm L650}$      & 39.7 & 1.0  & 650    & 0.6 - 3   & Cooling\\
C$_{\rm L1600}$     & 39.7 & 1.0  & 1\,600 & 0.6 - 3   & Cooling\\
C$_{\rm L8000}$     & 39.7 & 1.0  & 8\,000 & 0.6 - 3   & Cooling\\
C$_{\rm Lf}$        & 39.7 & 1.0  & 1\,600 & 0.6 - 3   & $f^2$$\sim$0.18\\
B$_{\rm ad}$ 	    & 32.4 & 1.5  & 100    & 1 - 5    & Adiabatic\\
A$_{\rm H} $        & 28.8 & 1.9  & 650    & 0.03 - 0.6 & Cooling\\
\hline
\end{tabular}
\end{spacing}
{\footnotesize {\bf Notes.} 1. The range $\bar{h}$ is an average of the smaller smoothing lengths in the bow shock to the typical ISM value.\\
 2. $f$ is the fraction of molecular to atomic hydrogen.}
\end{table}

\subsection{Model setup}

For numerical convenience, we choose a frame in which
the star is stationary and located at the origin
($x,y,z$ = 0, 0, 0) of a rectangular box. The ISM flows past
the star in the direction of the $x$ axis, interacting
with the stellar wind as it does so. Two different flags are
utilised to distinguish between ISM and wind particles.

\subsubsection{The ISM}
The ISM is assumed to be homogeneous. Although a cartesian
grid of particles is the simplest approach to producing such
a smooth medium, it also results in unwanted artefacts along
preferred grid directions.  To overcome this problem, we use a
`glass tile', where particles are randomly distributed in a periodic
box and evolved with the sign of gravity reversed until an
equilibrium configuration is reached \citep[see][for details]{Spr05}.
The relaxed glass tile is then scaled and cropped to the required
dimensions for the simulation.

We model a range of ISM densities, n$_{\rm H}$ = 0.3, 1.0, 1.5, 
and 1.9 cm$^{-3}$, with corresponding stellar velocities $v_*$ = 73, 40, 32,
and 29 km\,$^{-1}$, respectively (see Table.~\ref{tab: models}).
These number densities lie at the boundary between typical
values expected for either a warm or cold neutral ISM, so 
 we assume temperatures based on the phase diagram
of the standard model of \cite{Wolf95} (e.g.~their Fig.~3d). The
temperatures, T$_{\rm ISM}$, are 8\,000, 1\,600, 1\,000, and 650 K, 
respectively. The mass of each particle is determined by the volume of 
the simulation domain, $V_{\rm box}$, the number of ISM particles, 
$N_{\rm ISM}$, and the gas number density as follows:
\be
m_{\rm particle} = \frac{V_{\rm box} \mu m_{\rm H} n_{\rm H}}{N_{\rm ISM}}\,,
\ee
where $\mu = 1.4$ is the adopted mean molecular weight and $m_{\rm H}$ 
the mass of atomic hydrogen. The chosen particle mass effectively determines
the stellar wind resolution, i.e.~the number of particles required
to reproduce the observed mass-loss rate, as described in the section below.

\subsubsection{The stellar wind}

Although the extended envelope of Betelgeuse is thought to be
inhomogeneous and asymmetric close to the star, on much larger
scales, and more importantly on the scales that we are interested in,
the outflow is largely spherically symmetric \citep{Smi09}. Furthermore, as
discussed in \cite{Harp10}, the mass-loss mechanism
for M supergiant winds is not well understood. Given these
uncertainties, we do not model the wind acceleration in detail,
and instead assume that the wind is launched isotropically and
with constant velocity.

The particle mass calculated above then determines the number
of particles, $N_{\rm wind}$, that are injected in an interval
$\Delta t_{\rm wind}$ to produce a constant mass-loss rate of
$3.1 \times 10^{-6}\, \msun$yr$^{-1}$. The wind particles are
injected as spherical shells of typically a few hundred particles
``cut" from the glass tiles described above. Each shell is
randomly rotated in the $x, y$, and $z$ directions. Although this reduces
the smoothness of the outflow, it prevents the occurrence of numerical
artefacts. The particles are injected at a radius,
$R_{\rm inner}$$\sim$$10^{15}$\,cm, with a constant wind velocity,
$v_{\rm w}$$\sim$17\,km\,s$^{-1}$.  The initial temperature of
the particles is $T_{\rm wind}$$\sim$1\,000 K, the temperature at
$R_{\rm inner}$ derived from models of Betelgeuse's circumstellar
envelope, e.g.~\cite{Glas86} and \cite{Rod91}.

\subsubsection{Cooling}
\begin{table}
\begin{spacing}{1.5}
\caption{Summary of the combined cooling processes\label{tab: cool}.}
\begin{tabular}{|l|l|}
\hline
Label & Process$^1$\\
\hline
$\Lambda_{\rm atomic}$    & Atomic cooling \\
$\Lambda_{\rm CI, CII}$ & Carbon fine-structure cooling \\
$\Lambda_{\rm CO (H)}$ & CO vibrations excited by atomic hydrogen \\
$\Lambda_{\rm CO (H_2)}$ & CO vibrations excited by molecular hydrogen \\
$\Lambda_{\rm CO (r)}$ & CO rotational transitions collisionally \\
 & excited by molecular and atomic hydrogen \\
$\Lambda_{\rm CR}$ & Cosmic ray heating \\
$\Lambda_{\rm grain}$     & Dust grain cooling/heating  \\
$\Lambda_{\rm H_2(diss)}$ & Dissociation of molecular hydrogen cooling \\
$\Lambda_{\rm H_2(ref)}$ & H$_2$ reformation heating \\
$\Lambda_{\rm H_2(r-v)}$ & Molecular hydrogen rotational and \\
 & vibrational cooling \\
$\Lambda_{\rm H_2O(r)}$ & Water rotational cooling \\
$\Lambda_{\rm H_2O(H)}$ & Water vibrations excited by collisions \\
& with atomic hydrogen \\
$\Lambda_{\rm H_2O(H_2)}$ & Water vibrations excited by collisions \\
 & with molecular hydrogen \\
 $\Lambda_{\rm O (63\mu m)}$ & Oxygen non-LTE cooling \\
$\Lambda_{\rm OH}$ & OH cooling \\
\hline
\end{tabular}
\end{spacing}
\footnotesize{{\bf Notes.} 1.~The above processes are described in detail in \cite{Smi03} and references therein.}
\end{table}

Radiative cooling plays an important role in determining
the temperature structure and emission from the bow shock.
Given the large uncertainties in the processes involved, we adopt the
approach of \cite{Smi03} rather than model the cooling in detail. Their cooling curve consists
of analytic solutions and fits for various coolants based
on detailed calculations described extensively in the appendices
of \cite{Smi03} and references therein, e.g.~rotational and
vibrational transitions of H$_2$, CO and H$_2$O, H$_2$ dissociative
cooling and reformation heating, gas-grain cooling/heating,
and an atomic cooling function that includes non-equilibrium effects 
(see Table.~\ref{tab: cool}). 

First the non-equilibrium molecular and atomic fractions of hydrogen are 
calculated according to the prescription of \cite{Sut97}, and a
limited equilibrium C and O chemistry is used to calculate
the CO, OH, and H$_2$O abundances \citep{Smi03}.
We adopt standard solar abundances of hydrogen and helium,
and set the abundances of carbon and oxygen to
$\chi_{\rm C}$$\sim$2.5$\times10^{-4}$ and
$\chi_{\rm O}$$\sim$6.3$\times 10^{-4}$, respectively \citep{Lam84}.
We assume that the initial fraction of molecular to atomic hydrogen,
$f = n_{\rm H_2}/n_{\rm H}$, in the Betelgeuse outflow is small,
$f = 0.001$ (on a scale where 0 is atomic and 0.5 is fully
molecular).  The hydrogen in the ISM is assumed to be atomic.
The dust cooling function depends on the temperature of the grains, and we 
set this temperature to 40 K, the value derived by \cite{Ueta08}.
The change in specific internal energy, $\epsilon$, due to radiative
cooling is then calculated semi-implicitly for each particle, $i$:
\be
\epsilon^{(n+1)}_i = \epsilon^{(n)}_i + \frac{ \left(\Gamma - \Lambda\right)
\left[\rho^{(n)}_i, \epsilon^{(n+1)}_i \right]}{\rho^{(n)}_i} \Delta t \,,
\label{eq: cooling}
\ee
where $\Lambda$ and $\Gamma$ are the volumetric cooling and heating rates in
ergs\,cm$^{-3}$\,s$^{-1}$, respectively; $\Delta t$ is the timestep;
and $n$ and $n+1$ are the current and new timesteps, respectively \citep{Spr01}.
The change in specific internal energy is then converted to
a change in the entropy of each particle.

We do not include a treatment of dust formation and destruction, depletion of 
elements and molecules on to grains, the effects of magnetic or UV background 
radiation fields, density inhomogeneities in the stellar wind, and ISM.  Collisional 
ionisation, likely important at high temperatures is not explicitly 
included; instead, hot gas, $\gtrsim$10$^4$ K, cools primarily by atomic line emission  
\citep{Sut93} and thermal bremsstrahlung. Although 
these processes are likely to have an effect on the bow shock properties, both 
in terms of radiative transfer and cooling, the computational cost of their 
inclusion in 3D models is prohibitive.

In the following sections, we explore and discuss possible solutions 
for Betelgeuse's bow shock with 3D numerical models 
 (see Table.~\ref{tab: models}). We simulate four basic models, A-D,
with medium resolution, and with ISM number densities, 
n$_{\rm H}$ = 0.3, 1.0, 1.5, and 1.9 cm$^{-3}$, and  
stellar velocities $v_*$ = 73, 40, 32, and 29 km\,$^{-1}$, respectively. 
Several lower resolution models are also calculated in order to study 
the long-term evolution of model D (model D$_{\rm L}$) and 
the effect of varying the assumed ISM temperature (models 
C$_{\rm L650}$, C$_{\rm L1600}$, C$_{\rm L8000}$) and the 
initial molecular hydrogen fraction (model  C$_{\rm Lf}$).  Model 
A$_{\rm H}$ has the same parameters as model A, but is computed with 
higher resolution. Finally, we have also computed an adiabatic model 
(model B$_{\rm ad}$), discussed in the next section, which provides a 
useful basis both for comparison to previous studies and to our models 
with radiative cooling (Sect.~\ref{sec: cool}).

\section{Numerical test: Adiabatic model}
\label{sec: adiab}

We tested the numerical set-up with model B$_{\rm ad}$, a purely
adiabatic model, i.e.~no heat sources or sinks, and
$\gamma = 5/3$. The simulation begins at the start
of the red giant phase, with the star losing
$3.1 \times 10^{-6}\, \msun$yr$^{-1}$. Moving at
$\sim$32\,km\,s$^{-1}$, the star traverses a cold ISM
with $T_{\rm ISM} = 100$ K and density $n_{\rm H} =1.5$ cm$^{-3}$.

\begin{figure}
\centering
\includegraphics[scale=.37, angle=270,trim= 0 25 0 0]{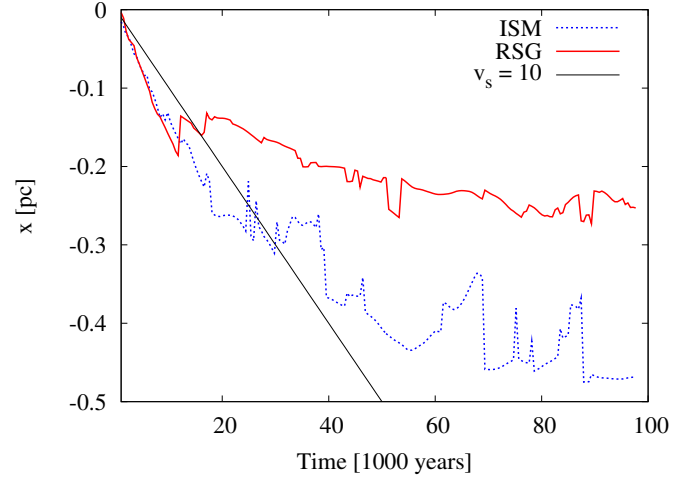}
\caption{Position of the shock along the $x$
axis (defined by the point with maximum temperature) as a
function of time for the ISM (dashed line) and
the stellar wind (dotted line). The black solid line
 demonstrates the trajectory for a $v_{\rm s}$ = 10 km\,s$^{-1}$
 shock velocity.  All velocities are measured with respect to the stationary
 frame of the star. 
\label{fig: adshockprof}}
\end{figure}

\begin{figure*}
\centering
\includegraphics[scale=.55, angle=0,trim= 10 35 0 0, clip=true]{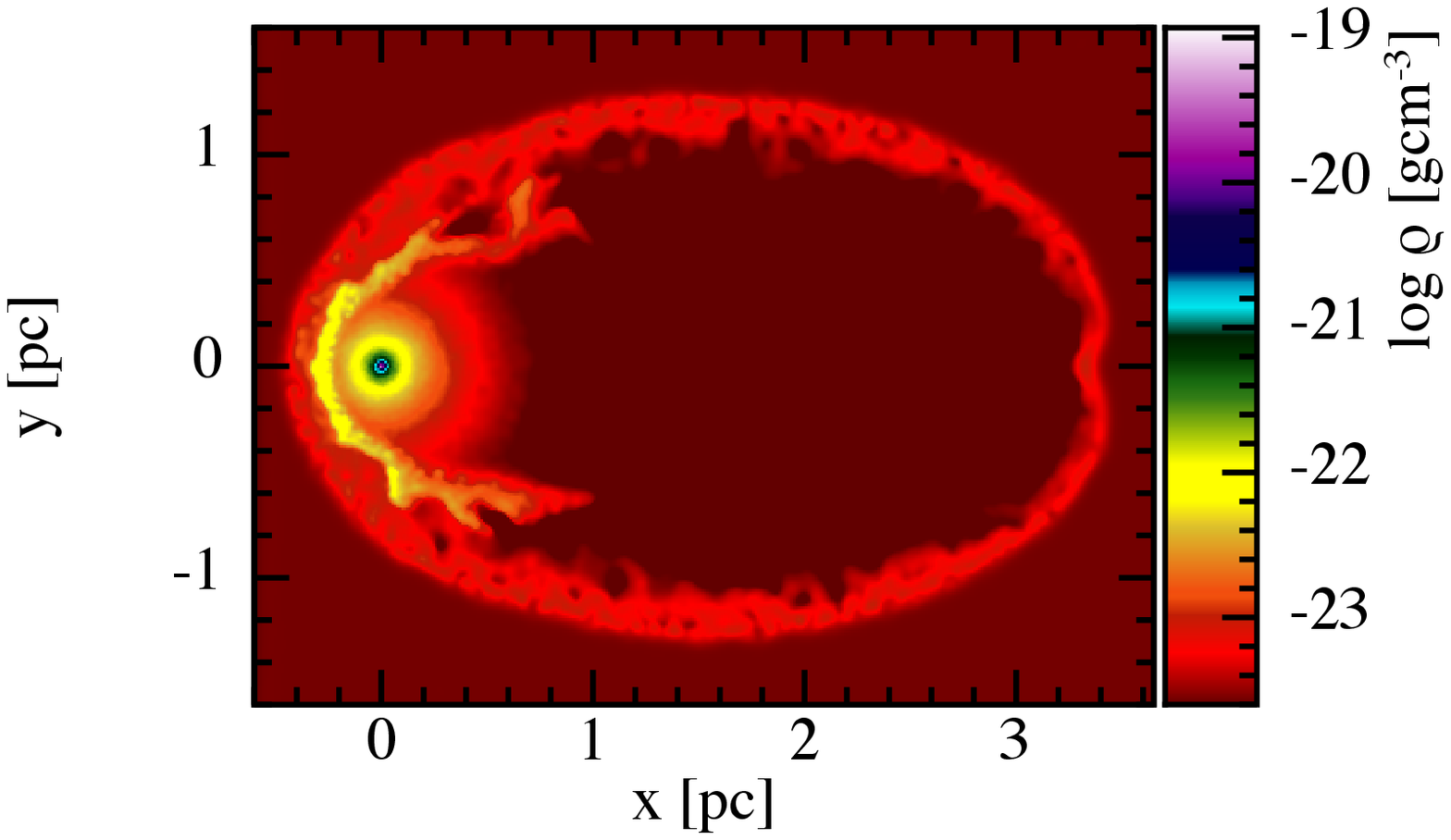}
\includegraphics[scale=.55, angle=0,trim= 50 35 0 0, clip=true]{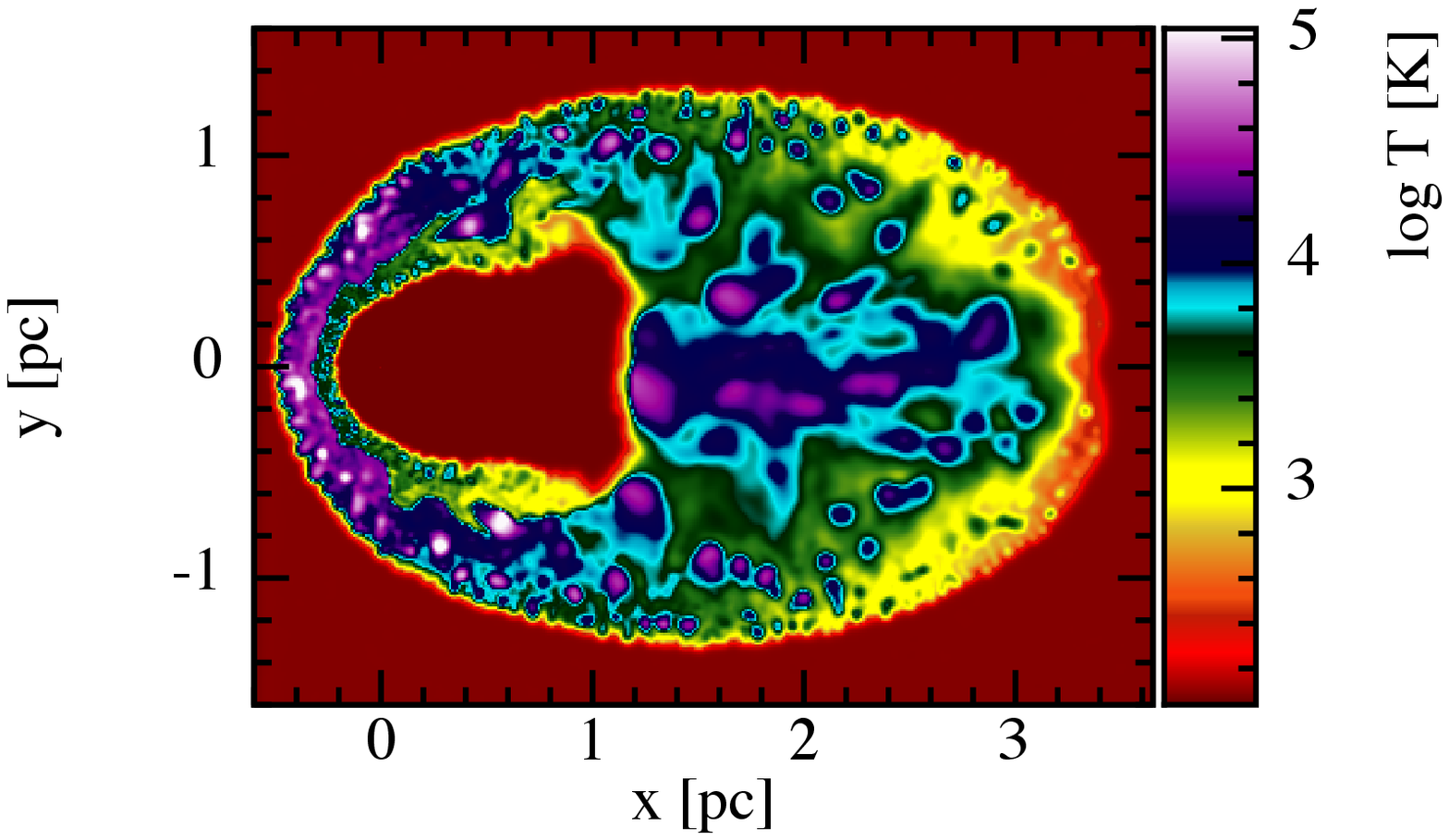}
\includegraphics[scale=.55, angle=0,trim= 10 10 0 0, clip=true]{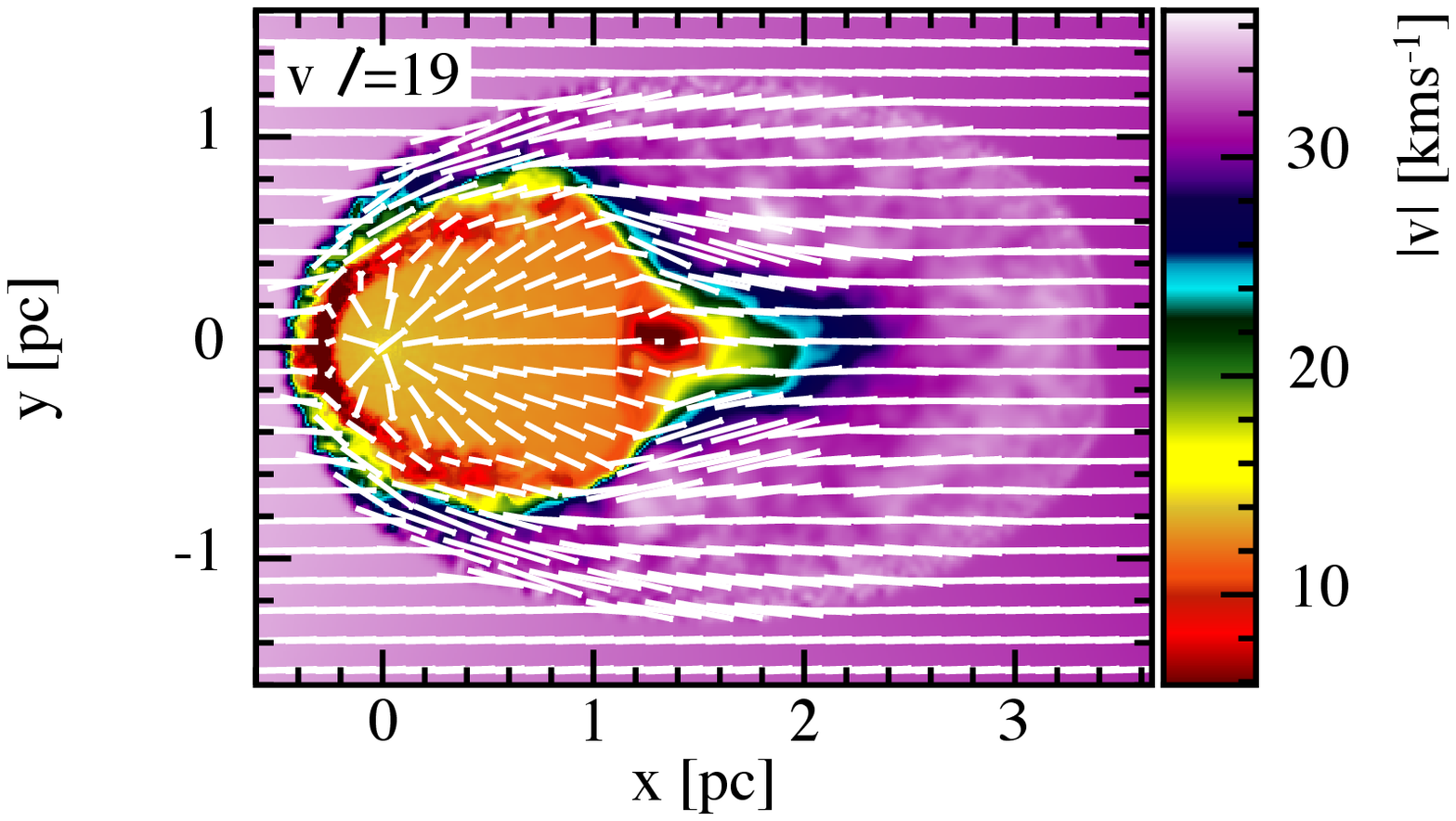}
\includegraphics[scale=.55, angle=0,trim= 50 10 0 0, clip=true]{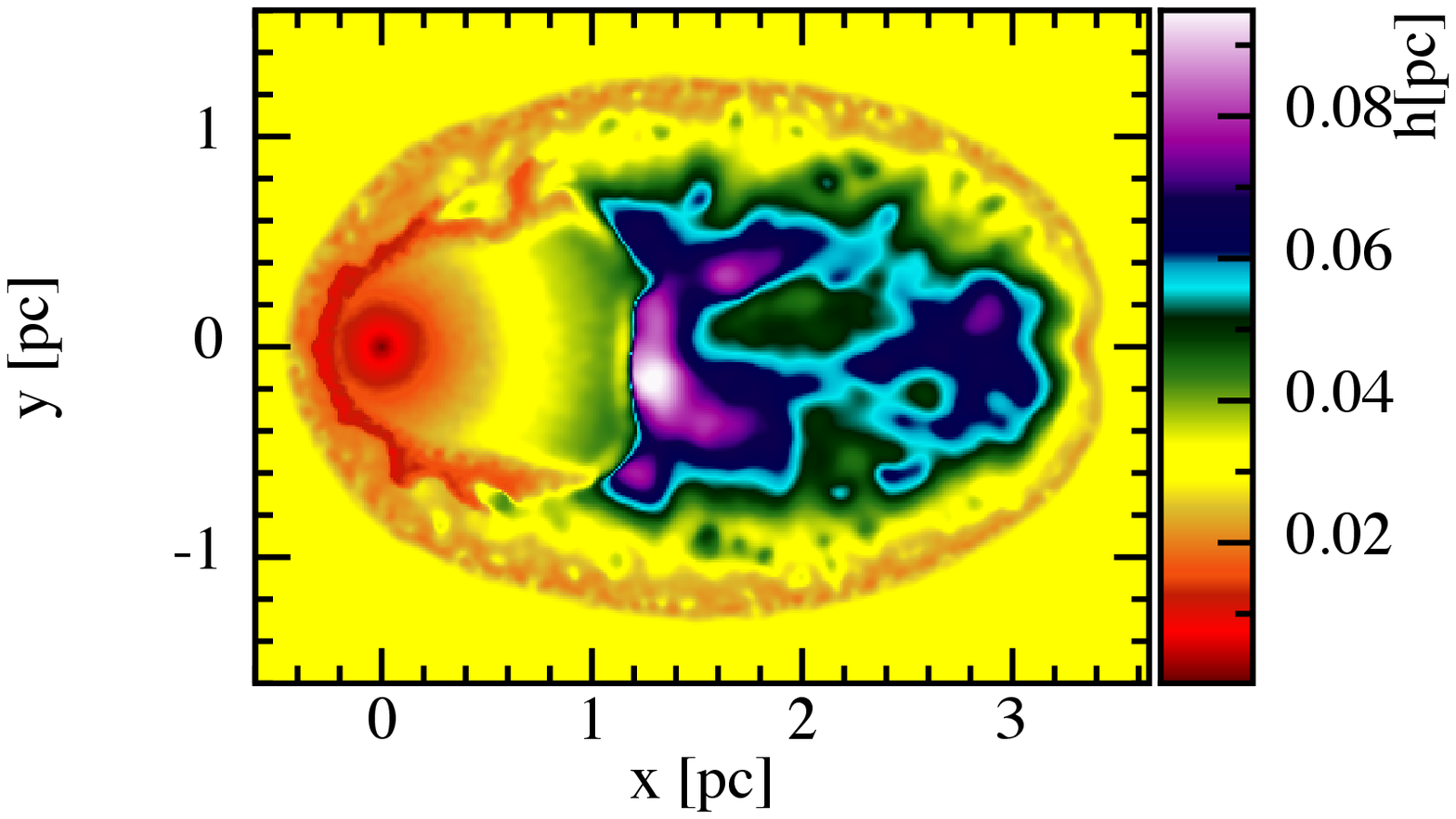}
\caption{Adiabatic model, B$_{\rm ad}$, of the
interaction of Betelgeuse's dense stellar wind with
the ISM after $t \approx100\,000$ years. We plot
cross-sectional profiles along the symmetry axis (the
$xy$ plane) for $\rho$, the gas density [top, left];
$T$, the temperature [top, right]; the velocity vector
field plotted over $|v|$, the gas velocity [bottom, left];
and the smoothing length, $h$ [bottom, right].
Approximately $1\times 10^7$ ISM particles filled the
simulation domain and $\sim$6$\times 10^5$ particles were
injected to produce the isotropic stellar wind.}
\label{fig: adiab}
\end{figure*}

The stellar wind collides with the ISM, and material accumulates 
at the contact discontinuity (the surface separating the two flows),  
 where part of the kinetic energy of the gas is thermalised. 
The heated ISM and wind material expand outwards from either side of
this surface, the former moves into the ISM and is called the
 forward shock, and the latter into the stellar wind and 
is known as the reverse shock. The expansion of these two 
regions is shown in Fig.~\ref{fig: adshockprof}.
The position of the reverse shock (defined as the point of maximum
temperature for the RSG wind particles along the $x$ axis) changes
slowly and reaches its average position, $\sim$-0.25 pc, after
50\,000 years. The forward shock (defined as the point of maximum 
temperature for the ISM particles along the $x$ axis) expands at 
10 km\,s$^{-1}$ during that time, and then decelerates 
to a position of -0.48 pc. The bow shock reaches a steady state 
after 80\,000 years, and has an average width of $\sim$0.2 pc.

The density, temperature, and velocity in the bow shock 
are shown in the cross-section in Fig.~\ref{fig: adiab}. After 
100\,000 years, the maximum width of the tail is 
approximately 3 pc (see Fig.~\ref{fig: adiab} [top, right]). 
The average temperature in the tail is approximately 4\,000 K;  
however, the cylindrical core of the tail is filled with 
much hotter, $\sim$15\,000 K, gas spanning $\sim$0.8 pc in
diameter. This hot gas is advected downstream from the
forward shock, and flows towards the central part of the tail, 
filling the low-pressure void (beyond the RSG stellar wind)
evacuated by the moving star (Fig.~\ref{fig: adiab} [bottom, left]).

\begin{figure}
\centering
\includegraphics[scale=.38, angle=270,trim= 0 40 0 0]{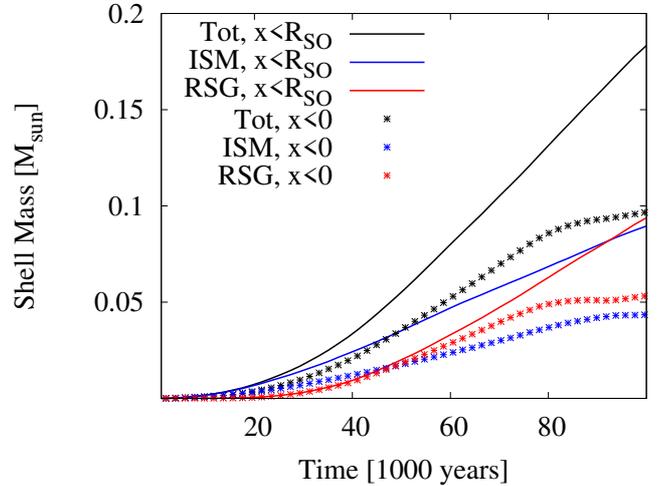}
\caption{Growth of the mass in the bow shock shell with time. 
The shell is defined as shocked regions with $T > 2\,000$ K
 and density above that in the ambient medium. The lines show the 
 total shell mass (black), the contributions from the ISM (blue) and the 
 RSG wind (red) for $x<R_{\rm SO}$ or $\theta \lesssim 115^\circ$, whereas 
  the points represent the same parameters except for shell material 
  with  $x< 0$ or $\theta \lesssim 90^\circ$.
\label{fig: adshellmass}}
\end{figure}

While the cometary tail is largely composed of ISM material at early times, 
both the RSG wind and ISM contribute similar amounts of mass to the bow shock 
head. Defining material in the bow shock shell as shocked regions
with $T > 2\,000$ K and density above that in the ambient medium, we see in 
Fig.~\ref{fig: adshellmass} that after $\sim$40\,000 years, the RSG wind  
contributes slightly more mass  to the bow shock head ($\theta \lesssim 90^\circ$,  
i.e.~$x < 0$) than the ISM, and it takes $\sim$90\,000 years for this to 
occur for $\theta \lesssim 115^\circ$, i.e.~for $x<R_{\rm SO}$. 
The mass accumulates in the bow shock almost linearly 
after $\sim$30\,000 years and only reaches a steady state for 
$\theta \lesssim 90^\circ$ after $\sim$80\,000 years. 
An estimate for a typical shell mass can be obtained from
\be
{\cal M}_{\rm shell} = \frac{\dot{M}_{\rm w} R_{\rm SO}}{v_{\rm w}}\,,
\label{eq: Mshell}
\ee
which yields 0.05 $\msun$. This expression assumes the mass is distributed over
 4$\pi$ sterad in a spherical shell and only takes the contribution from the stellar 
 wind into account. Thus for $\theta \lesssim 90^\circ$, 
 the expected RSG shell mass would be 0.025 $\msun$, i.e.~half of 
 ${\cal M}_{\rm shell}$. With a significant contribution from the wings of the 
 paraboloid (the shell radius is greater than $R_{\rm SO}$ for all 
 $\theta > 0^\circ$), the RSG shell mass obtained from the simulations 
 is twice the theoretical value (red points in Fig.~\ref{fig: adshellmass}). Including the mass from the ISM, the
 total mass in the bow shock head after 100\,000 years is $\approx$0.1 $\msun$.

\subsection{Numerical resolution and smoothing}
\label{sec: res}

The smoothing length, $h$, determines the scale on which the 
above fluid properties are smoothed, so it effectively defines the resolution
of a simulation. In {\small GADGET-2} the smoothing length of each
particle is allowed to vary in both time and space,
thus the full extent of SPH's powerful Lagrangian adaptivity is
 achieved. As shown in Fig.~\ref{fig: adiab} [bottom, right], the resolution
in high-density regions (e.g.~in the bow shock)
is naturally increased with large particle numbers, hence
smaller smoothing lengths ($h$$\lesssim$0.01\,pc), and
little computational time is wasted evolving voids, which
can be described by fewer particles with longer smoothing
lengths ($h$$\sim$0.08\,pc).

\begin{figure}
\centering
\includegraphics[scale=.56, angle=0,trim= 5 50 0 0, clip=true]{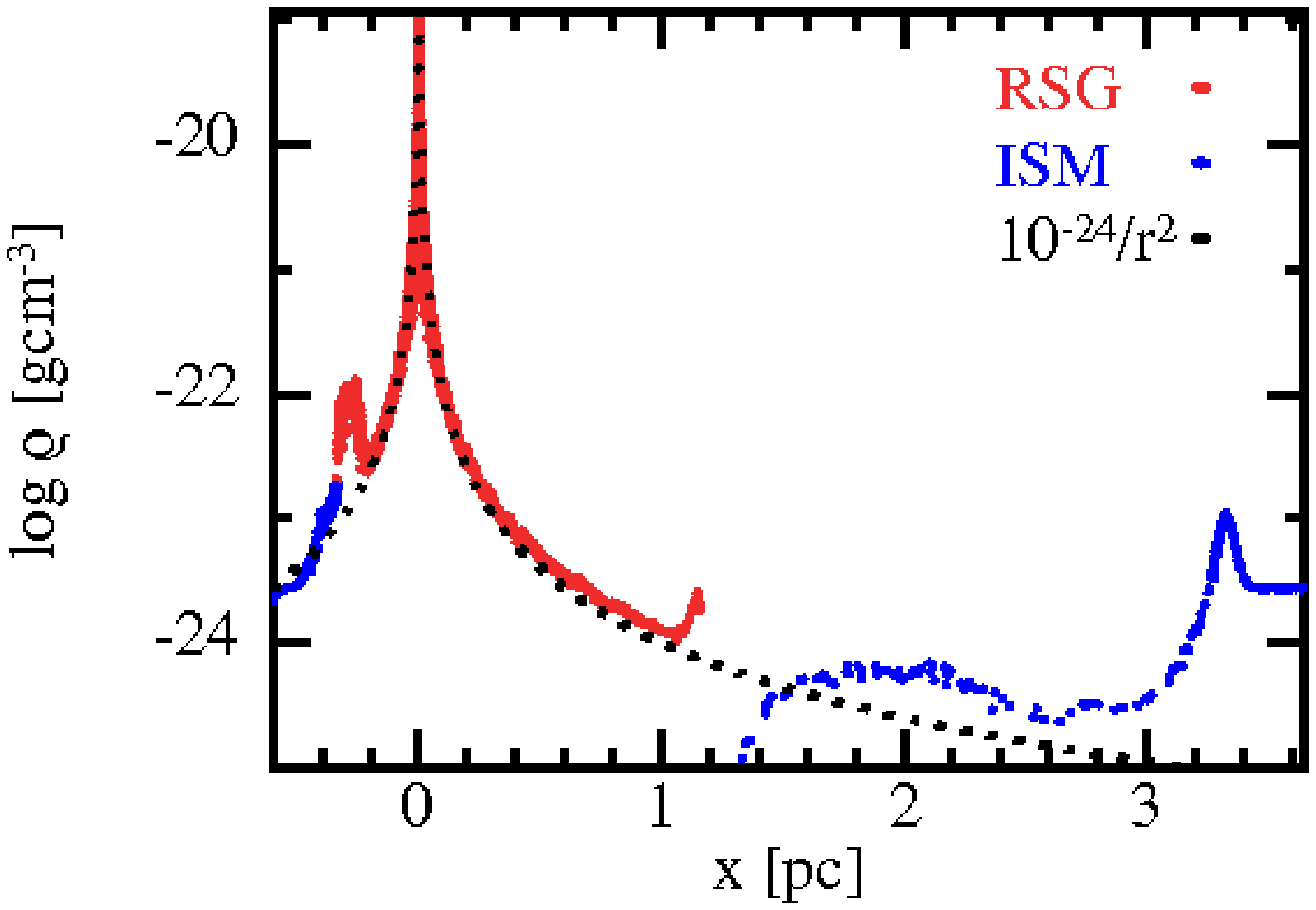}
\includegraphics[scale=.56, angle=0,trim= 5 0 0 0, clip=false]{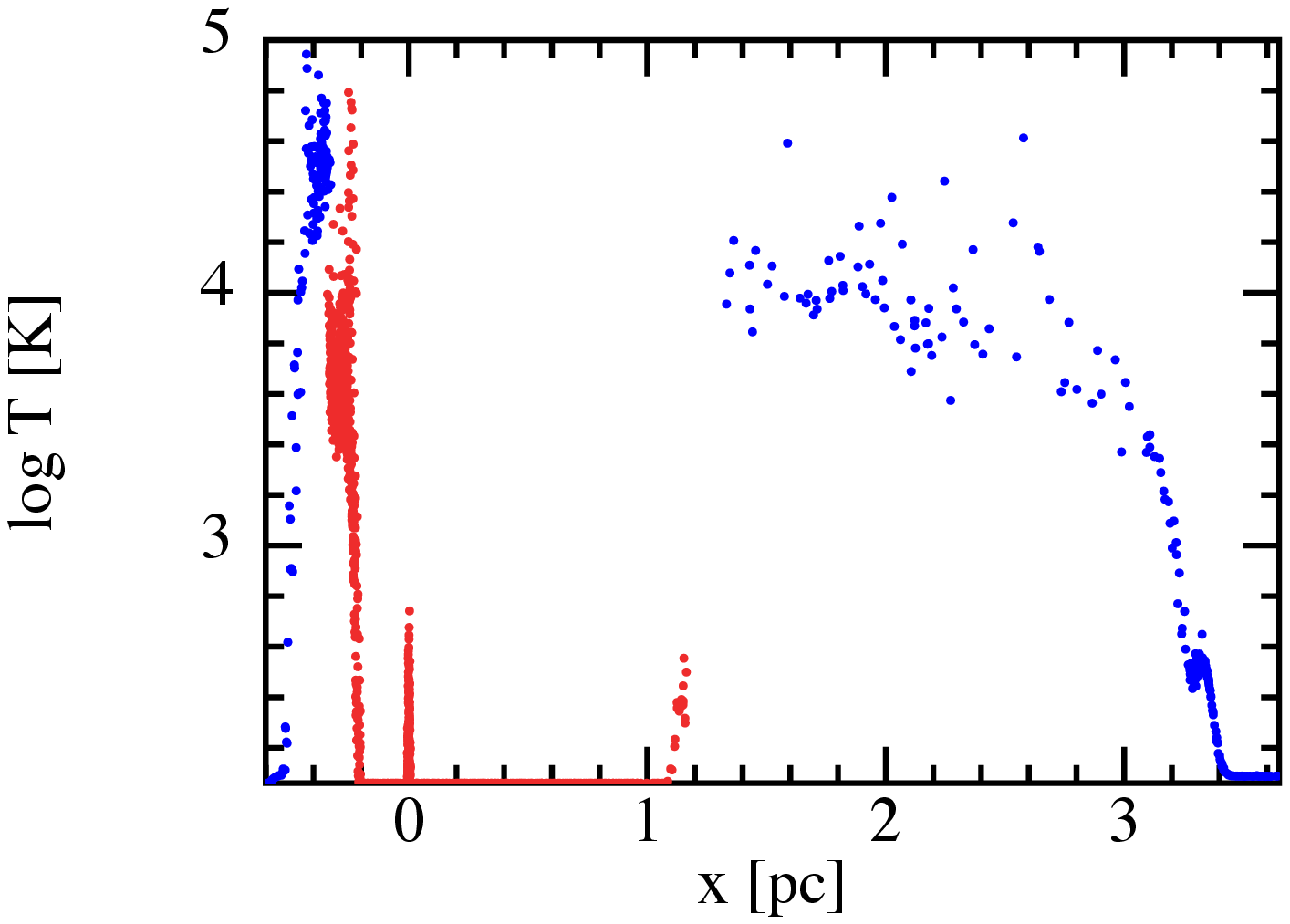}
\caption{Particle density [top] and temperature
[bottom] profiles for model B$_{\rm ad}$ in a thin
slice, $-0.03 < z < 0.03$ and $-0.03 < y < 0.03$,
along the $x$ axis. ISM and RSG particles are blue
and red, respectively.
\label{fig: adprof}}
\end{figure}

Unfortunately, this smoothing process also means that
the shock that results from the collision of the ISM
and the stellar wind is broadened over a few particle
smoothing lengths. To minimise this broadening, it is
essential that simulations employ as large a number of
particles as possible to achieve shorter  
smoothing lengths and thus sharper discontinuities.
In this simulation, we model the ISM with $\sim$$10^7$ particles
and the stellar wind with $\sim$$10^5$ particles, achieving a
typical smoothing length of 0.03 pc or better in the
regions of interest. More important, however, despite
the shock broadening, the post-shock fluid properties are
still described well; e.g.~as expected from the
Rankine-Hugoniot jump conditions, the density jump
is a factor of 4 (see Fig.~\ref{fig: adprof} [top]).
Assuming that all the kinetic energy in these winds
is thermalised, we can derive an estimate for the
expected temperature of the shocked gas,
 \be
T_{\rm s} = \frac{1}{2} \frac{ \mu m_{\rm H} v^2}{ k_{\rm B}} \,,
 \ee
where $k_{\rm B}$ is the Boltzmann constant.  For the
parameters assumed here, this equation yields
19\,000 K for the shocked stellar wind and 92\,000 K
for the shocked ISM temperature, so very similar to
the results obtained in the simulation (see
Fig.~\ref{fig: adprof} [bottom]).

\begin{figure}
\centering
\includegraphics[scale=.36, angle=270,trim= 0 30 0 10]{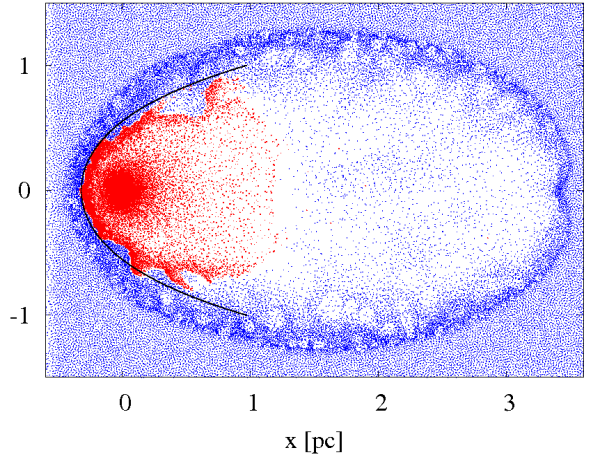}
\caption{Particles in the $xy$ plane with
the ISM in blue, and the stellar wind in red. The
analytic solution (Eq.~\ref{eq: shape})
is shown in black. 
\label{fig: adanal}}
\end{figure}

\subsection{Development of instabilities}
\label{sec: adinstab}

The shear produced by the relative motion of the
shocked ISM and shocked stellar wind regions (as shown in
the vector plot in Fig.~\ref{fig: adiab} [bottom, left]),
excites Kelvin-Helmholtz (K-H) instabilities at the
contact discontinuity. Utilising the prescriptions
described by \cite{Bri95} (Eqs.~on p55), we find that the 
K-H growth time scale is at least an order of magnitude smaller
than the one required for the Rayleigh-Taylor (R-T)
instability. Thus, it is the K-H rolls that are advected
downstream, and the characteristic R-T `fingers' that appear
in \cite{Bri95} (their Fig.~2 [top panel]) are absent from
our simulation. The main reason for the different behaviour
is that they utilise an order of magnitude higher mass-loss
rate and a much slower wind velocity, both of which lead to a
much stronger density contrast between the ISM and wind, which
favours shorter growth scales for R-T instabilities.

\begin{figure*}[!ht]
\centering
\includegraphics[scale=.58, trim= 0 90 30 0]{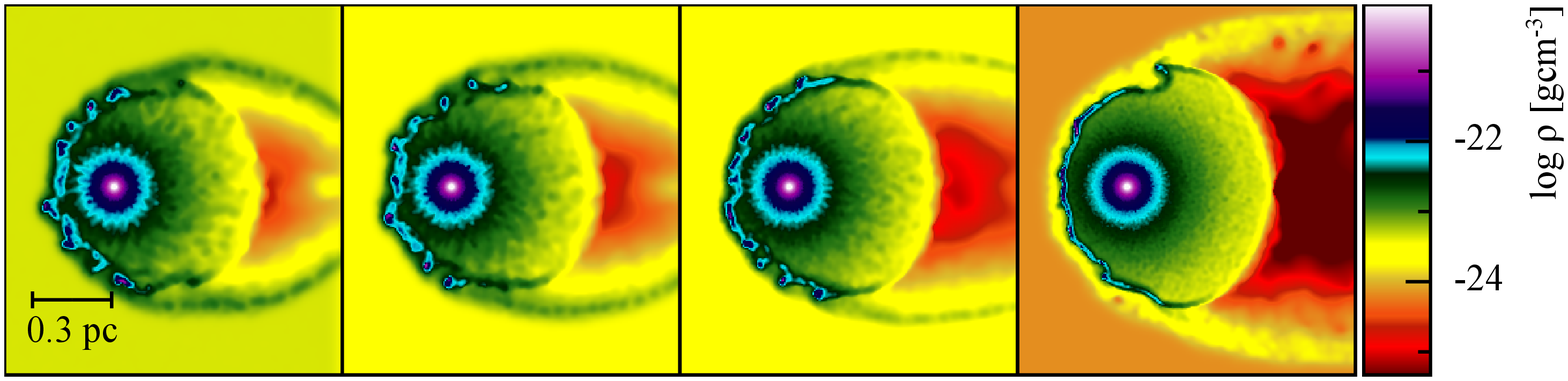}
\includegraphics[scale=.58, trim= 0 90 30 0]{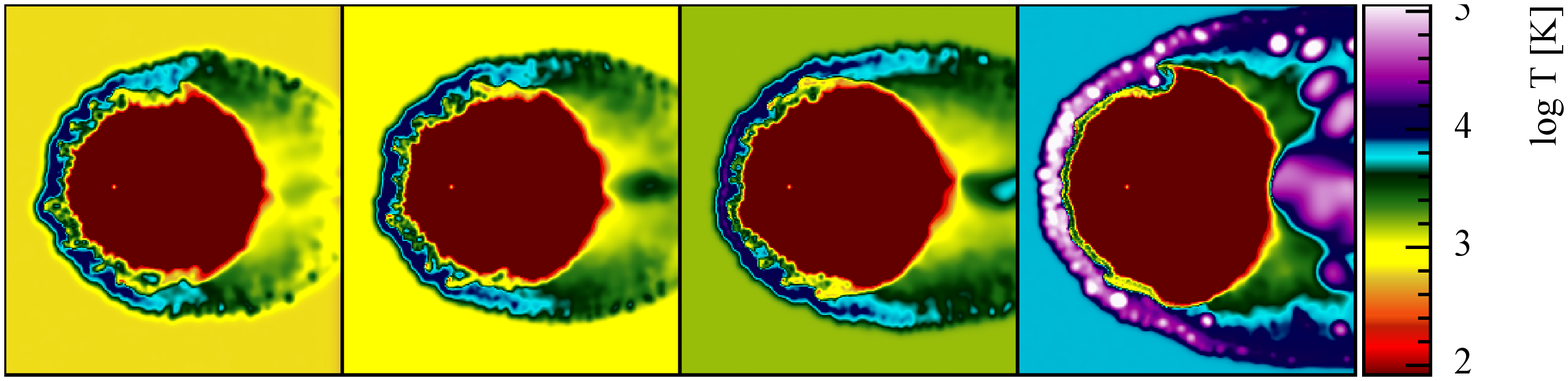}
\includegraphics[scale=.58, trim= 0 90 30 0]{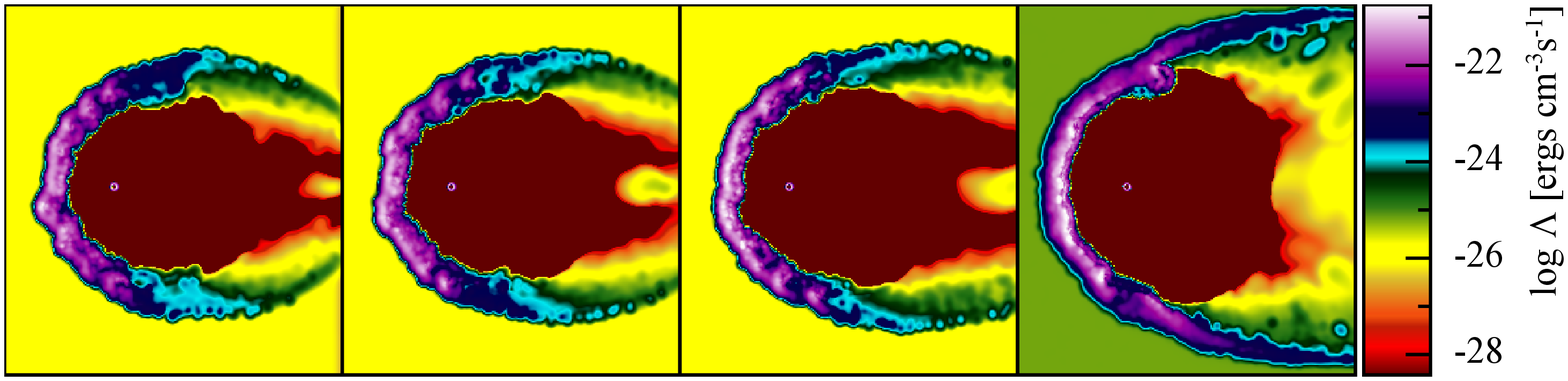}
\caption{Density [top],  temperature [middle], and total emissivity [bottom]
cross-sections in the symmetry plane for models A-D (left to right) after
32\,000 years. 
\label{fig: series}}
\end{figure*}

The development of K-H instabilities in our model may be
surprising. Previous studies, e.g.~\cite{Agertz07}, have found
that instabilities are suppressed in `standard'
SPH, owing to difficulty estimating gradients in regions
with strong density contrasts. Several methods have been
proposed to address this issue. One approach is to include
artificial conductivity, which acts as an additional pressure
across the contact discontinuity, facilitating mixing between
the two layers \citep{Pri08, Wad08}. However, at present the
formulation is not consistent with including self-gravity
and tends to be highly dissipative \citep{Val10}. The other
approach, which was the result of a detailed study by \cite{Val10}, 
is to minimise particle disorder, which prevents the `oily'
problem (the development of an artificial gap between the fluids)
and allows the gas to mix. They achieve this by using a higher order
smoothing kernel to prevent particle clumping, but this kernel does not
conserve energy as well as the cubic spline kernel. Given the caveats
to the above solutions, we opted to use glass initial conditions
instead, i.e. to start from a very smooth particle distribution with
minimal particle disorder. Thus, strong K-H instabilities do develop,
despite the factor of 10 in the density contrast.

\subsection{Comparison to analytic models}

Although the contact discontinuity is distorted by the
K-H instability, the overall shape of this surface is
in fairly good agreement with the one predicted by the
\cite{Wil96} solution (Eq.~\ref{eq:  ram}) for 
small polar angles ($\theta$),  and the stand-off distance matches 
the position of the contact discontinuity well (see Fig.~\ref{fig: adanal} [top]).
In the thin-shell limit, the ISM and wind material are assumed 
to cool instantaneously and mix fully. In contrast, the forward 
and reverse shock of the adiabatic model have finite width, 
$\sim$0.2\,pc, and as expected, there is little mixing between 
these two layers of gas.  Overall, the adiabatic model demonstrated 
that the set-up and code can achieve results that are consistent with both 
theoretical expectations and  previous studies.

\section{Models with cooling}
\label{sec: cool}

The models that include radiative cooling exhibit a
 global structure that is similar to the structure of the adiabatic
model. In Fig.~\ref{fig: series}, we show the density, temperature,  
and emissivity for models A-D. The four regions  described above  
can easily be distinguished: the unshocked ISM 
and similarly unshocked,  free-flowing stellar wind, separated by 
a layer of hot, shocked ISM (the forward shock)  and shocked stellar 
wind (the reverse shock). Based on their flow characteristics and 
morphology, models A-D can be grouped into two classes: the 
`slow' models, A-C, with ISM densities $\gtrsim$1\,cm$^{-3}$ 
and peculiar velocities $\lesssim$40\,km\,s$^{-1}$, and 
the `fast' model, D, with an ISM density of 0.3 cm$^{-3}$
and a stellar velocity $\sim$73\,km\,s$^{-1}$.

\begin{figure}
\centering
\includegraphics[scale=.44, angle=0,trim= 5 50 9 0, clip=true]{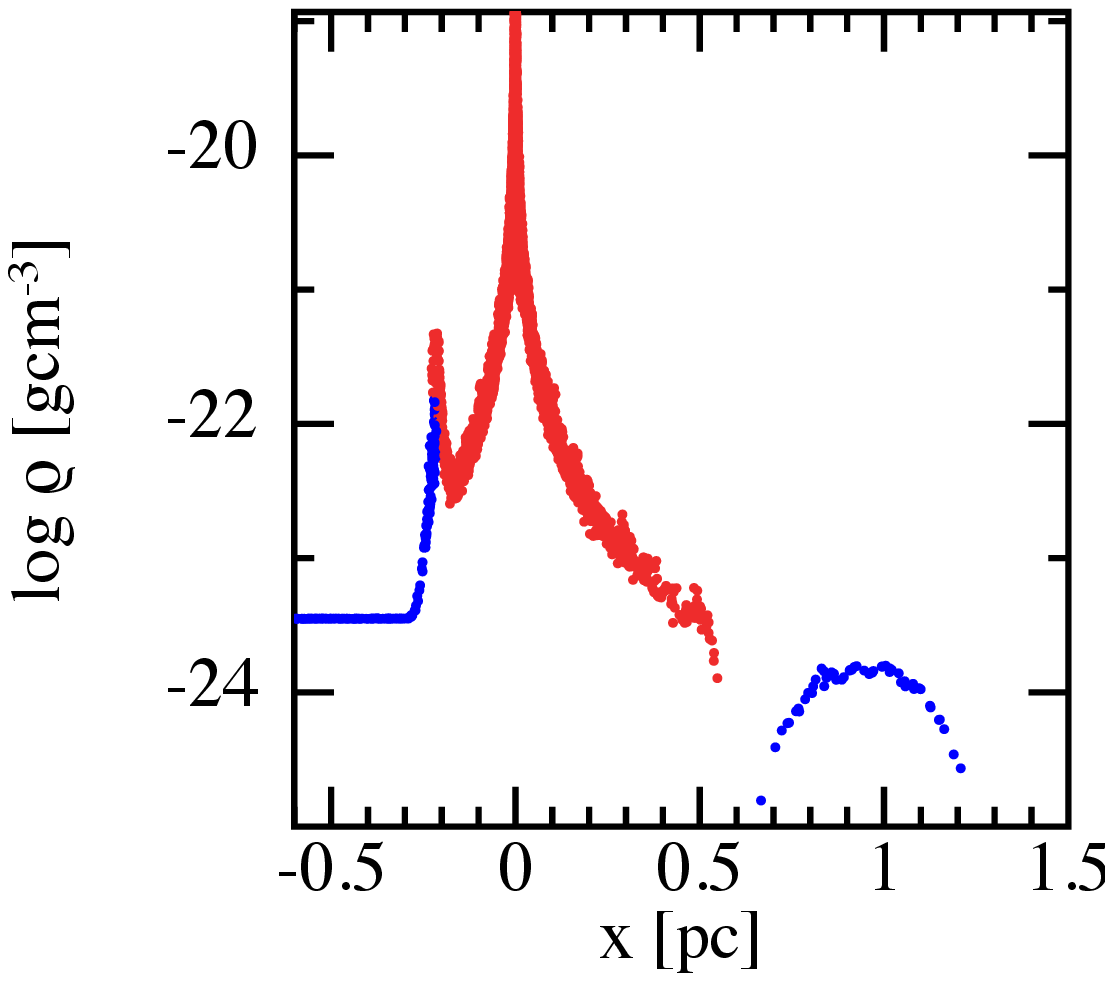}
\includegraphics[scale=.44, angle=0,trim= 88 50 0 0, clip=true]{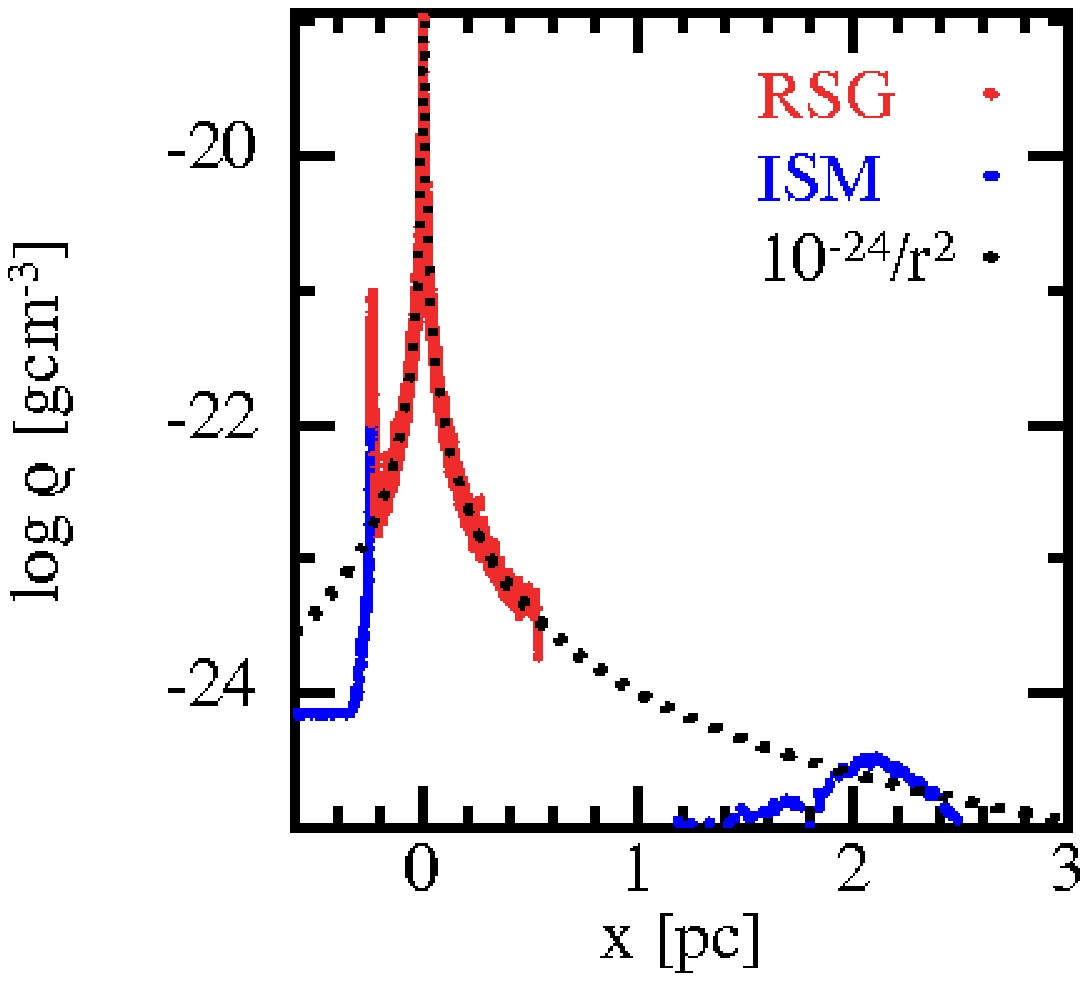}\\
\includegraphics[scale=.44, angle=0,trim= 7 50 7 10, clip=true]{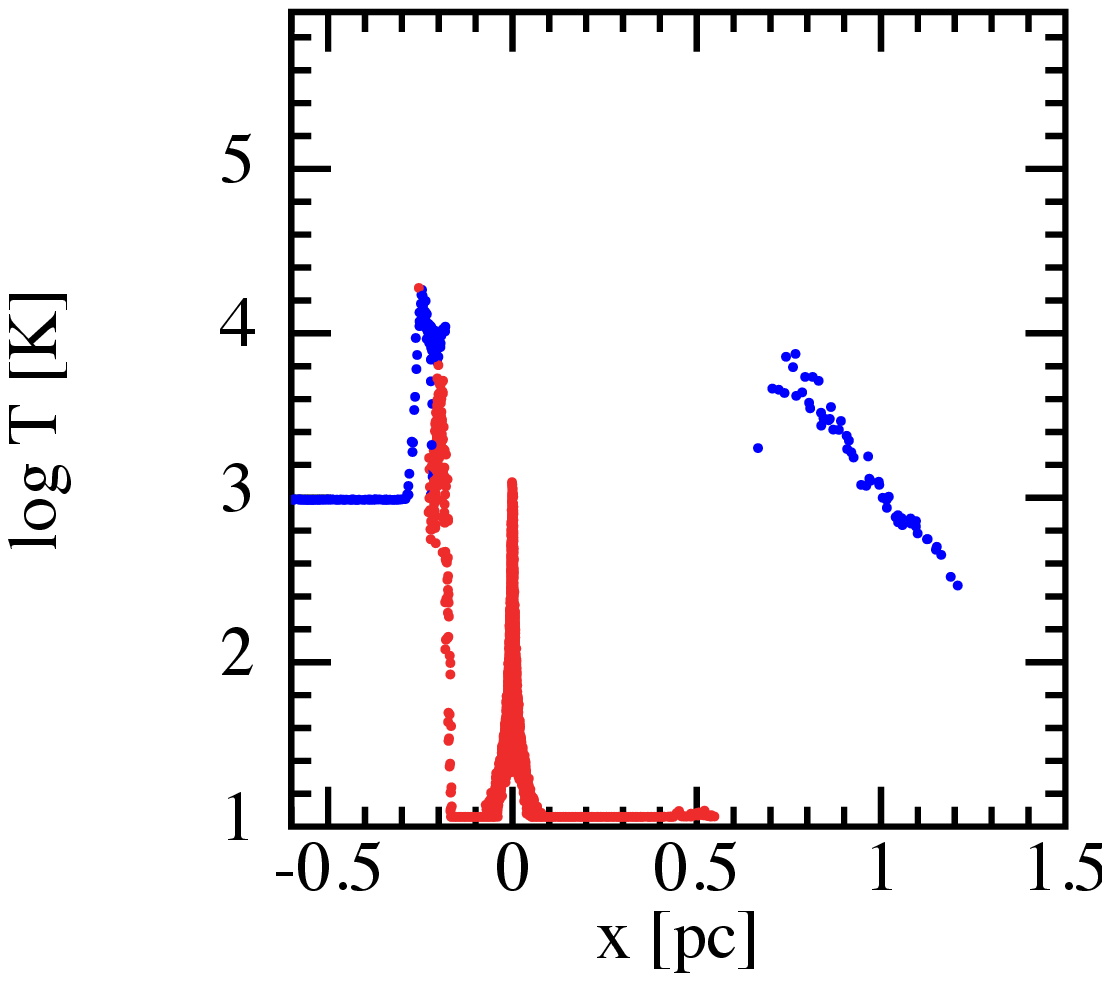}
\includegraphics[scale=.472, angle=0,trim= 111 62 8 10, clip=true]{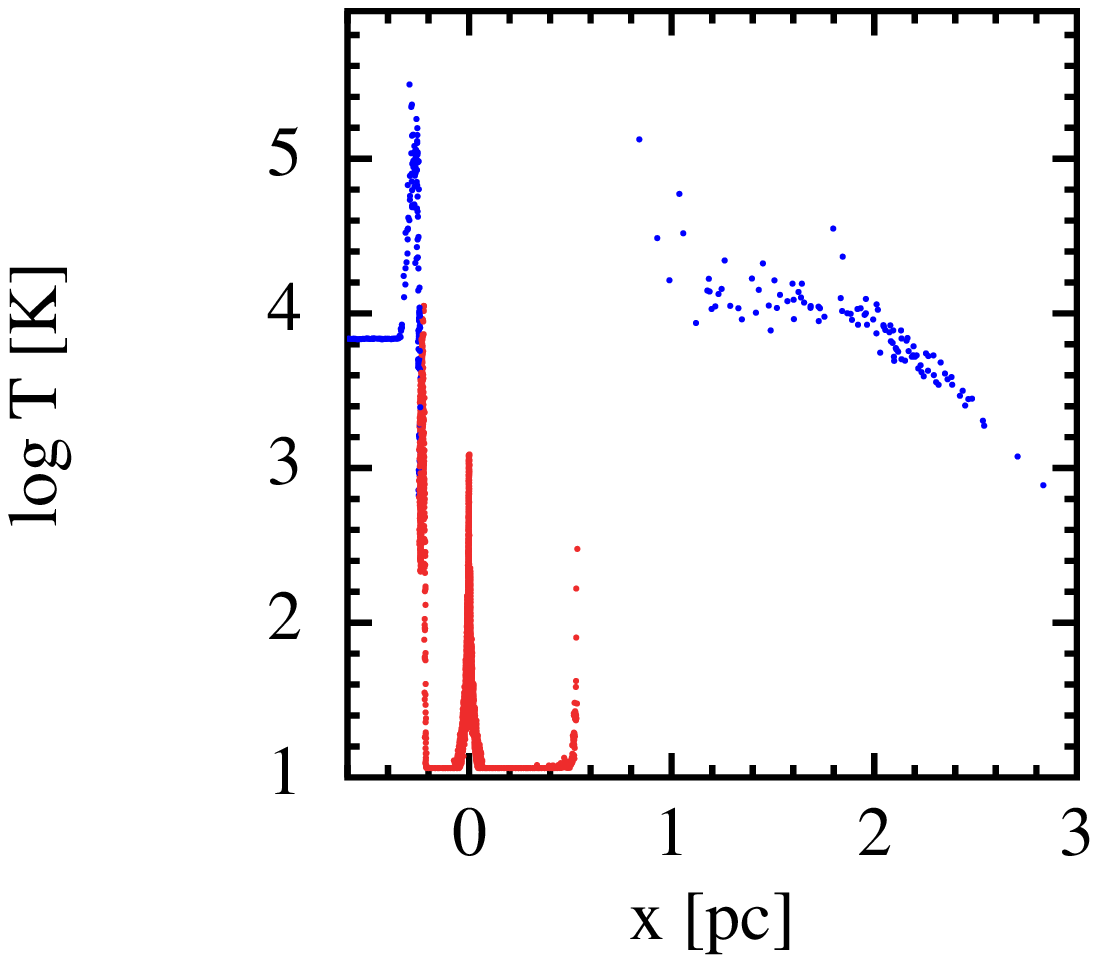}\\
\includegraphics[scale=.44, angle=0,trim= 5 0 9 10, clip=true]{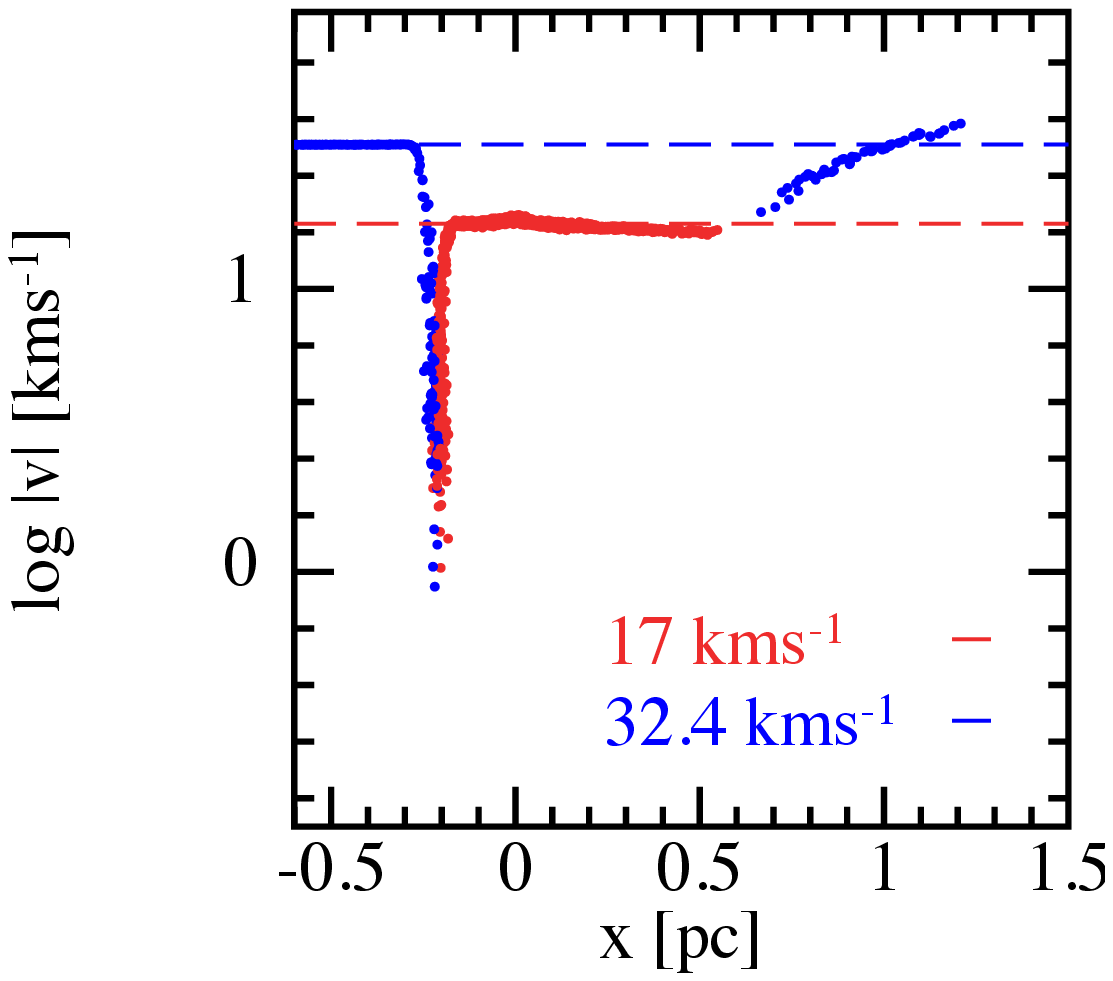}
\includegraphics[scale=.44, angle=0,trim= 88 0 0 10, clip=true]{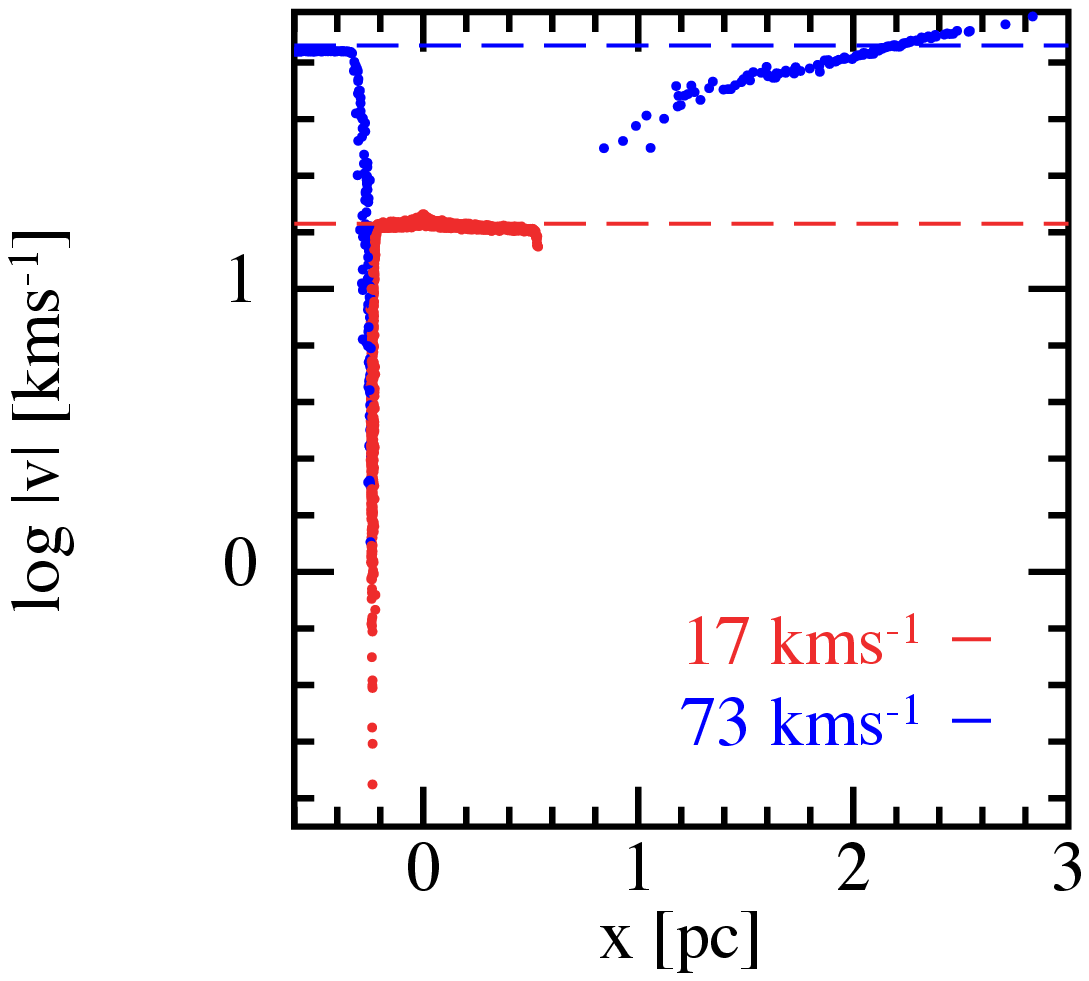}
\caption{Particle density [top], temperature [middle], and velocity
[bottom] profiles for models B [left] and D [right] in a thin slice, $-0.02 < z < 0.02$ and
$-0.02 < y < 0.02$, along the $x$ axis. In all plots, the ISM and RSG particles are blue
and red, respectively.
\label{fig: nprof}}
\end{figure}

\begin{figure}
\centering
\includegraphics[scale=.34, angle=270,trim= 0 20 0 20]{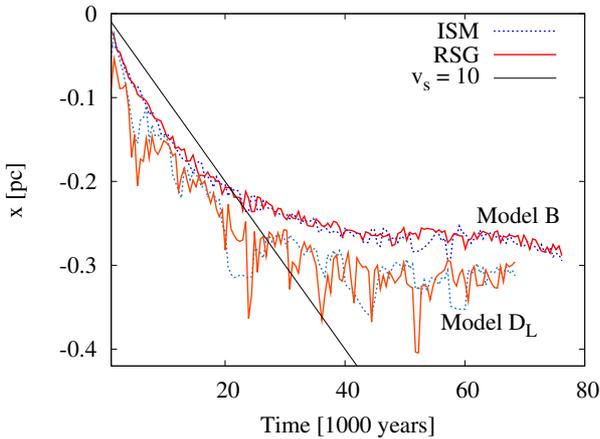}
\caption{Same as Fig.~\ref{fig: adshockprof} except for models B and D$_{\rm L}$.
\label{fig: shockpos}}
\end{figure}

The fluid properties of the slow models show significant departures 
from the adiabatic case (see Fig.~\ref{fig: nprof}); the most obvious being that 
the radiative cooling results in lower post-shock temperatures, e.g.~$T_{\rm s}$ 
 for model B is an order of magnitude lower than in model B$_{\rm ad}$  
 for both the shocked ISM and RSG wind (see Fig.~\ref{fig: nprof} [top, middle]). 
 Furthermore, by reducing the thermal pressure of the gas, the strong 
 cooling also enables further compression in the shock, resulting in greater 
 post-shock densities (see Fig.~\ref{fig: nprof} [top, left]) than in model B$_{\rm ad}$. 
  Consequently, the time scale for the growth of R-T instabilities is reduced and, similar 
  to the models of \cite{Bri95} (their Fig.~2.), R-T `fingers' develop 
  faster than the K-H rolls that characterised the adiabatic case 
  (see Secs.~\ref{sec: adinstab} and \ref{sec: instab}).   
Owing to the lower thermal pressure and mixing of the ISM
and RSG gas, both the bow shock and the tail are narrower in the 
slow models; e.g.,~the model B width is 
approximately half that of model B$_{\rm ad}$. 

The forward shock of the fast model shows similar characteristics to 
model $B_{\rm ad}$;  the combination of a large stellar velocity 
through the ISM, and an initially hot and lower density ISM, means that 
the fast-moving gas does not have enough time to cool beyond its initial condition. 
The fast model achieves the same shocked ISM temperature, $T_{\rm ISM}$,  
as the adiabatic case in which the star was moving approximately 
half as fast (see Fig.~\ref{fig: nprof} [middle panel]).  Furthermore, this 
hot gas also flows along the bow shock  and eventually settles in the 
core of the tail along the $x$ axis, as it did in the adiabatic model.  However,
the shocked RSG wind temperature is approximately 1\,000 K,  
much lower than the adiabatic counterpart and similar to the value for the slow 
models.

The evolution of the forward and reverse shock positions for models 
B and D$_{\rm L}$, i.e.~the position of maximum temperature along the 
symmetry axis,  is shown in Fig.~\ref{fig: shockpos}. Not only are the oscillations 
in the shock position less violent than the adiabatic case, an equilibrium
shock position is attained more rapidly for the models with radiative cooling, 
after only 40\,000 years. The strong cooling in both the forward and reverse shocks in 
the slow models results in the excitation of R-T type instabilities (see below), which causes the 
gas to mix,  hence the small difference between the positions of the 
RSG and ISM components. In contrast, there is little mixing between the 
two layers in the fast model's bow shock,  the forward shock has significant width,  
thus the hottest gas extends much further into the impinging ISM flow. 

\begin{figure}
\centering
\includegraphics[scale=.37, angle=270,trim= 0 0 0 72, clip=true]{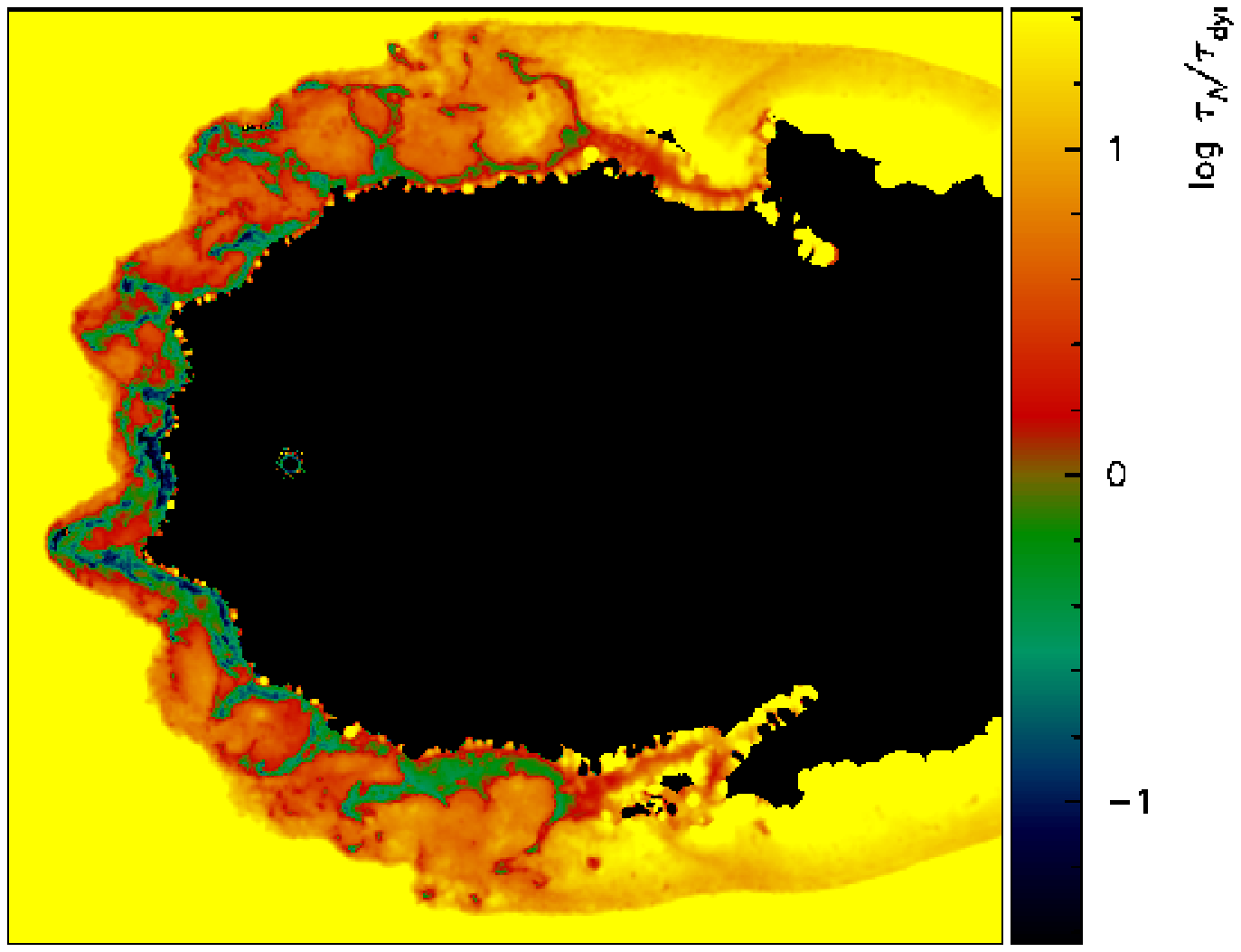}
\includegraphics[scale=.37, angle=270,trim= 0 0 0 72, clip=true]{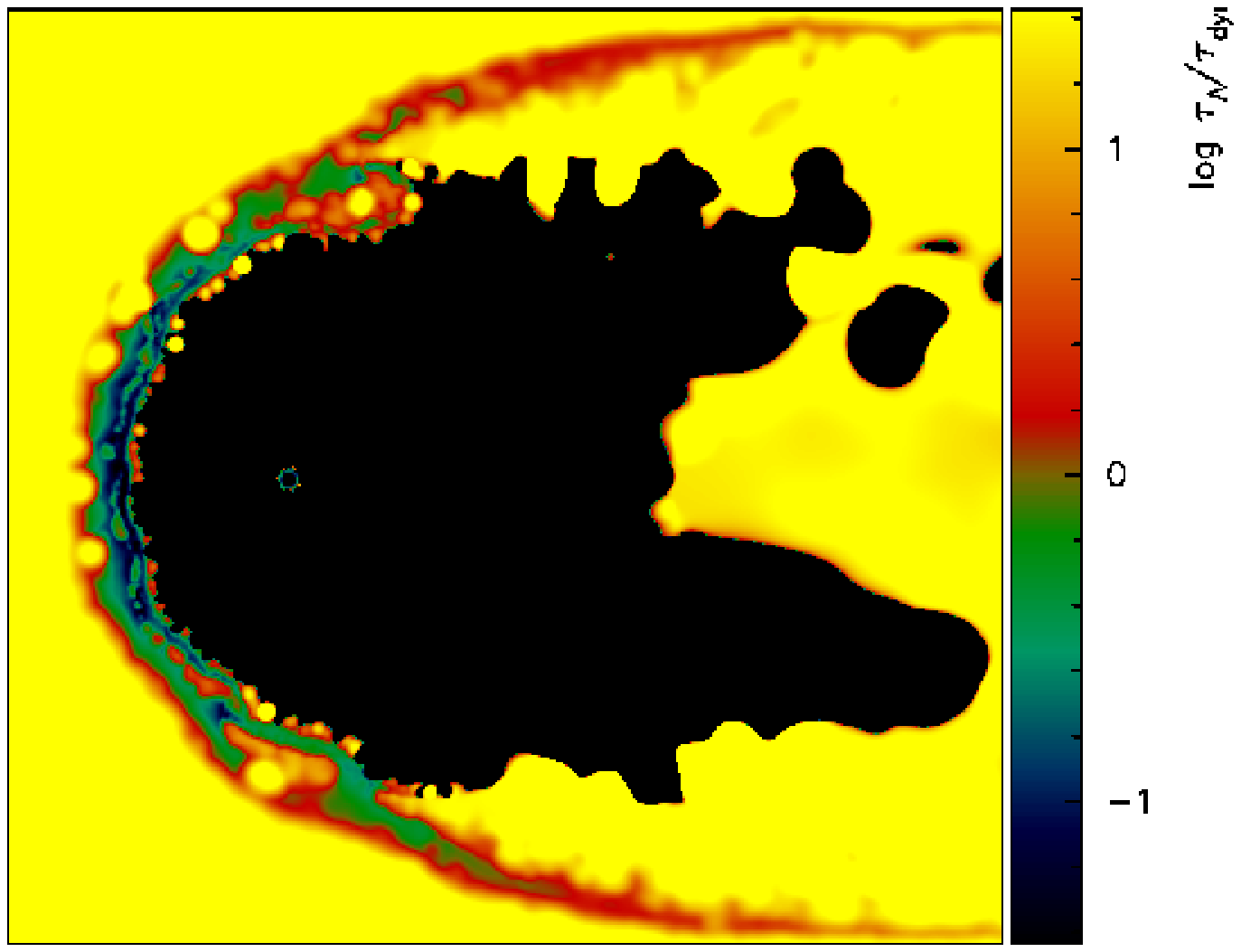}
\includegraphics[scale=.3, angle=0,trim= 0 0 0 175, clip=true]{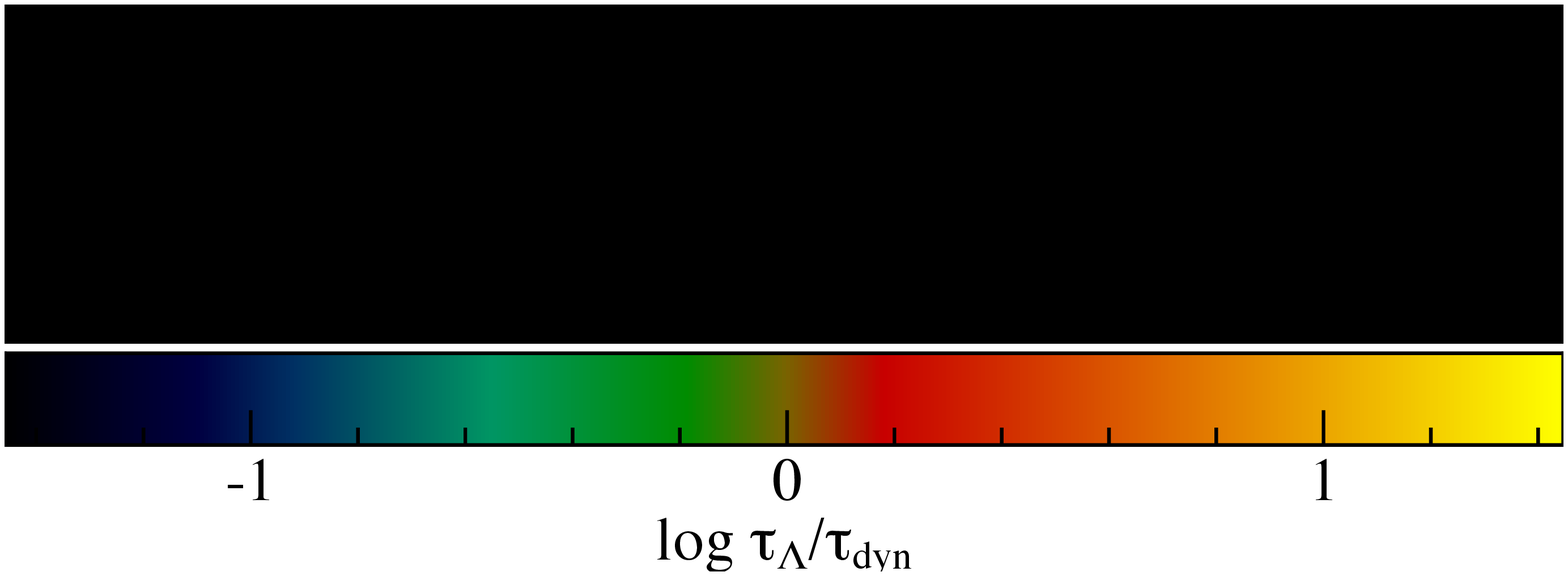}
\caption{Ratio of the cooling and dynamical
time scales for models A$_{\rm H}$ [left] and D [right]. If the cooling time
scale is much greater than the dynamical time scale,
i.e.~$\tau_\Lambda >> \tau_{\rm dyn}$, the gas behaves
 adiabatically, and if $\tau_\Lambda << \tau_{\rm dyn}$, the
 gas is approximately isothermal.
\label{fig: tau}}
\end{figure}

The typical cooling time scale in the bow shock for $x < 0$
is $\sim$10\,000 years, and is of the order of both
the characteristic dynamical time scale for the shocked ISM,
 $R_{\rm SO} / v_*$, which is $\sim$9\,400, 8\,300, 6\,800, 3\,700 
 years for models A-D, respectively, and the characteristic time
 scale for the shocked wind, $R_{\rm SO} / v_{\rm w}$, which is
 approximately 17\,200 years. In Fig.~\ref{fig: tau} we
 compare the radiative cooling and characteristic dynamical 
 time scales of the gas:
\be
\tau_{\Lambda} = \left(\Lambda/\rho \epsilon\right)^{-1} \,,
\ee
and 
\be
\tau_{\rm dyn} =  R_{\rm SO} / v \,,
\ee
respectively. For regions with $\tau_\Lambda << \tau_{\rm dyn}$, e.g.~the bow
shock arc where the gas is strongly decelerated, the gas cools strongly and 
behaves almost isothermally. The gas in the tail has a much longer 
cooling time scale, $\sim$30\,000 years (much greater than the cooling 
time scale for the shocked ISM) resulting in a largely adiabatic flow there.

\subsection{Bow shock morphology}
\label{sec: morph}

\begin{figure}
\centering
\includegraphics[scale=.37, angle=270,trim= 0 30 -20 20]{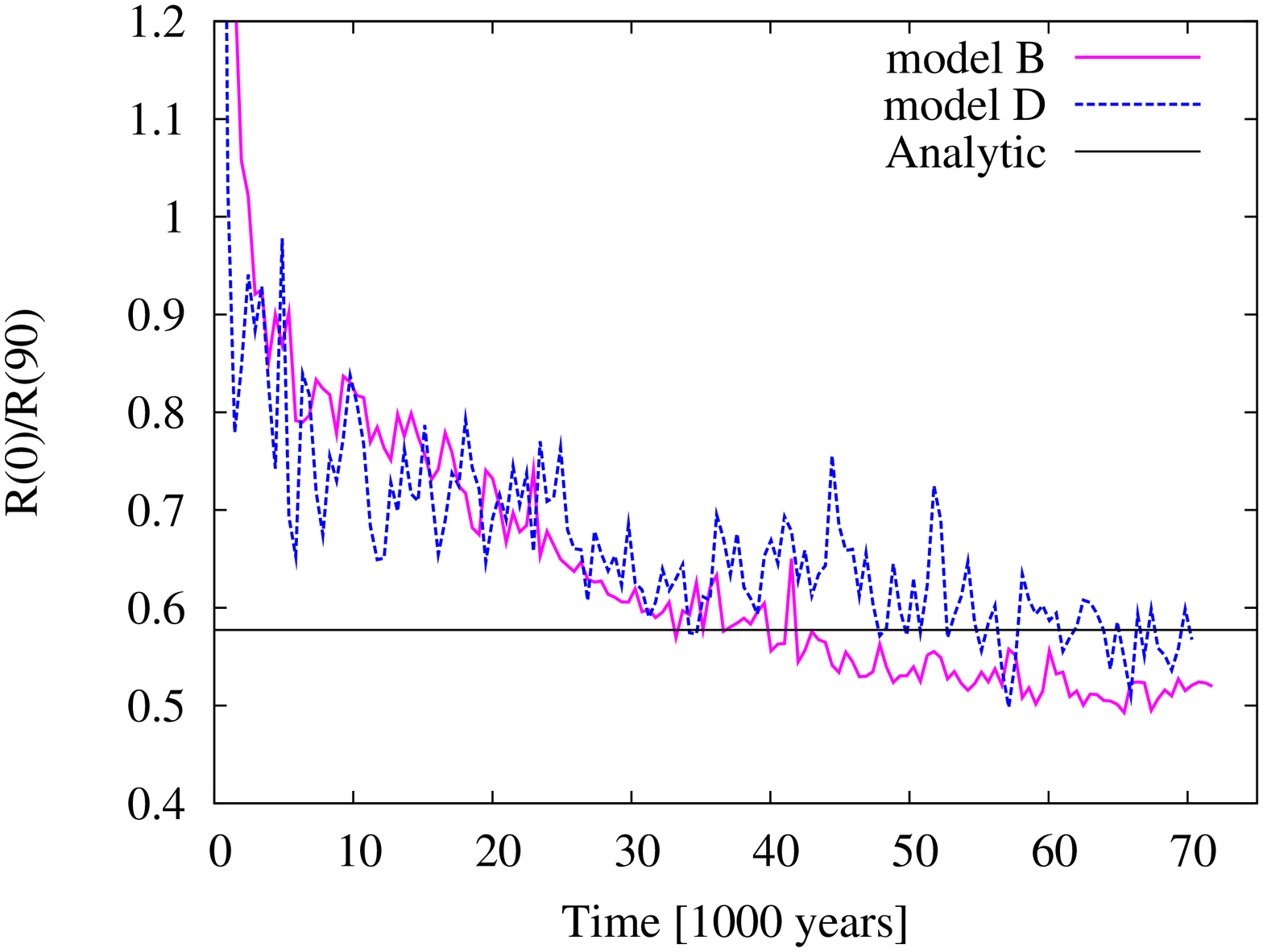}
\includegraphics[scale=.35, angle=0,trim= 0 0 170 0, clip=true]{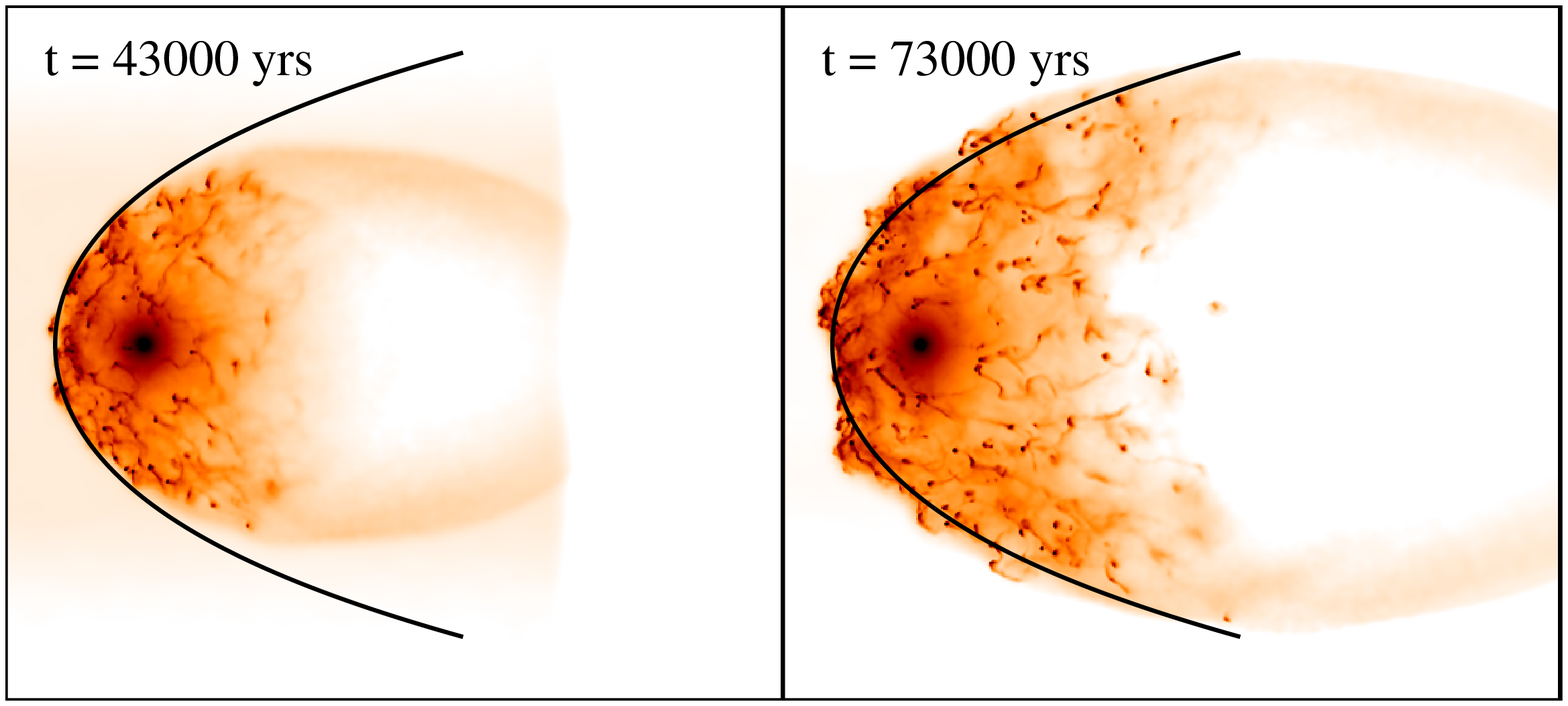}
\caption[figure11]{[Top] Ratio of $R(0^\circ)/R(90^\circ)$ as 
a function of time for models B (magenta) and D (blue) compared to the 
analytic value in black. 
[Bottom]  Hydrogen column density (on a logarithmic scale) 
for model B after 43\,000 years [left] and 73\,000 years [right]. The analytic
solution for the shape of the bow shock, given by Eq.~\ref{eq: shape},
is shown in black.\protect\footnotemark 
\label{fig: shape}}
\end{figure}
\footnotetext{Movies demonstrating the evolving bow shock shape are included in the electronic version.}

The shape of the bow shock changes with time: it is initially circular and becomes 
increasingly `parabolic' until it reaches a steady state, at which point the global 
morphology is described by Eq.~\ref{eq: shape}, the analytic \cite{Wil96} solution.  
The ratio of the shock position at angles $\theta=0^\circ$ and  $\theta=90^\circ$, 
$r_{\rm s}(0^\circ)/r_{\rm s}(90^\circ)$, as a function of time, is shown in Fig.~\ref{fig: shape} 
[top]. It takes several dynamical time scales for the bow shock
 to achieve the steady state value $R(0^\circ)/R(90^\circ)$=$1/\sqrt{3}$. The wind takes 
 at least $t = R(\theta)/v_{\rm w}$ to reach the thin-shell position, ignoring the drag due to the 
 moving environment, thus it takes $\sim$16\,000 years for 
 $\theta=0^\circ$ ($R(0^\circ)$=$R_{\rm SO}$), while for $\theta=90^\circ$ ($R(90^\circ)$ = $\sqrt{3} R_{\rm SO}$) 
it takes $\sim$28\,000 years. In reality, the above time scales are longer. The drag is maximal for 
$\theta=0^\circ$,  so it takes about 30\,000 years to reach $R_{\rm SO}$ 
(e.g.~Fig.~\ref{fig: shockpos}), and coincidentally it takes about the same time 
to reach the solution position for $\theta=90^\circ$ since the distance traversed by 
the wind is greater, but the drag is reduced. 

It takes even longer for a steady state to be achieved for larger angles, $\theta > 90^\circ$. 
In Fig.~\ref{fig: shape} [bottom] we compare the thin-shell solution with model B at 43\,000 
years and at 73\,000 years. 
Although the simulations coincide with the analytic solution in the latter, as expected, the 
agreement for large angles of $\theta$ at earlier times is not good, and  the same effect is  
found for the fast model. The shape of the bow shock tail also changes with time, and younger 
systems show strong curvature in the bow shock tail. This is the case for both the slow 
and fast models; however, for the latter the curvature is quickly left far downstream due 
to the large space motion.

\begin{figure*}
\centering
\includegraphics[scale=.38, angle=0,trim= 50 0 50 0, clip=true]{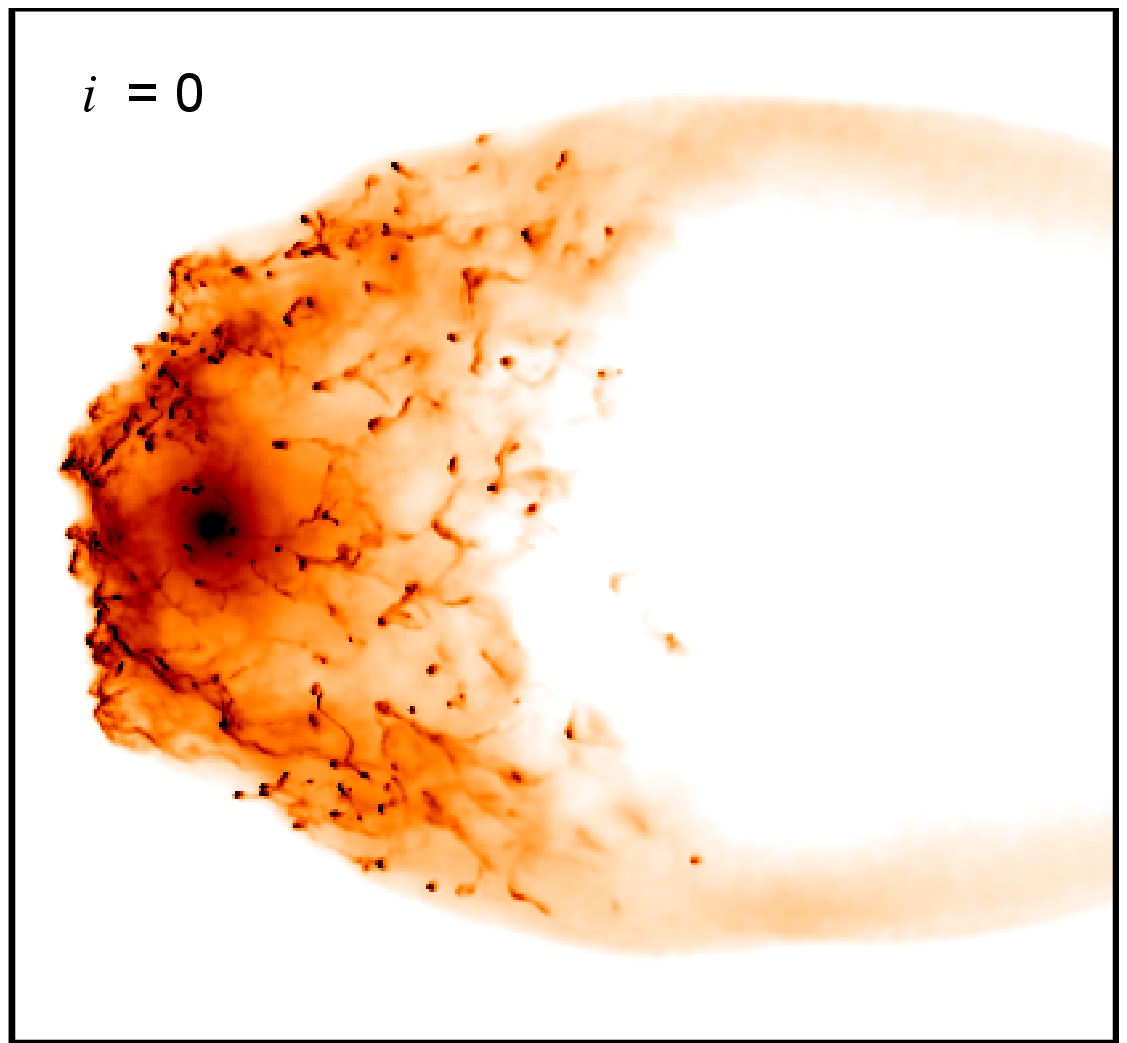}
\includegraphics[scale=.38, angle=0,trim= 50 0 50 0, clip=true]{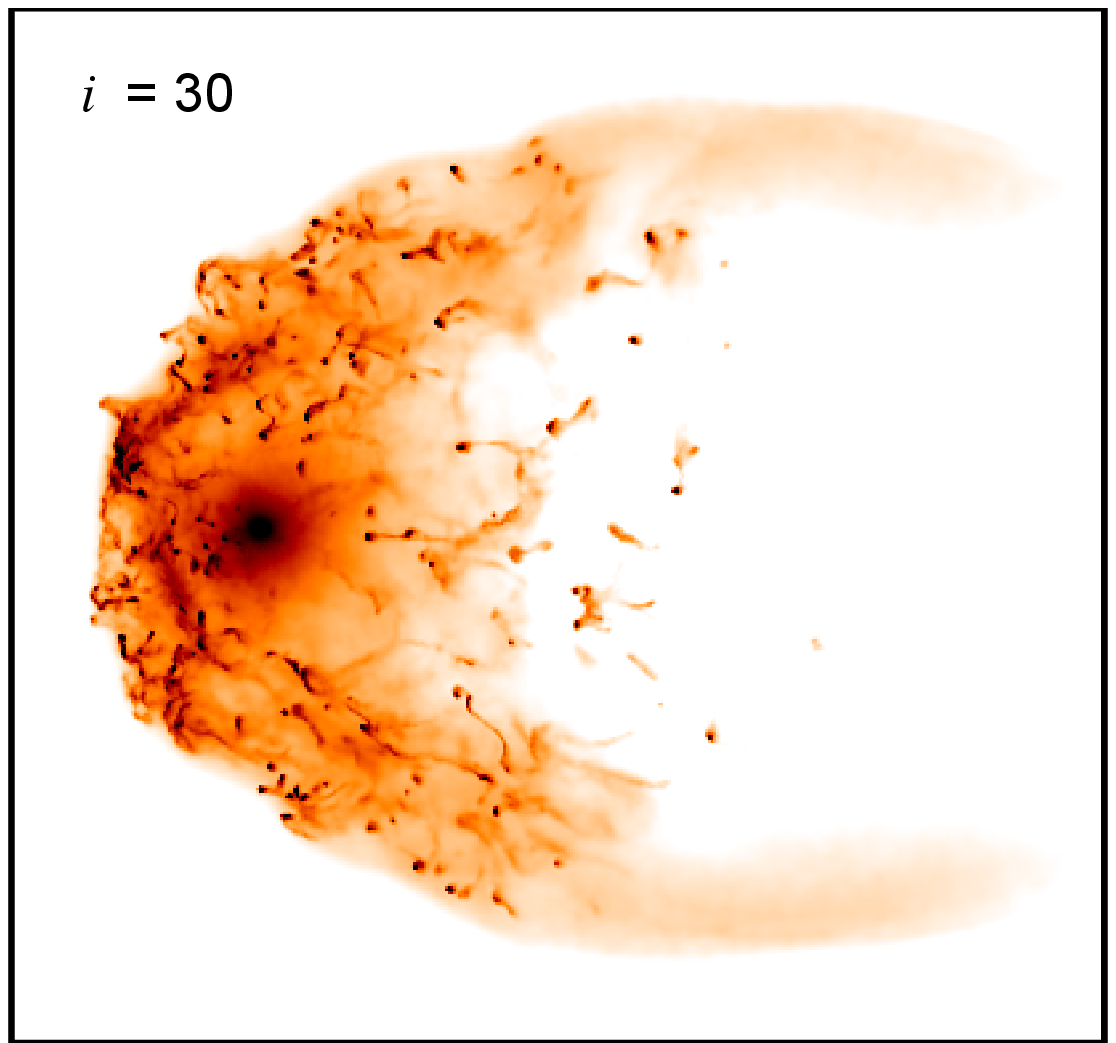}
\includegraphics[scale=.38, angle=0,trim= 45 0 45 0, clip=true]{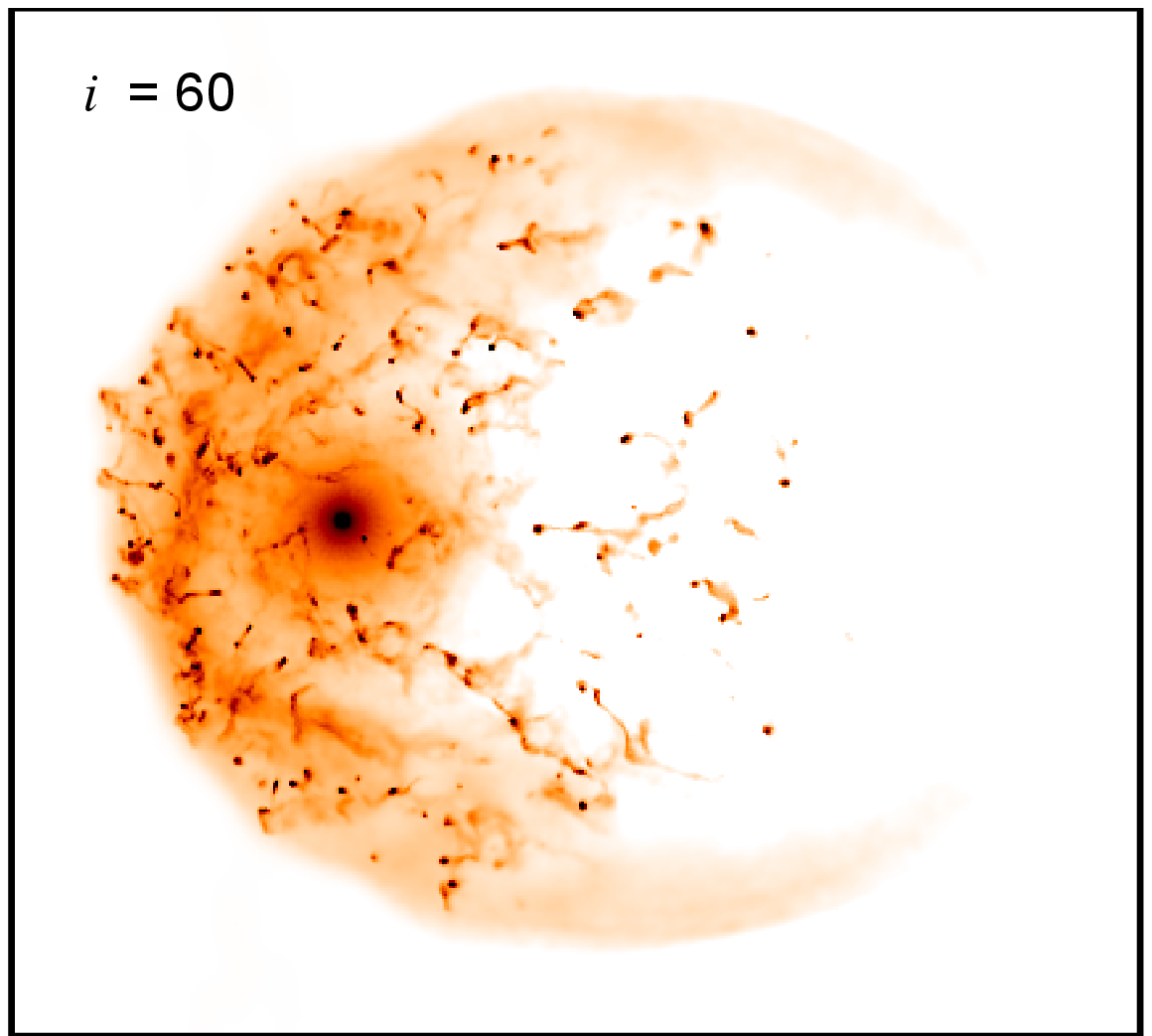}
\includegraphics[scale=.38, angle=0,trim= 42 0 45 0, clip=true]{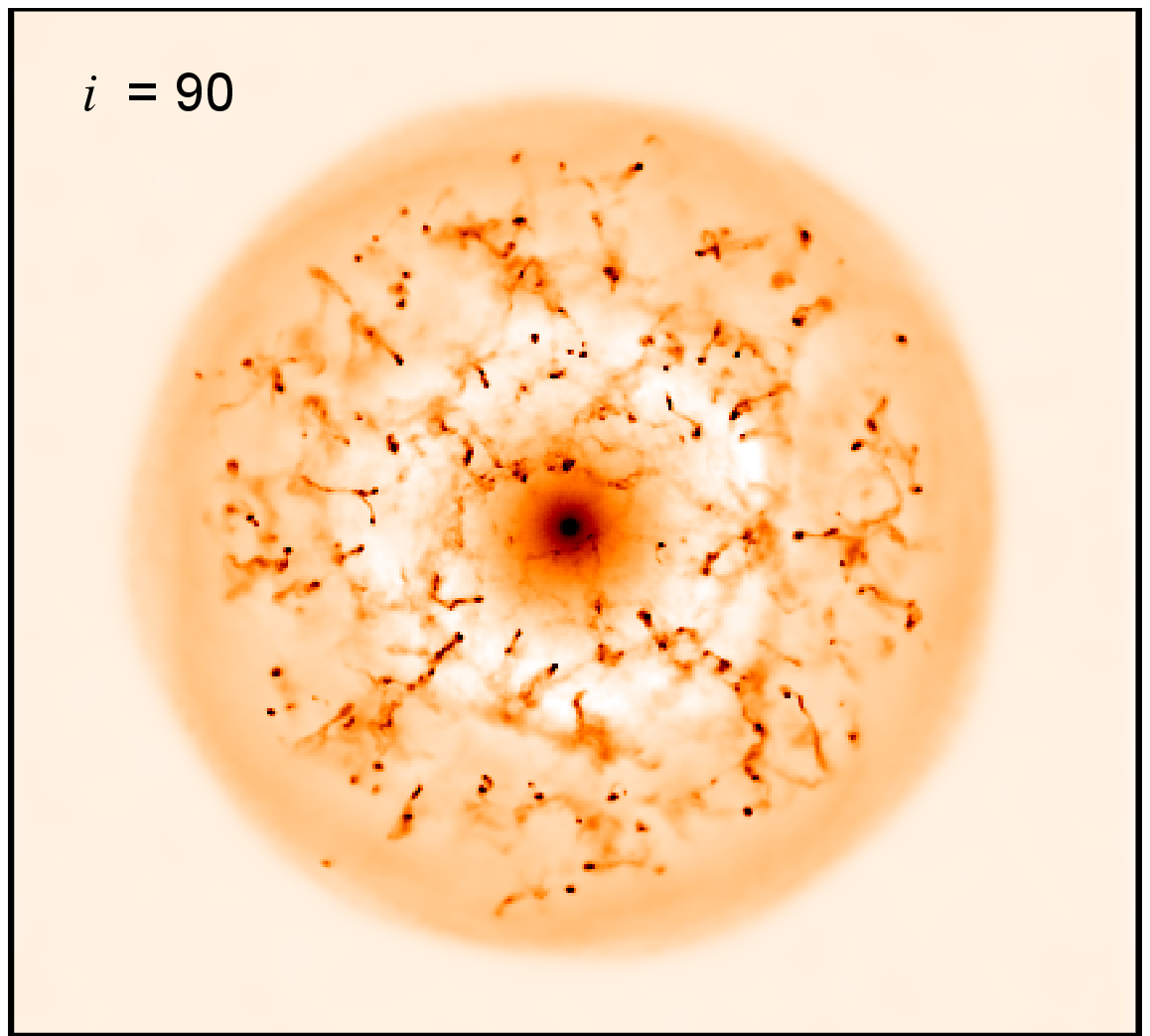}
\includegraphics[scale=.37, angle=0,trim= 45 0 45 -7 , clip=true]{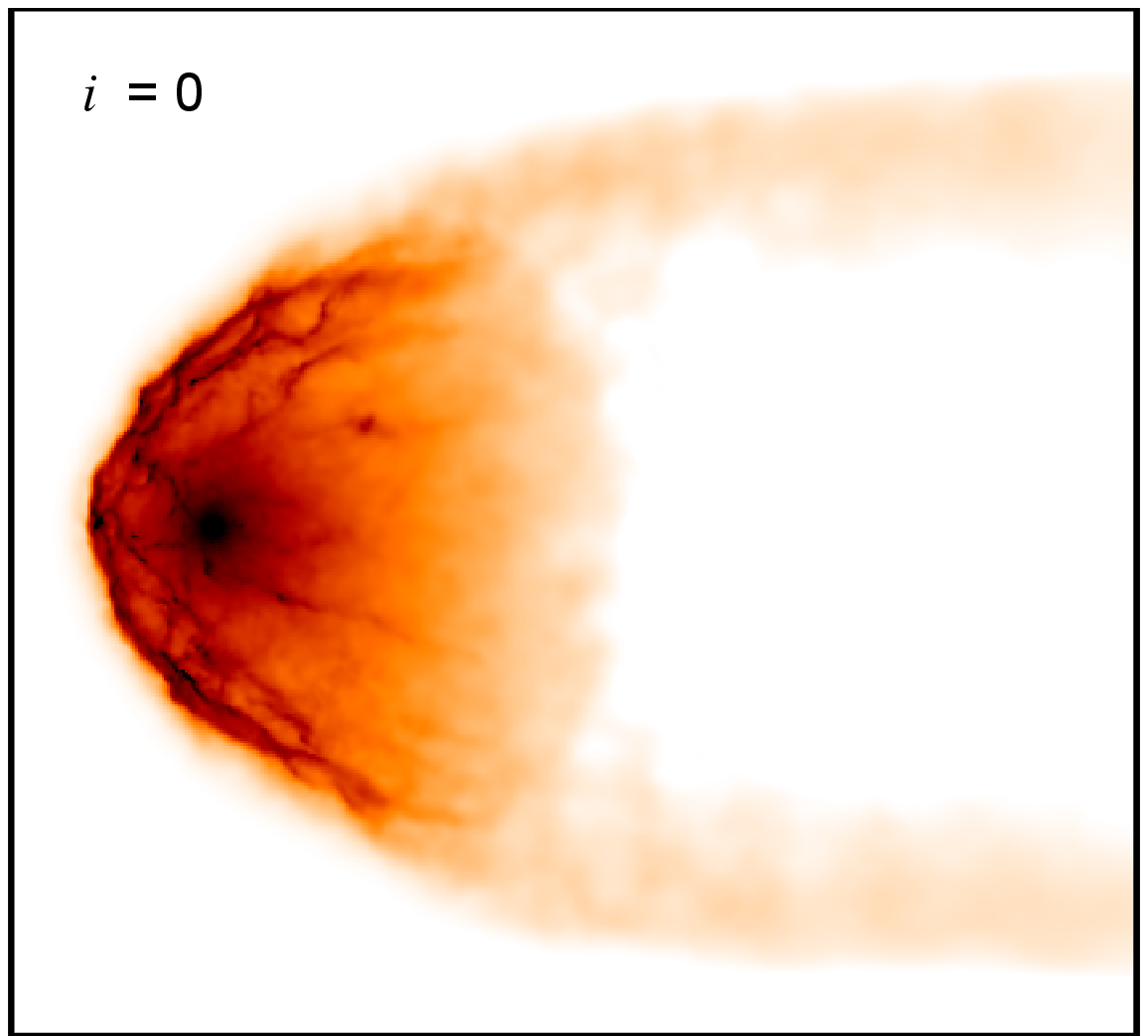}
\includegraphics[scale=.37, angle=0,trim= 45 0 45 0, clip=true]{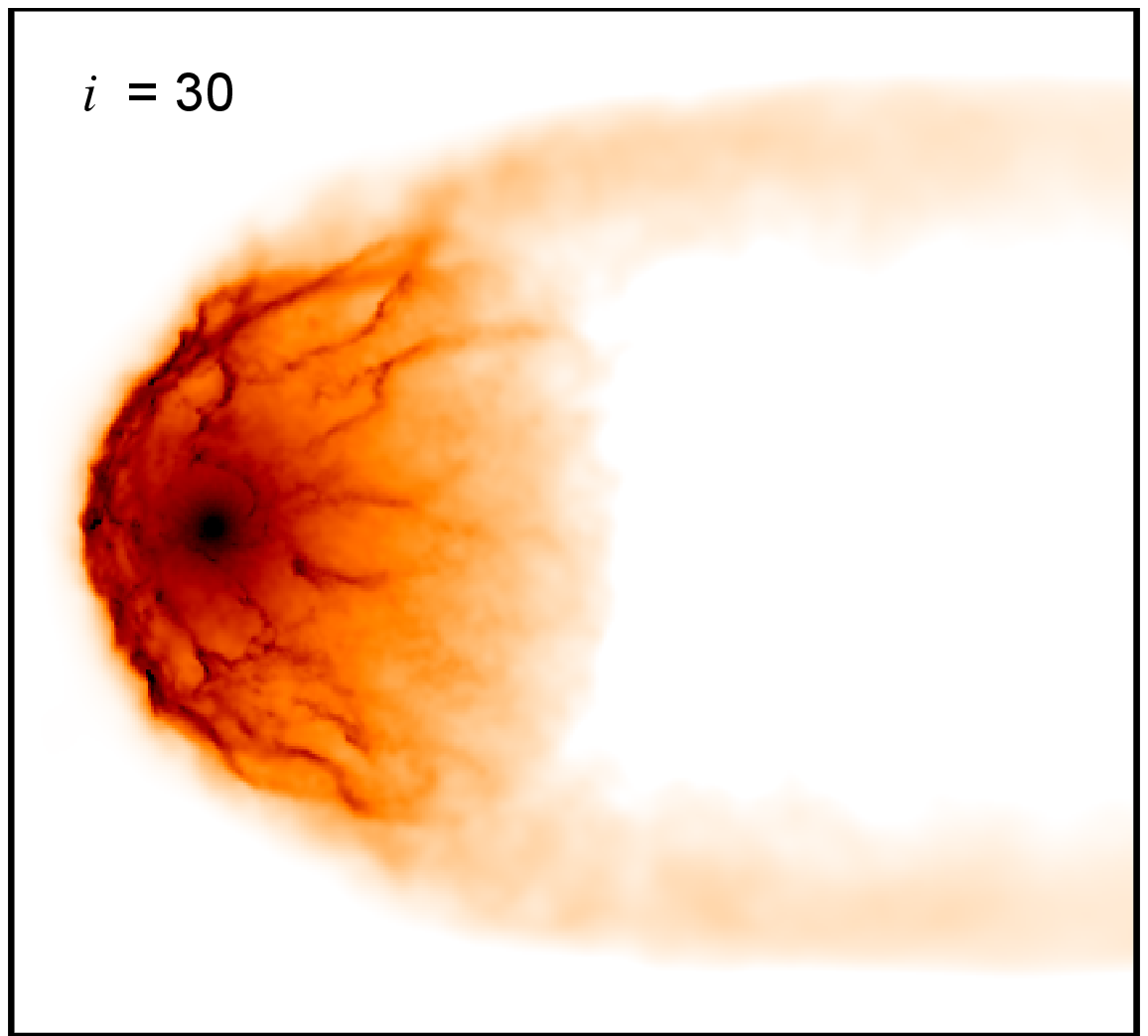}
\includegraphics[scale=.37, angle=0,trim= 40 0 37 0, clip=true]{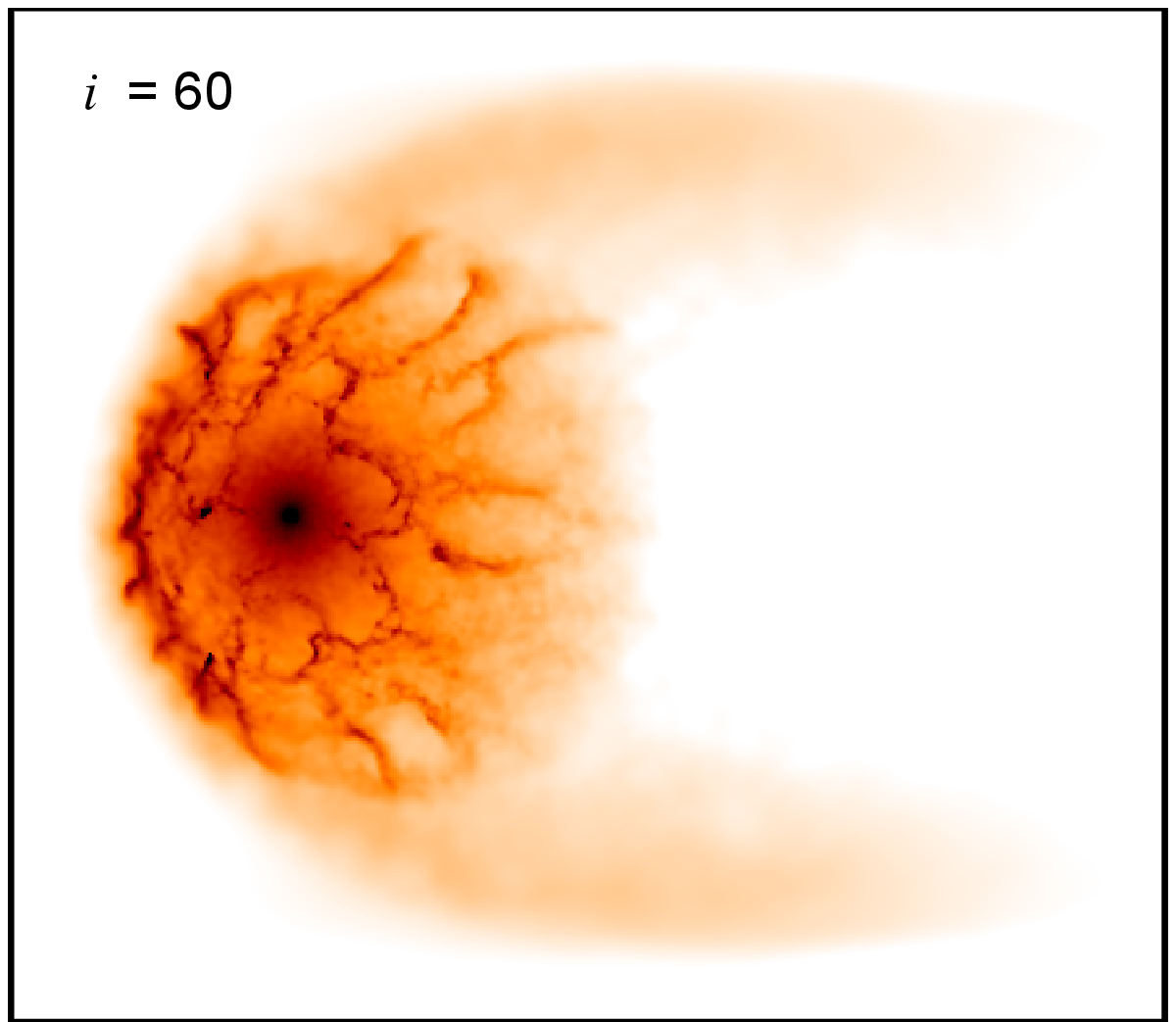}
\includegraphics[scale=.37, angle=0,trim= 40 0 40 0, clip=true]{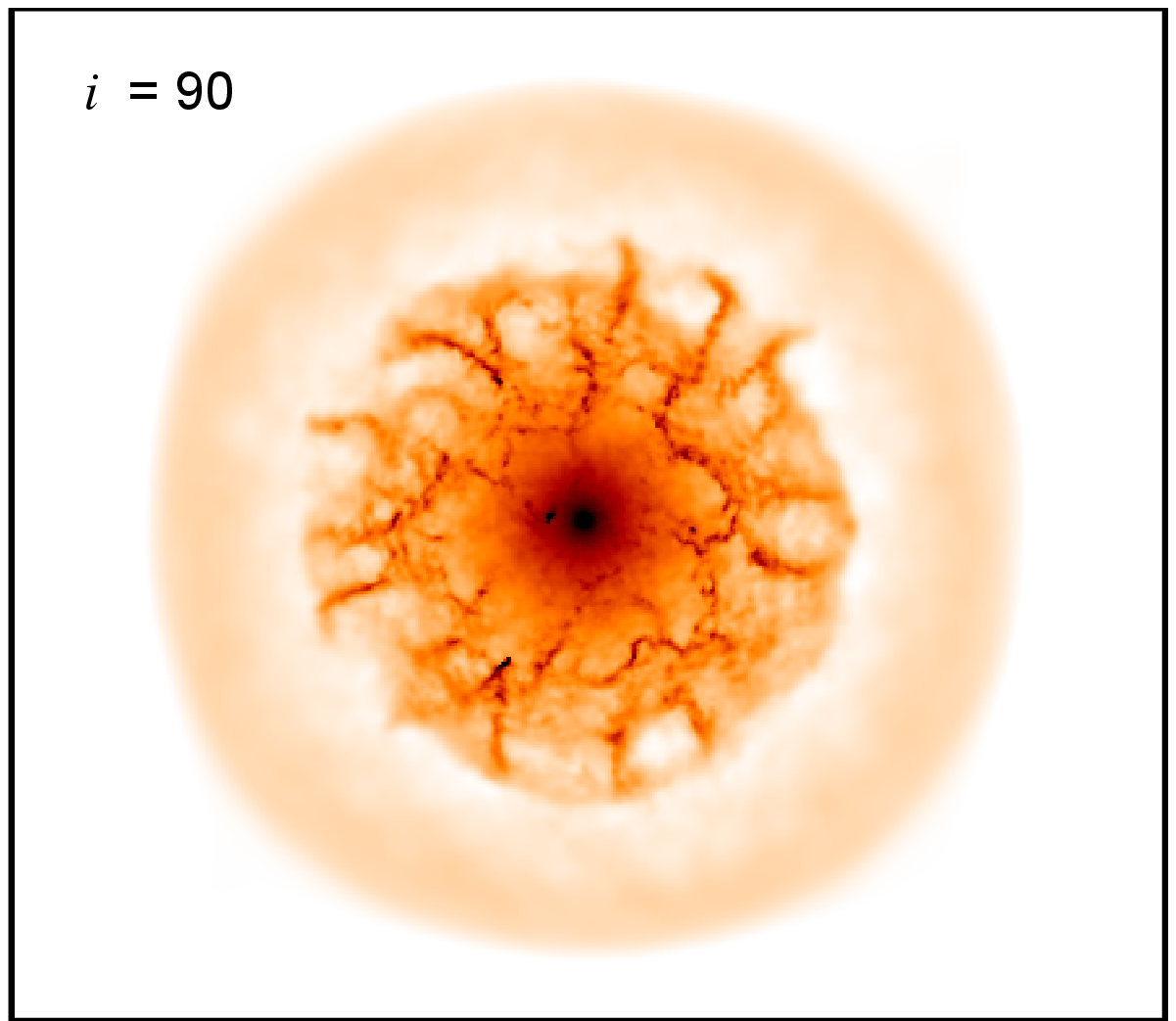}
\caption{Hydrogen column density (on a logarithmic scale) at 76\,000 years for models B [top] and 
D$_{\rm L}$ [bottom] seen at inclination angles $0\,^{\circ}$ [left]  to 
90$\,^{\circ}$ [right] in 30$\,^{\circ}$ intervals.\protect\footnotemark
\label{fig: rot}}
\end{figure*}

The inclination angle, the angle between the apex of the bow 
shock and the plane of the sky, also has a marked effect on the 
shape of the bow shock. In Fig.~\ref{fig: rot} we show the column densities 
for both the fast and slow models at different inclination angles. The bow 
shock becomes increasingly circular and the curvature in the tail also 
increases with larger inclination angles. At the same time, increasing 
the inclination angle reduces the density contrast between the peak 
at the apex of the bow shock and the rest of the cometary structure, making 
the detection and identification of highly inclined systems more difficult.

\subsubsection{Substructure: Instabilities}
\label{sec: instab}

 The bow shock is unstable for all models, however, the slow and fast 
models clearly exhibit very different bow shock 
substructure. The slow models show a thin, rather smooth outer
shock and a contact discontinuity that is highly distorted  (on a scale of 
$\sim$0.1\,pc) by prominent R-T `fingers' of the RSG wind (see 
Figs.~\ref{fig: nprof}, \ref{fig: rot},  \ref{fig: Lproj15}, \ref{fig: Lcross},  and App.~\ref{sec: appA}).  
The protrusions are even more developed in the high-resolution model 
A$_{\rm H}$ (see Figs.~\ref{fig: Lproj15} and \ref{fig: resol}).  In the column 
density plots, the small-scale R-T instabilities result in  a 
clumpy/knot-like structure that becomes one of the dominant features 
of the bow shock at large inclination angles  (see Fig.~\ref{fig: rot}). 

In contrast, the bow shock in the fast model is much smoother overall,   
which results in a more layered/string-like appearance (see 
Figs.~\ref{fig: nprof}, \ref{fig: rot}, \ref{fig: Lproj03},  \ref{fig: Lcross},   
and App.~\ref{sec: appA}). As the bow shock becomes increasingly circular 
with larger $i$ (Fig.~\ref{fig: rot}), the instabilities in the slow and fast models appear more, and more distinct, with 
the fluctuations from the latter occurring at much larger wavelengths, $\sim$0.2\,pc. 
The vortex on the upper right of the bow shock is similar to those 
formed as a result of the K-H instability and dragged downstream in the adiabatic case. 
Indeed, K-H instabilities are likely to arise in the fast model due to the shear between 
the 73 km\,s$^{-1}$ ISM and the 17 km\,s$^{-1}$ RSG wind.

\subsubsection{Emission maps}

\begin{figure*}
\centering
\includegraphics[scale=.55, angle=0,trim= 0 30 0 0, clip=true]{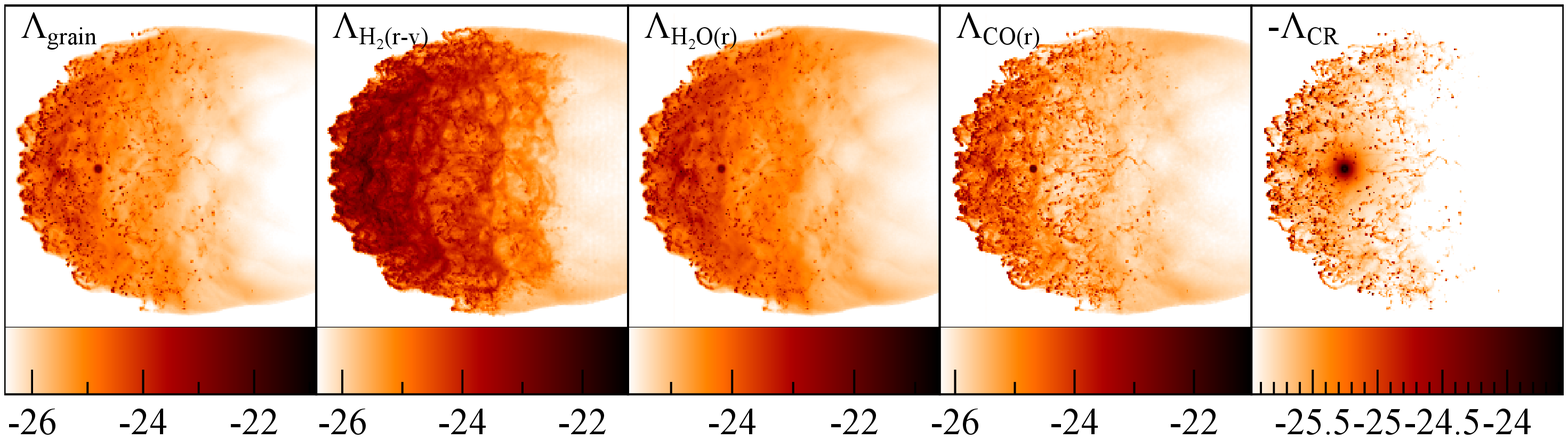}
\includegraphics[scale=.55, angle=0,trim= 0 30 0 0, clip=true]{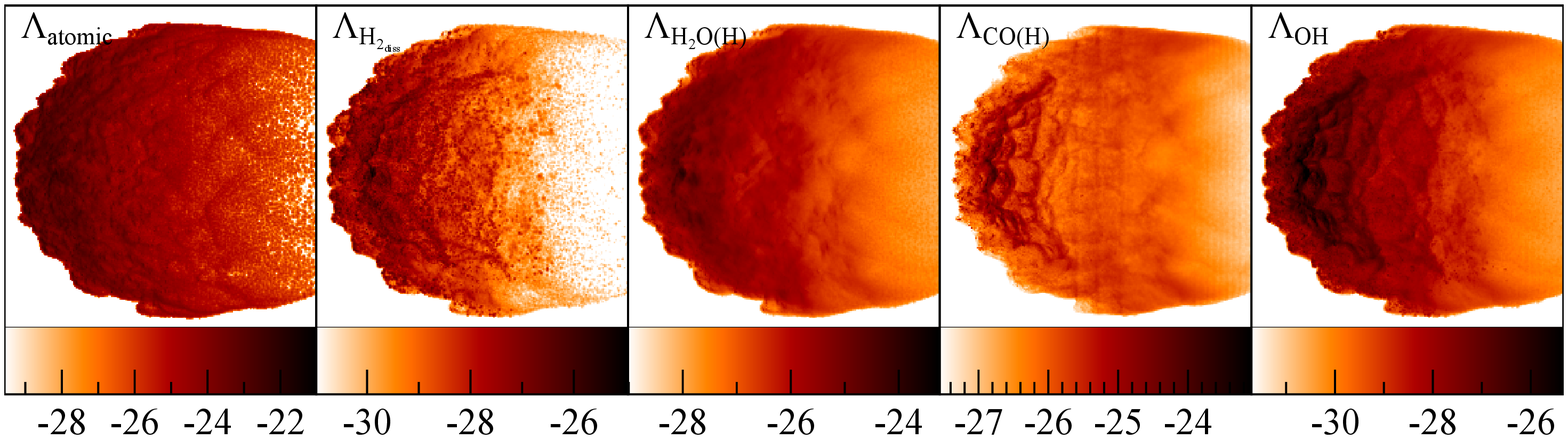}
\includegraphics[scale=.55, angle=0,trim= 0 30 0 0, clip=true]{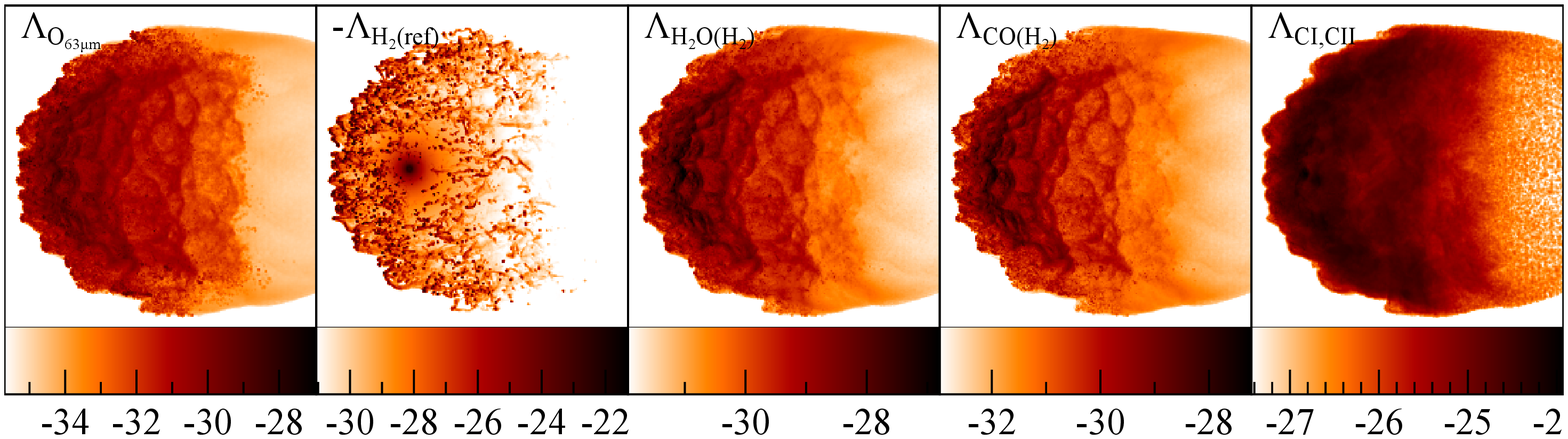}
\caption{Projected emissivity (ergs\,cm$^{-2}$\,s$^{-1}$)
 in the symmetry plane for model A$_{\rm H}$ for the various
molecular, atomic and dust species. 
 \label{fig: Lproj15}}
\end{figure*}

\begin{figure*}
\centering
\includegraphics[scale=.55, angle=0,trim= 0 30 0 0, clip=true]{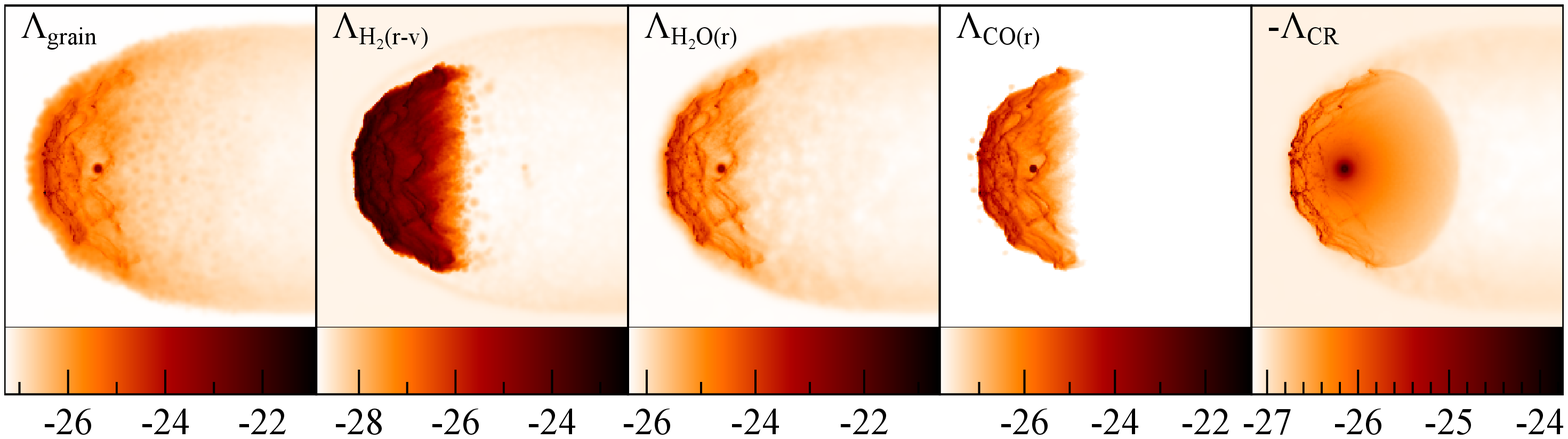}
\includegraphics[scale=.55, angle=0,trim= 0 30 0 0, clip=true]{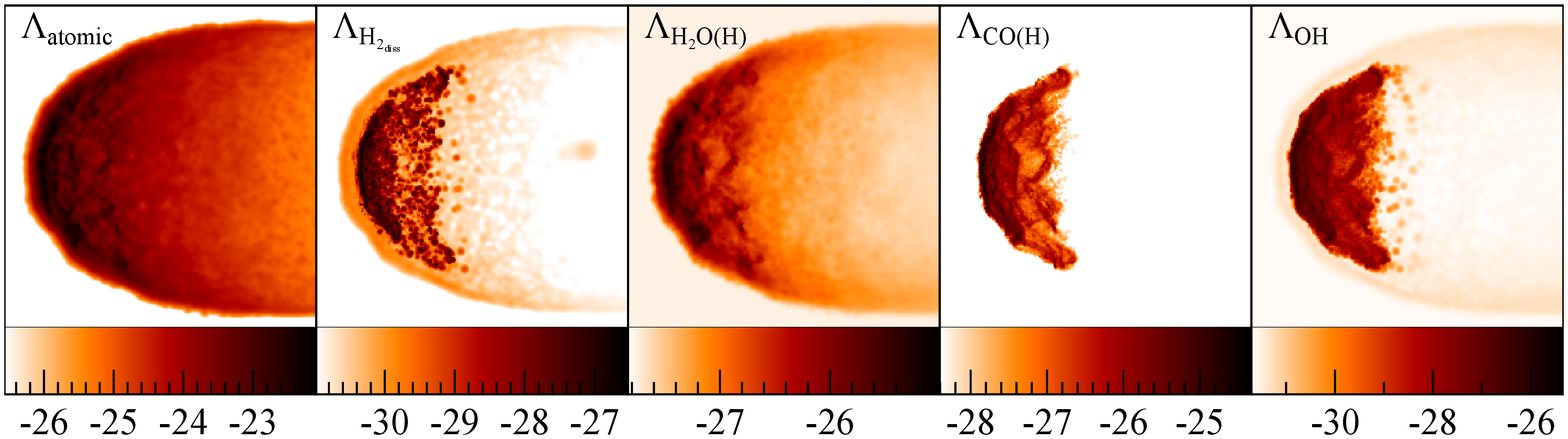}
\includegraphics[scale=.55, angle=0,trim= 0 30 0 0, clip=true]{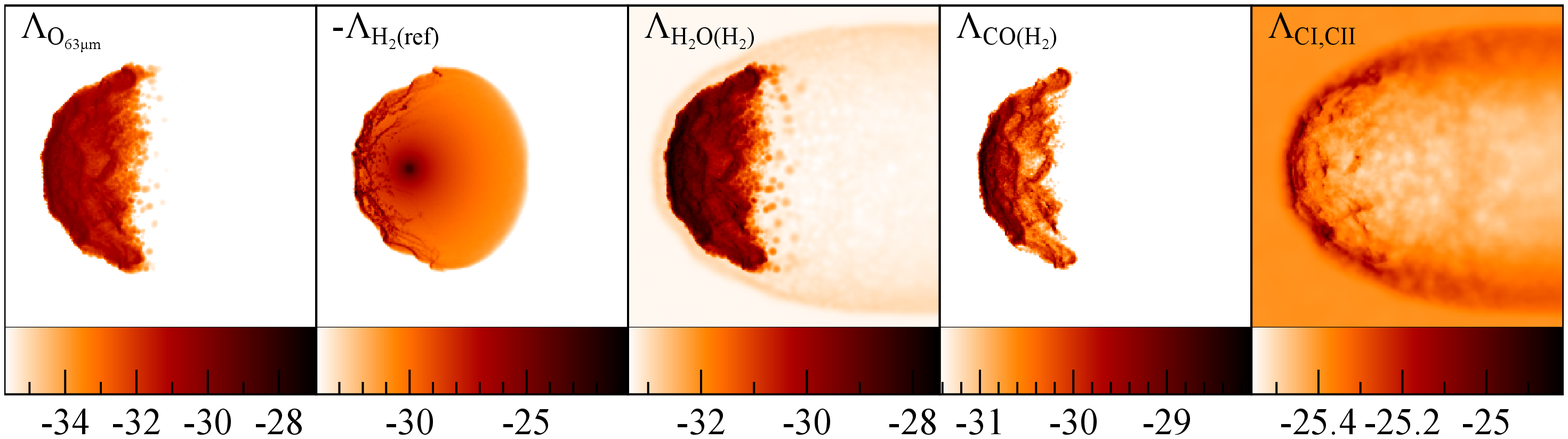}
\caption{Same as Fig.~\ref{fig: Lproj15} except for model D.
\label{fig: Lproj03}}
\end{figure*}

The strongest radiative emission for both the fast and slow models
originates just ahead of the contact discontinuity, forming a bright
ridge along that surface (see Fig.~\ref{fig: series} [bottom]).
This overall emissivity is a sum of the contributions from 15 different
coolants (see App.~\ref{sec: appC} for more details). While
some species radiate from the  entire bow shock surface,
e.g.~$\Lambda_{\rm H_2O (r)}$, others like $\Lambda_{\rm CO (r)}$
are almost entirely confined to the reverse shock or forward shock.
This has important consequences for the appearance of the bow shock
shell as shown in Figs.~\ref{fig: Lproj15} and \ref{fig: Lproj03}. For example, 
the emission from the forward shock (hotter gas) results in a much smoother 
shell accentuated by small gentle `hill-like' ripples, e.g.~$\Lambda_{\rm atomic}$,
whereas emission from the reverse shock (the shocked RSG wind)
produces a layered, more `cloud-like' appearance, e.g.~$\Lambda_{\rm H_2O (H_2)}$. 
Several coolants, such as gas-grain, rotational transitions of CO and H$_2$O, and the
heating species produce a more `finger-like', clumpy structure.

\subsection{Luminosity of the bow shock}
\label{sec: lum}
\footnotetext{Movies showing the bow shock 
rotated through different inclination angles are included in the electronic version.}
\begin{table}
\begin{spacing}{1.5}
\caption{Bow shock luminosities in ${\rm ergs\,s}^{-1}$. \label{tab: lum}}
\begin{tabular}{|l|ll|}
\hline
Species            & Model B & Fast model\\
\hline
1. $L_{\rm total}$         & 3.8$\times10^{32}$  & 9.2$\times10^{32}$ \\
2. $L_{\rm grain}$         & 4.1$\times10^{29}$  & 2.5$\times10^{29}$ \\
3. $L_{\rm O(63\mu m)}$    & 2.7$\times10^{25}$  & 2.1$\times10^{26}$ \\
4. $L_{\rm CI,II}$         & 9.7$\times10^{30}$  & 2.7$\times10^{30}$ \\
5. $L_{\rm H_2 (r-v)}$     & 8.1$\times10^{31}$  & 1.0$\times10^{32}$ \\
6. $L_{\rm atomic}$        & 2.2$\times10^{32}$  & 7.9$\times10^{32}$ \\
7. $L_{\rm H_2O (r)}$      & 6.2$\times10^{31}$  & 2.4$\times10^{31}$ \\
8. $L_{\rm H_2O (H_2)}$  & 7.5$\times10^{26}$  & 3.0$\times10^{27}$ \\
9. $L_{\rm H_2O (H)}$    & 2.2$\times10^{30}$  & 9.5$\times10^{29}$ \\
10. $L_{\rm CO (r)}$       & 5.2$\times10^{30}$  & 4.8$\times10^{30}$ \\
11. $L_{\rm CO (H_2)}$   & 1.3$\times10^{26}$  & 4.8$\times10^{26}$ \\
12. $L_{\rm CO (H)}$     & 8.5$\times10^{29}$  & 1.3$\times10^{30}$ \\
13. $L_{\rm OH}$           & 2.3$\times10^{28}$  & 5.1$\times10^{28}$ \\
\hline
\multicolumn{3}{|c|}{Theory}\\
\hline
$\dot{E}_{\rm tot}$   & 1.3$\times10^{33}$ & 5.6$\times10^{33}$  \\
\hline
\multicolumn{3}{|c|}{Observations}\\
\hline
$L_{\rm AKARI}$  &   6.9$\times 10^{33}$  ($65\mu$m) & 2.5$\times 10^{33}$\ ($90\mu$m)\\
($F_{\rm AKARI}^{1}$  &  10.7 Jy ($65\mu$m) &  6.9 Jy ($90\mu$m))\\
$L_{\rm IRAS}$  & 4.7$\times 10^{34}$ ($60\mu$m) &  1.87$\times 10^{34}$ ($100\mu$m)\\
($F_{\rm IRAS}^2$   & 110$\pm$20 Jy ($60\mu$m) & 40$\pm$10 Jy ($100\mu$m))\\
\hline
\end{tabular}
\end{spacing}
{\footnotesize{{\bf Notes.} [1] T. Ueta, private comm. [2] \citealt{Nor97}.}}
\end{table}

When the ISM and RSG winds collide, most of their  
kinetic energy is thermalised. An upper limit for total radiative luminosity  
in this case (assuming no other energy sources are present) is 
$\dot{E}_{\rm tot} \approx \dot{E}_{\rm wind} + \dot{E}_{\rm ISM}$ \citep{Wil97}, where the kinetic luminosity 
of the RSG wind is 
\be
\dot{E}_{\rm wind} = \frac{1}{2}\dot{M}_{\rm w} v_{\rm w}^2 = 2.8\times10^{32} {\rm ergs\,s}^{-1}\,,
\ee
and the ISM luminosity is 
\be
\dot{E}_{\rm ISM} =\frac{1}{2}\dot{M}_{\rm w} v_{\rm *}^2 = 1.6\times10^{33} n_{\rm H}^{-1}{\rm ergs\,s}^{-1}\,. 
\ee

In reality, only a fraction of the kinetic energy will be converted, 
therefore $\dot{E}_{\rm tot}$ is an upper limit for the radiative 
luminosity. $\dot{E}_{\rm ISM}$ for the fast model is 
5.3$\times10^{33}$ ergs\,s$^{-1}$, and is 1.1$\times10^{33}$ 
ergs\,s$^{-1}$ for the slow models assuming an average ISM density of 
$n_{\rm H} = 1.5$ cm$^{-3}$. The combined mechanical luminosity 
$\dot{E}_{\rm tot}$ (see Table.~\ref{tab: lum}) therefore 
does not exceed $\sim$$6\times 10^{33}$\,ergs\,s$^{-1}$ for Betelgeuse's 
parameters. This is a fairly robust upper limit because (a) for the most 
energetic case (model D) the wind kinetic luminosity is less than the 
ISM contribution;  (b) uncertainties in the observed Betelgeuse mass-loss rate 
are towards lower values, e.g.~\cite{You93b}; and (c) the stellar velocity 
cannot be a factor of two greater since this scenario would produce a very bright 
forward shock at shorter wavelengths (cf.~Mira and IRC 10216 -- see 
Sect.~\ref{sec: disc}).

\begin{figure}
\centering
\includegraphics[scale=.37, angle=270,trim= 50 35 195 0, clip=true]{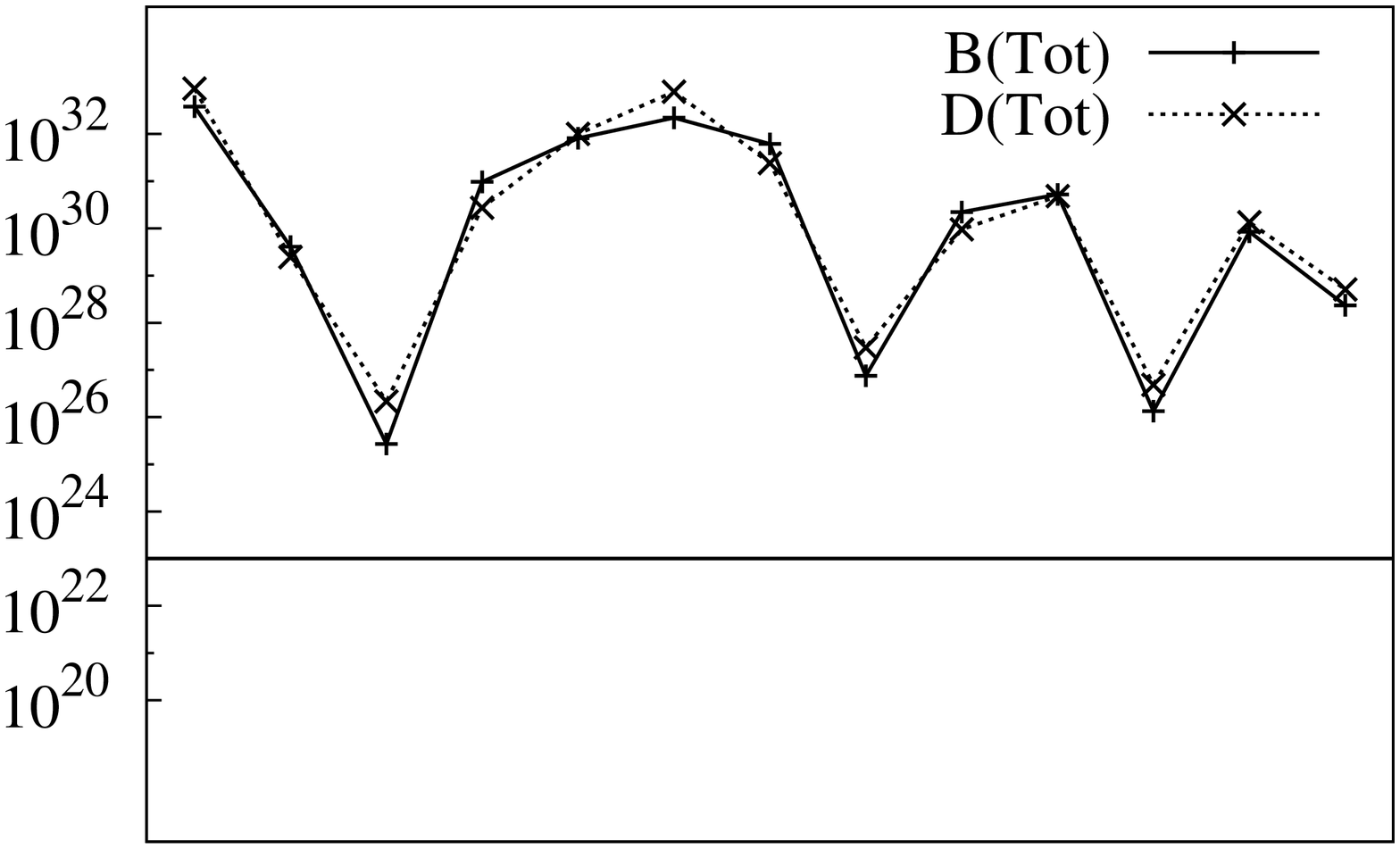}
\includegraphics[scale=.37, angle=270,trim= 49 35 20 0, clip=true]{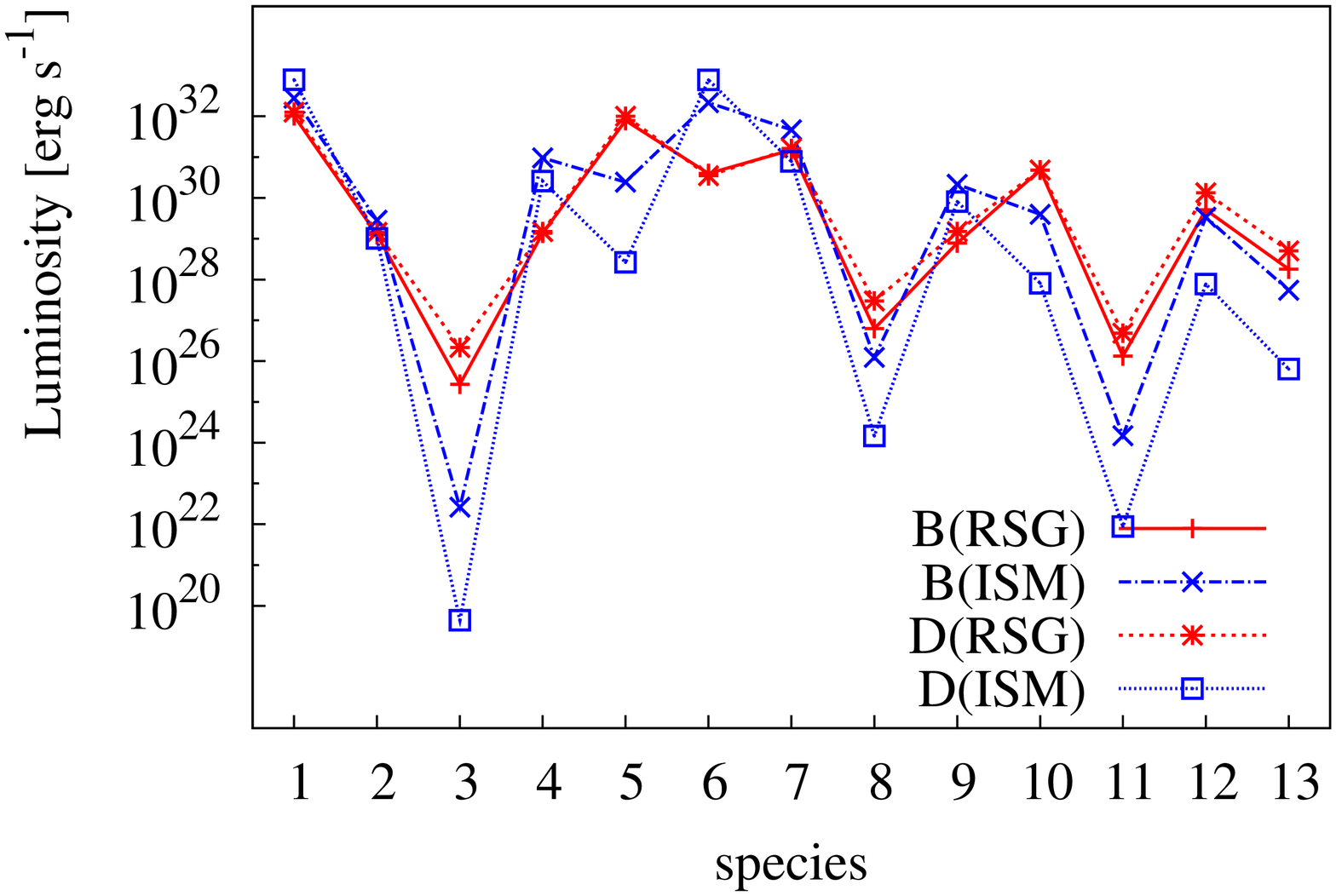}
\caption{ Models B and D radiative luminosity contributions for the various 
molecular and atomic species ordered on the $x$-axis as they are listed in 
Table.~\ref{tab: lum}. The total luminosity is shown in black [top], with the RSG 
and ISM components shown in red and blue [bottom], respectively.
\label{fig: lum}}
\end{figure}

We compare these theoretical values with estimates of the radiative luminosities 
from the simulations also given in Table.~\ref{tab: lum}. We used the following procedure 
to isolate only the contribution from the bow shock. For each model, a ($1/r^2$) 
density profile was utilised to subtract the unshocked RSG wind, then only particles 
with both densities and temperatures greater than the ambient ISM were selected. We 
further restricted the integration to within $\theta \lesssim 115^\circ$ ($x<R_{\rm SO}$) 
the dominant emission region in the {\it AKARI} and {\it IRAS} images (the cut-off also 
serves to clearly delineate an emission region that can be directly compared with future 
high-resolution observations). The luminosities for the slow models do not differ 
significantly (the increasing stellar velocity largely compensates for the decreasing ISM 
 density from model A to C), thus we only include the values for model B in 
 Table.~\ref{tab: lum}. The total bow shock luminosity for the fast and slow models is 
 approximately 16\% and 29 \% of the theoretical luminosity, respectively. Since $\dot{E}_{\rm tot}$ 
 is the expected luminosity over 4$\pi$ sterad and we only include emission from 
 $x < R_{\rm SO}$, these values are a lower limit estimate for the fraction of  kinetic energy 
 that is thermalised. In Fig.~\ref{fig: lum}, we show both the total radiative luminosities given 
 in  Table.~\ref{tab: lum} and the individual ISM and RSG contributions of each species. 
 Since the stellar wind properties are the same for all the models, 
 it is not surprising that the luminosities produced by the RSG wind component  
 are similar for both the fast and slow models. In contrast, the ISM contribution in 
 the fast and slow models differs significantly: the higher density in the slow models 
 leads to greater luminosities for all species except for the atomic lines, and the higher 
 shock temperature in the fast model results in higher $L_{\rm atomic}$ 
 (Fig.~\ref{fig: lum} [bottom]).

We can estimate the luminosity in the {\it IRAS} and {\it AKARI} wavelength 
bands from the observed fluxes:
\be
 L_{\nu} = 4 \pi D^2 F_\nu \Delta\nu \,,
\ee
where $\Delta \nu$ is the instrument bandwidth at the given wavelength and $F_\nu$ 
 the flux measured in that wavelength band. This yields an underestimate of 
the bolometric luminosity. The {\it AKARI} flux is 10.7 Jy and 6.9 Jy at 
$65\,\mu$m and $90\,\mu$m, respectively \citep[private comm.]{Ueta08}, 
and the {\it IRAS} fluxes are significantly greater with 110$\pm$20 Jy and 
40$\pm$10 Jy at $60\,\mu$m and $100\,\mu$m, respectively \citep{Nor97}. The derived 
luminosities are given in Table.~\ref{tab: lum} and show that, while 
the {\it IRAS}  luminosities exceed the theoretical upper limit, the {\it AKARI} values are
consistent given the uncertainties. Contamination by flux from the bar and 
Betelgeuse itself, which is very luminous in the infrared, may play a role 
and are likely to account for the difference between the {\it AKARI} and 
{\it IRAS} values. Higher resolution observations, together with a careful 
subtraction of the bar and central source, are essential for resolving this 
discrepancy.

 The most important coolants at low gas temperatures, $\sim$300\,K, are the 
gas-grain cooling, rotational modes of CO and H$_2$O, and the rotational 
and vibrational transitions of H$_2$. The dust, and C and O fine 
structure lines are often invoked to account for the infrared emission 
from bow shocks.  It is possible that including non-LTE chemistry could 
change the abundance of OI, however it appears to contribute little to the 
far-infrared emission. In contrast, the dust and CI, CII fine structure lines 
 appear to be the dominant far-infrared emitters in the bow shock shell, although 
 it should be noted that the cooling rate of the latter is very uncertain 
 \citep[private comm.]{Smi03}. The combined luminosity derived from the simulations 
 due to grains, OI, CI, and CII is at least three orders of magnitude lower than inferred from the 
{\it AKARI} observations. A possible explanation is that there is an additional source of 
infrared flux, e.g.~some of Betelgeuse's radiation, or radiation produced by hot gas in 
the bow shock itself, is absorbed and reemitted by the gas and dust in the far-infrared. 

\section{Discussion}
\label{sec: disc}

\subsection{Comparison with previous studies}

Our results are largely consistent with the previous hydrodynamic 
studies highlighted in Sect.~\ref{sec: intro}.  The slow models 
are similar  to the 2D models of \cite{Bri95}  and  \cite{Van11} in their flow characteristics 
and their R-T substructure.  However, the bow shocks in 
the 3D models of \cite{War07a} are generally smoother, and where 
instabilities are present, they occur on much larger scales than in our models. 
The difference is likely due to our including low-temperature cooling (below 
$10\,000$ K -- the temperature they assume for the stellar wind) 
and may also be a resolution effect.    

The smoother bow shock and the prominent vortex in the
upper right of the fast model appear to be more consistent with 
the {\it AKARI} observations of Betelgeuse's bow shock than the 
highly unstable bow shock in the slow models. However, 
with such high ISM shock temperatures (10$^5$ K), the 
bow shock should be visible at shorter wavelengths, e.g.~in the UV. 
To our knowledge, no such emission has been detected for Betelgeuse. 
For Mira (o Ceti), an AGB star, $5\times10^5$ K gas was detected
serendipitously with the {\it GALEX} UV satellite in both the bow shock 
and a 4 pc long tail \citep{Mar07}. The star is estimated to be moving
through a low density, $n_{\rm H}$$\sim$0.03 - 0.8 cm$^{-3}$, ISM at 130 km\,s$^{-1}$ \citep{War07b,Mar07}.
 The emission mechanism for this UV radiation is uncertain,
 but is thought to be the interaction of hot electrons (heated in the forward
 shock) with molecular hydrogen \citep{Mar07}. The hot gas in the core of
 the tail for the fast model is certainly reminiscent of Mira's narrow tail, and 
 would provide a ready supply of electrons. Mira's bow shock has also more 
 recently been detected in the infrared with {\it Herschel} \citep{May11}. Another 
 system with both UV and confirmed infrared emission from the bow shock is 
 IRC 10216, an AGB star moving through the ISM at $\sim$90\,km\,s$^{-1}$ 
 \citep{Sah10,Lad10}.

High-resolution observations, e.g.~with {\it Herschel} and {\it SOFIA}, should be 
able to distinguish between the fast and slow models based on the 
morphology of the bow shock and the presence/length scale of the instabilities.

\subsection{The age of Betelgeuse's bow shock}

\subsubsection{Bow shock shell mass}

\begin{figure}
\centering
\includegraphics[scale=.3, angle=270,trim= 0 130 0 120, clip=false]{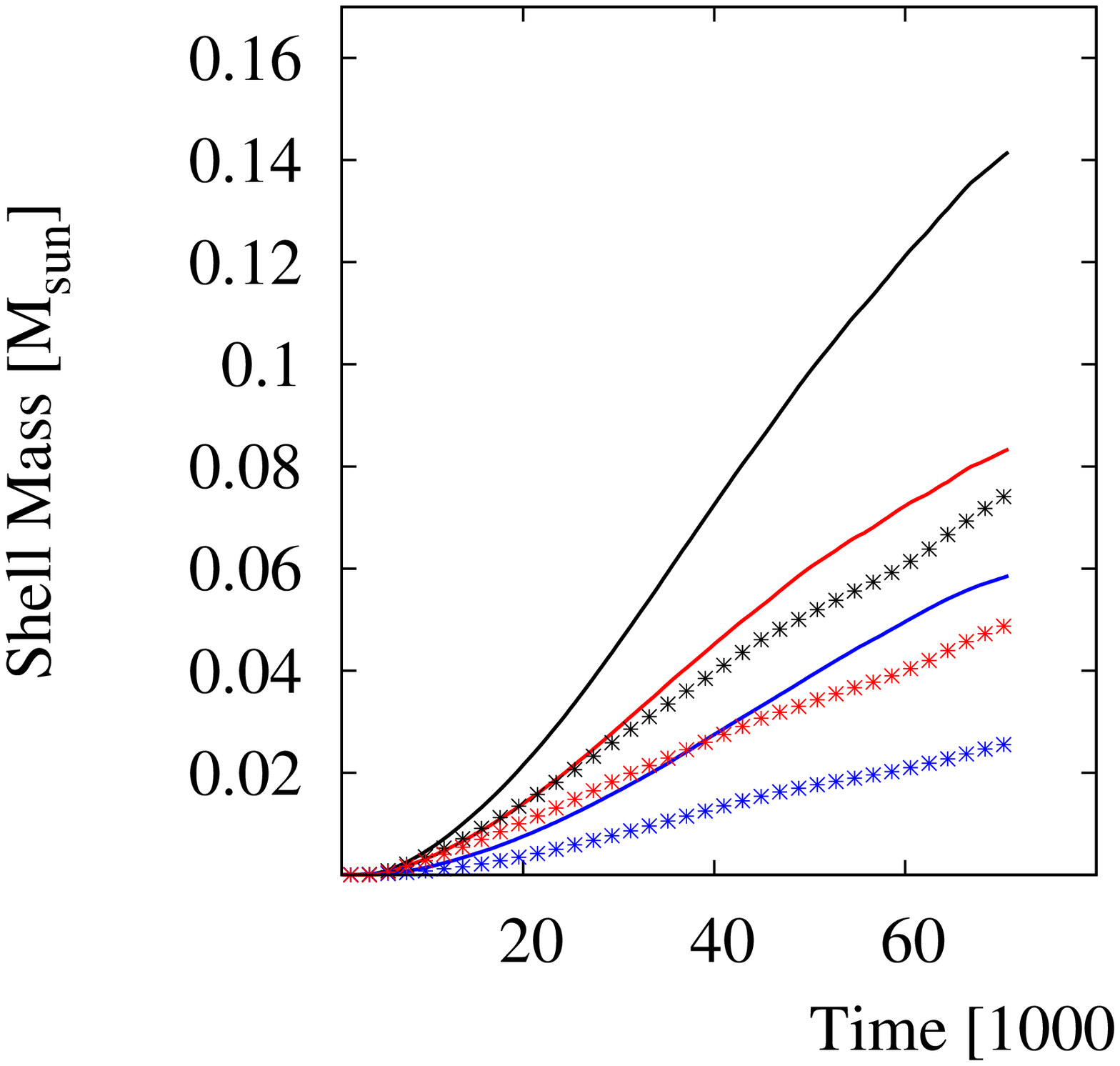}
\includegraphics[scale=.3, angle=270,trim= 0 270 0 90, clip=true]{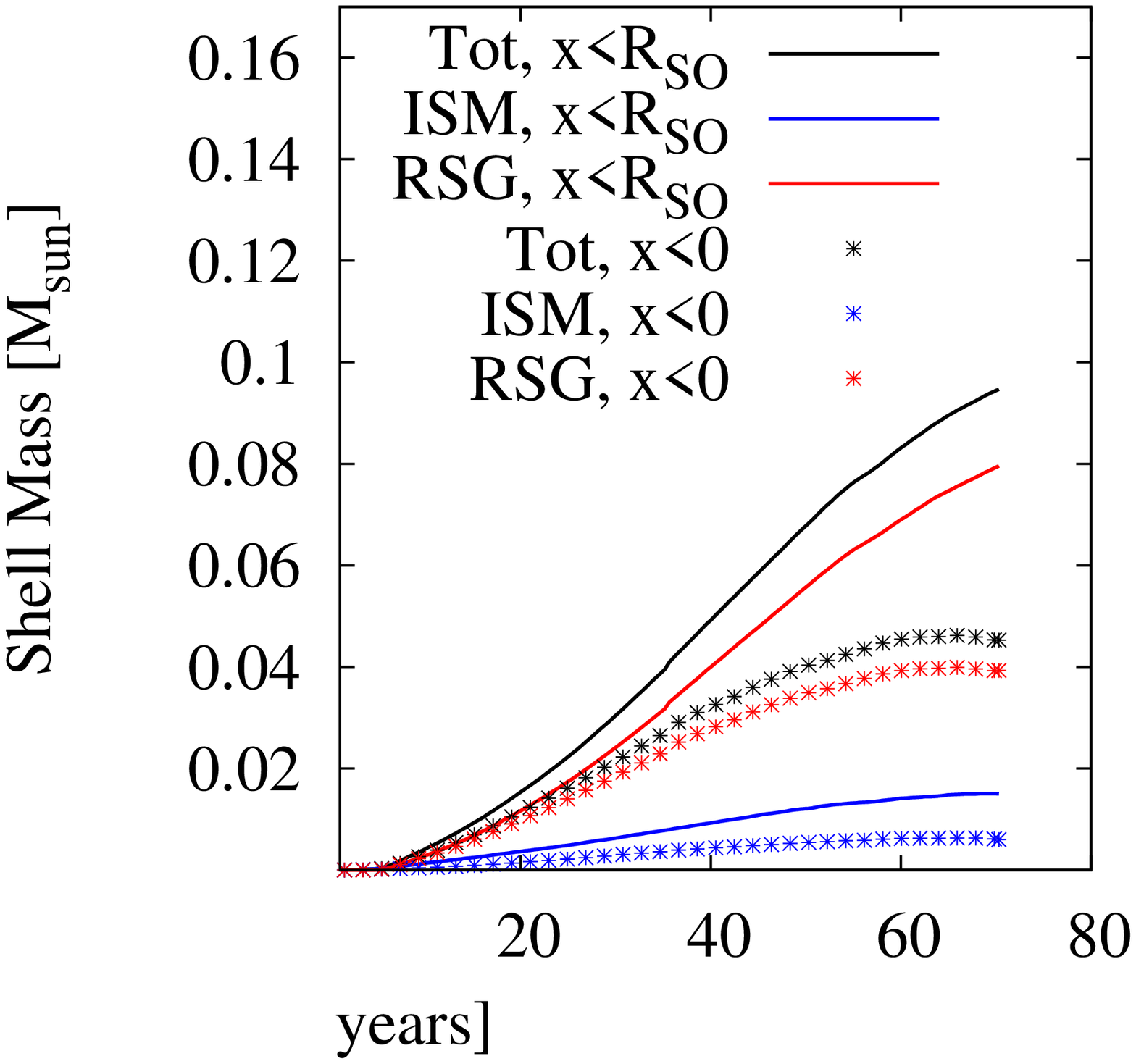}
\caption{ Evolution of the bow shock shell mass for models B [left] 
and D [right]. For both models, the RSG and ISM contributions are 
in red and blue, respectively, and their combined mass is plotted in black.   
The solid lines trace mass for $x<R_{\rm SO}$, while the points trace 
$x<0$ gas.
\label{fig: shellmass}}
\end{figure}

By comparing the bow shock shell mass derived from the 
models with observational estimates, we can 
constrain the age of the bow shock shell.
In Fig.~\ref{fig: shellmass} we show the evolution of the bow 
shock shell mass for models  B [left] and D [right]. Contributions 
from material in the bow shock were obtained using the same procedure  as 
for the luminosity calculations described above, essentially only 
shocked particles  in the bow shock head (i.e.~with $x<0$, points in 
Fig.~\ref{fig: shellmass}) and $x<R_{\rm SO}$ (solid lines in  Fig.~\ref{fig: shellmass})
were included.  The former is useful for comparison with the theoretical estimates, while 
the latter is utilised when discussing the observations. The mass in the bow shock 
increases almost linearly at early times and only reaches an equilibrium state 
 for the bow shock head after at least 60\,000 years. The slow models take longer 
to achieve this steady state because the gas dragging material downstream has much lower velocity 
than in the fast models. Most of the mass in the bow shock shell is from the RSG wind, 
and as expected, its contribution does not differ significantly between the fast 
and slow models, except at later times  when more wind material is able to accumulate at 
the bow shock  due to mixing caused by R-T instabilities in the latter. As was 
the case for model B$_{\rm ad}$, the RSG shell mass for $x<0$ is approximately $\sim$$0.05\,\msun$, 
twice that of the theoretical estimate derived using 0.5$\times{\cal M}_{\rm shell}$.
Owing to the higher ISM density assumed, the ISM contribution in model B 
is larger (half  as much as the RSG wind) than that in model D (only about a quarter), 
and as a result the total mass in the bow shock head, $\gtrsim$0.08\,$\msun$, is also greater 
than in model D, $\gtrsim$0.048\,$\msun$.

The mass in the bow shock based on the observed $60\,\mu$m flux is given by
\be
{\cal M}_{\rm shell} ({\rm obs}) = 0.042 \msun (F_{60}/135\, {\rm Jy}) (D/200\, {\rm pc})^2 \,,
\ee
\citep{Nor97}. Again, assuming the \cite{Harp08} distance
 of 197 pc, and the {\it IRAS} flux of 110 Jy in 7.5 arcmin 
 yields a shell mass of 0.033 $\msun$. (\citealt{Nor97} derived 
 a shell mass of 0.14 $\msun$, but assumed $D$ = 400 pc 
   for the distance to Betelgeuse.) The corresponding 
age of the bow shock is approximately 30\,000 years 
(see Fig.~\ref{fig: shellmass}, solid lines). However, as discussed 
in Sect.~\ref{sec: lum}, this {\it IRAS} flux is most likely an 
overestimate, thus the age derived based on that value is an upper limit. 

The shell mass from the {\it AKARI} fluxes is an order of 
magnitude lower ($\sim$$0.0033\,\msun$), which would imply an even 
younger bow shock age. If this is the case, however, the wind would not have 
sufficient time to expand to the stand-off distance. One possible solution is
that the observed shell mass is underestimated due to uncertainties in the 
conversion from flux to mass (e.g.~due to dust properties). In our models the 
shell takes $\sim$20\,000 years to reach the correct $R_{\rm SO}$ 
(see Fig.~\ref{fig: shockpos}), by which time the mass in the bow shock is 
approximately $0.02\,\msun$. This is higher than the value obtained from {\it AKARI} 
observations, but may be consistent within the uncertainties and  
with the consensus estimates for Betelgeuse's wind parameters. However, 
at this age none of our models are close to reaching a steady state.

Another possibility is that the shell is older; however, this requires extreme
stellar wind properties.  Utilising Eq.~\ref{eq: Mshell}, we see that decreasing
the mass-loss rate and increasing the wind velocity will both decrease the shell
mass.  However, the largest line widths observed in the wind are $40\,$km\,s$^{-1}$  
\citep[Fig.~1]{Hug94}, thus the wind velocity, which is usually taken to be half 
of this linewidth, must be approximately 20 km\,s$^{-1}$ or less. Thus, to decrease the 
shell mass from a steady state value of $0.05\,\msun$ to $0.0033\,\msun$ requires that 
we decrease the mass-loss rate by a factor of 15, i.e.~$\dot{M}=2\times10^{-7} \,\msun\,$yr$^{-1}$,
somewhat below the lowest estimate of \citet{You93b}.  The latter was derived
from CO observations, which tends to underestimate the true mass-loss rate due to 
incomplete CO synthesis in Betelgeuse's stellar wind \citep{Nor97}. Although this lower 
mass-loss rate and a higher wind velocity cannot be excluded, they are outlying values in 
the range of observations so must be considered unlikely.

\subsubsection{The shape of the bow shock}

Based on these arguments, our results suggest that Betelgeuse's bow shock is 
young and may not have reached a steady state yet. The circular nature and 
smoothness of bow shock can be naturally accounted for if this is the case, 
and need not be due solely to the inclination of the bow shock with respect 
to the plane of the sky;  however, Betelgeuse's tangential velocity, 
$V_{\rm t}$=25 km\,s$^{-1}$, and radial velocity, $V_{\rm rad}$=20.7 km\,s$^{-1}$ 
(assuming the distance to the star is 197 pc, see Sect.~\ref{sec: intro}) yield an 
inclination angle of $\arctan(V_{\rm t}/V_{\rm rad})$$\sim$50$^\circ$. Given the 
uncertainty in the distance to the star, this inclination angle is consistent with the  
 \cite{Ueta08} value (56$^\circ$) derived by fitting the shape of the observed arc 
 with the analytic \cite{Wil96} solution, even though as discussed in Sect.~\ref{sec: morph}, 
 the latter is only valid if the bow shock has reached a steady state.

Utilising Betelgeuse's radial and tangential motions, the stellar velocity is ($V_{\rm t}^2 +V_{\rm rad}^2)^{1/2}$$\sim$32.5 km\,s$^{-1}$, the same as model B. Although the smooth appearance 
of the shell would seem to rule out the slow models, if the bow shock is young, the strong 
instabilities that characterise those simulations may not have had enough time to grow. 
We can estimate the age of the bow shock by comparing the observed shape 
with what is predicted by the simulations in Fig.~\ref{fig: shape} [top]. 
From the {\it AKARI} observations, the ratio of $R(0^\circ)/R(90^\circ)$ is approximately 0.7,    
which corresponds to an age of $\lesssim$30\,000 years. 

According to our simulations, if the bow shock is young, the bow shock tail 
should show strong curvature and should not be too distant from 
the head of the bow shock. Although the emission from such a tail 
is likely to be weak, deep, high-resolution observations may be able 
to detect it. Indeed, there appears to be faint emission extending 
from the bow shock head towards the tail in the {\it AKARI} observations, 
e.g.~the structure located at (RA,DEC) offsets of (-2,9) arcmin in the WIDE-S 
image of Fig.~1 of \cite{Ueta08}. A curved tail has already been observed for 
the massive O9.5 supergiant $\alpha$ Camelopardalis (S.~Mandel\footnote{See 
image on APOD at http://apod.nasa.gov/apod/ap061124.html}) and likely for the 
O6 I(n)f runaway star $\lambda$ Cep \citep{Gva11}. Although these are much hotter 
stars with faster stellar winds, the physical mechanism responsible for producing 
a curved tail is the same. In light of our simulations, such a feature implies 
either a very small space motion for the star relative to ambient ISM (unlikely 
given the strength of the bow shock emission) or that the bow shock is young. 
Thus, modelling the shape of the bow shock (head and tail, if present) is a promising avenue 
for constraining the age and the evolutionary stage of these systems.  

\subsubsection{Implications}

Estimates for Betelgeuse's age are highly uncertain, 
ranging from 8-13 million years depending on the stellar 
evolution model and distance assumed \citep{Harp08}. The core helium-burning phase
 lasts about 10\% of the total lifetime, thus $\tau_{\rm He} = 0.8 -1.3$ Myr. 
The star is expected to start and spend a significant fraction of 
this phase as an RSG. According to current stellar models, the star
may or may not have undergone a so-called blue loop in the HR diagram, 
spending a fraction of core helium burning as a blue supergiant, 
before returning to the RSG stage at core helium exhaustion \citep{Mey00,Heg00}. 

A young age of Betelgeuse's bow shock might imply that the star has entered
the RSG stage only recently. It may be interesting to relate this
idea with the recent finding that the diameter of Betelgeuse at 11\,$\mu$m
has systematically decreased by about 15\% over the past 15 years \citep{Tow09}.
While \cite{Rav10} and \cite{Ohn11} 
debate whether this implies a real radius change, or rather a change
in the density of certain layers in the envelope of the star,
it is remarkable that the time scale of these changes is of the order of
100 years. This might imply that the envelope of Betelgeuse is not in thermal
equilibrium, as might be the case when a star is entering the RSG stage.

On the other hand, the radius of a star entering the RSG phase  for the first time after core
hydrogen exhaustion only increases, with time scales 
for the radius change ($R/\dot R$) down to about 1\,000 years. 
This is different, however, when the star enters the RSG stage after core helium 
exhaustion, because the igniting helium shell leads to an intermittent episode of shrinkage.
Since the star might have lost a substantial fraction of its hydrogen-rich
envelope by that time, its thermal time scale might also have decreased.
Thus, while more observational and theoretical efforts are required to
give this speculation more substance, a consistent picture might involve
Betelgeuse recently having finished core helium burning and returning
to the RSG stage from a previous blue supergiant excursion.

If Betelgeuse has only recently entered the RSG stage, 
it may not have had enough time to travel beyond 
its main sequence or blue supergiant wind bubble. 
Assuming a wind mass-loss rate 
of $10^{-7}\,\msun\,$yr$^{-1}$, and a wind velocity of $\sim$$10^3$\,km\,s$^{-1}$, 
the stand-off distance for Betelgeuse's main sequence or blue supergiant
bow shock shell was around  
1 pc. A RSG phase of a $\sim$few$\times$10\,000 years would bring the star 
close to the edge of such a bubble. This raises the possibility that the 
mysterious `bar' ahead of the bow shock could be a remnant of this earlier 
phase of evolution. A theoretical investigation of such a scenario is currently 
underway.

\section{Conclusions}

We presented the first 3D models of the interaction of Betelgeuse's RSG 
wind with the ISM. We took  dust, atomic-, molecular-, and metal-line 
cooling into account. The models cover a range of plausible ISM densities of 0.3 - 1.9 cm$^{-3}$ and stellar velocities  of 28 - 73 km\,s$^{-1}$. We showed that the 
flow dynamics and morphology of the bow shocks in the models differed 
due to the growth of Rayleigh-Taylor or Kelvin-Helmholtz instabilities. 
The former dominate the slow models,  resulting in a clumpy substructure, whereas the 
latter are characteristic of the fast model and produce a more layered substructure. 
In the fast model, gas is shocked to high temperatures. If gas as hot as this were to be detected at 
short wavelengths (e.g.~UV), this would exclude the slow models as an explanation for Betelgeuse's bow shock. 
High spatial and spectral resolution observations (e.g.~{\it Herschel}, {\it SOFIA} 
and {\it ALMA}) particularly of the more dominant cooling/emitting species, e.g.~rotational 
lines of H$_2$O or CO, could also be used to further constrain the physical 
characteristics of the system. In addition, better determination of the molecular 
to atomic hydrogen fraction in the RSG wind, the abundances of various species, and 
temperature of the ISM would also reduce the number of free parameters in the model. 

The large fluxes in the infrared compared to the theoretical limit for the bolometric 
luminosity suggest that the stellar flux and/or flux from hotter gas in the bow shock 
is reprocessed by the dust and reemitted in the far-infrared. We showed that, if the bow shock shell 
mass is low, as is suggested by the {\it AKARI} fluxes, then Betelgeuse's 
bow shock is young. The smoothness and circular nature of the bow shock would be 
consistent with this conclusion. Furthermore, if the bow shock has not yet reached 
a steady state, we are less able to constrain the physical parameters 
of the system, e.g.~the ram pressure of the ISM. This also raises the 
intriguing possibility that Betelgeuse has only recently become an RSG  and that the mysterious `bar' 
ahead of the bow shock is a remnant of a wind shell created during an earlier phase of 
evolution. 

\begin{acknowledgements}
The authors are grateful to  Vasilii Gvaramadze,
Michael Smith, Toshiya Ueta, and the referee, Alejandro Esquivel, for 
many helpful and insightful comments and discussions. We thank 
the John von Neumann Institute for Computing for a grant for computing time on the JUROPA supercomputer at Juelich Supercomputing Centre and Lorne Nelson for computing time on 
the Centre de Calcul Sci{\'e}ntifique de l'Universit{\'e} de Sherbrooke Usherbrooke.
The rendered figures in this paper were made using a modified
version of the {\small SPLASH} visualisation toolkit \citep{Pri07}. 
\end{acknowledgements}

\bibliographystyle{aa}
\bibliography{betelv5}

\begin{thebibliography}{56}
\expandafter\ifx\csname natexlab\endcsname\relax\def\natexlab#1{#1}\fi

\bibitem[{{Agertz} {et~al.}(2007){Agertz}, {Moore}, {Stadel}, {Potter},
  {Miniati}, {Read}, {Mayer}, {Gawryszczak}, {Kravtsov}, {Nordlund}, {Pearce},
  {Quilis}, {Rudd}, {Springel}, {Stone}, {Tasker}, {Teyssier}, {Wadsley}, \&
  {Walder}}]{Agertz07}
{Agertz}, O., {Moore}, B., {Stadel}, J., {et~al.} 2007, \mnras, 380, 963

\bibitem[{{Barnes} \& {Hut}(1986)}]{BH86}
{Barnes}, J. \& {Hut}, P. 1986, \nat, 324, 446

\bibitem[{{Bernat} {et~al.}(1979){Bernat}, {Hall}, {Hinkle}, \&
  {Ridgway}}]{Ber79}
{Bernat}, A.~P., {Hall}, D.~N.~B., {Hinkle}, K.~H., \& {Ridgway}, S.~T. 1979,
  \apjl, 233, L135

\bibitem[{{Brighenti} \& {D'Ercole}(1995)}]{Bri95}
{Brighenti}, F. \& {D'Ercole}, A. 1995, \mnras, 277, 53

\bibitem[{{Cao} {et~al.}(1997){Cao}, {Terebey}, {Prince}, \&
  {Beichman}}]{Cao97}
{Cao}, Y., {Terebey}, S., {Prince}, T.~A., \& {Beichman}, C.~A. 1997, \apjs,
  111, 387

\bibitem[{{Eldridge} {et~al.}(2011){Eldridge}, {Langer}, \& {Tout}}]{Eld11}
{Eldridge}, J.~J., {Langer}, N., \& {Tout}, C.~A. 2011, \mnras, 414, 3501

\bibitem[{{Esquivel} {et~al.}(2010){Esquivel}, {Raga}, {Cant{\'o}},
  {Rodr{\'{\i}}guez-Gonz{\'a}lez}, {L{\'o}pez-C{\'a}mara}, {Vel{\'a}zquez}, \&
  {De Colle}}]{Esq10}
{Esquivel}, A., {Raga}, A.~C., {Cant{\'o}}, J., {et~al.} 2010, \apj, 725, 1466

\bibitem[{{Frisch} {et~al.}(1990){Frisch}, {Sembach}, \& {York}}]{Fri90}
{Frisch}, P.~C., {Sembach}, K., \& {York}, D.~G. 1990, \apj, 364, 540

\bibitem[{{Garcia-Segura} {et~al.}(1996){Garcia-Segura}, {Langer}, \& {Mac
  Low}}]{Gar96}
{Garcia-Segura}, G., {Langer}, N., \& {Mac Low}, M.-M. 1996, \aap, 316, 133

\bibitem[{{Glassgold} \& {Huggins}(1986)}]{Glas86}
{Glassgold}, A.~E. \& {Huggins}, P.~J. 1986, \apj, 306, 605

\bibitem[{{Gvaramadze} \& {Gualandris}(2011)}]{Gva11}
{Gvaramadze}, V.~V. \& {Gualandris}, A. 2011, \mnras, 410, 304

\bibitem[{{Harper}(2010)}]{Harp10}
{Harper}, G.~M. 2010, in Astronomical Society of the Pacific Conference Series,
  Vol. 425, Hot and Cool: Bridging Gaps in Massive Star Evolution, ed.
  {C.~Leitherer, P.~Bennett, P.~Morris, \& J.~van Loon}, 152--+

\bibitem[{{Harper} {et~al.}(2008){Harper}, {Brown}, \& {Guinan}}]{Harp08}
{Harper}, G.~M., {Brown}, A., \& {Guinan}, E.~F. 2008, \aj, 135, 1430

\bibitem[{{Hartmann} \& {Avrett}(1984)}]{Hart84}
{Hartmann}, L. \& {Avrett}, E.~H. 1984, \apj, 284, 238

\bibitem[{{Heger} \& {Langer}(2000)}]{Heg00}
{Heger}, A. \& {Langer}, N. 2000, \apj, 544, 1016

\bibitem[{{Huggins} {et~al.}(1994){Huggins}, {Bachiller}, {Cox}, \&
  {Forveille}}]{Hug94}
{Huggins}, P.~J., {Bachiller}, R., {Cox}, P., \& {Forveille}, T. 1994, \apjl,
  424, L127

\bibitem[{{Ladjal} {et~al.}(2010){Ladjal}, {Barlow}, {Groenewegen}, {Ueta},
  {Blommaert}, {Cohen}, {Decin}, {De Meester}, {Exter}, \& {et al.,}}]{Lad10}
{Ladjal}, D., {Barlow}, M.~J., {Groenewegen}, M.~A.~T., {et~al.} 2010, \aap,
  518, L141+

\bibitem[{{Lambert} {et~al.}(1984){Lambert}, {Brown}, {Hinkle}, \&
  {Johnson}}]{Lam84}
{Lambert}, D.~L., {Brown}, J.~A., {Hinkle}, K.~H., \& {Johnson}, H.~R. 1984,
  \apj, 284, 223

\bibitem[{{Martin} {et~al.}(2007){Martin}, {Seibert}, {Neill}, {Schiminovich},
  {Forster}, {Rich}, {Welsh}, {Madore}, {Wheatley}, {Morrissey}, \&
  {Barlow}}]{Mar07}
{Martin}, D.~C., {Seibert}, M., {Neill}, J.~D., {et~al.} 2007, \nat, 448, 780

\bibitem[{{Mayer} {et~al.}(2011){Mayer}, {Jorissen}, {Kerschbaum}, {Mohamed},
  {van Eck}, {Ottensamer}, {Blommaert}, {Decin}, {Groenewegen}, {Posch},
  {Vandenbussche}, \& {Waelkens}}]{May11}
{Mayer}, A., {Jorissen}, A., {Kerschbaum}, F., {et~al.} 2011, \aap, 531, L4+

\bibitem[{{Meynet} \& {Maeder}(2000)}]{Mey00}
{Meynet}, G. \& {Maeder}, A. 2000, \aap, 361, 101

\bibitem[{{Monaghan}(1992)}]{Mon92}
{Monaghan}, J.~J. 1992, \araa, 30, 543

\bibitem[{{Neilson} {et~al.}(2011){Neilson}, {Lester}, \& {Haubois}}]{Hil11}
{Neilson}, H., {Lester}, J., \& {Haubois}, X. 2011, in The Ninth Pacific Rim
  Conference on Stellar Astrophysics, ed. {S.~Qian}, Astronomical Society of
  the Pacific Conference Series

\bibitem[{{Noriega-Crespo} {et~al.}(1997){Noriega-Crespo}, {van Buren}, {Cao},
  \& {Dgani}}]{Nor97}
{Noriega-Crespo}, A., {van Buren}, D., {Cao}, Y., \& {Dgani}, R. 1997, \aj,
  114, 837

\bibitem[{{Ohnaka} {et~al.}(2011){Ohnaka}, {Weigelt}, {Millour}, {Hofmann},
  {Driebe}, {Schertl}, {Chelli}, {Massi}, {Petrov}, \& {Stee}}]{Ohn11}
{Ohnaka}, K., {Weigelt}, G., {Millour}, F., {et~al.} 2011, \aap, 529, A163+

\bibitem[{{Price}(2007)}]{Pri07}
{Price}, D.~J. 2007, Publications of the Astronomical Society of Australia, 24,
  159

\bibitem[{{Price}(2008)}]{Pri08}
{Price}, D.~J. 2008, Journal of Computational Physics, 227, 10040

\bibitem[{{Raga} {et~al.}(2008){Raga}, {Cant{\'o}}, {De Colle}, {Esquivel},
  {Kajdic}, {Rodr{\'{\i}}guez-Gonz{\'a}lez}, \& {Vel{\'a}zquez}}]{Rag08}
{Raga}, A.~C., {Cant{\'o}}, J., {De Colle}, F., {et~al.} 2008, \apjl, 680, L45

\bibitem[{{Ravi} {et~al.}(2010){Ravi}, {Wishnow}, {Lockwood}, \&
  {Townes}}]{Rav10}
{Ravi}, V., {Wishnow}, E.~H., {Lockwood}, S., \& {Townes}, C.~H. 2010, ArXiv
  e-prints

\bibitem[{{Rodgers} \& {Glassgold}(1991)}]{Rod91}
{Rodgers}, B. \& {Glassgold}, A.~E. 1991, \apj, 382, 606

\bibitem[{{Rosswog}(2009)}]{Ros09}
{Rosswog}, S. 2009, \nar, 53, 78

\bibitem[{{Sahai} \& {Chronopoulos}(2010)}]{Sah10}
{Sahai}, R. \& {Chronopoulos}, C.~K. 2010, \apjl, 711, L53

\bibitem[{{Smith} \& {Rosen}(2003)}]{Smi03}
{Smith}, M.~D. \& {Rosen}, A. 2003, \mnras, 339, 133

\bibitem[{{Smith} {et~al.}(2009){Smith}, {Hinkle}, \& {Ryde}}]{Smi09}
{Smith}, N., {Hinkle}, K.~H., \& {Ryde}, N. 2009, \aj, 137, 3558

\bibitem[{{Springel}(2005)}]{Spr05}
{Springel}, V. 2005, \mnras, 364, 1105

\bibitem[{{Springel}(2010)}]{Spr10}
{Springel}, V. 2010, \araa, 48, 391

\bibitem[{{Springel} \& {Hernquist}(2002)}]{Spr02}
{Springel}, V. \& {Hernquist}, L. 2002, \mnras, 333, 649

\bibitem[{{Springel} {et~al.}(2001){Springel}, {Yoshida}, \& {White}}]{Spr01}
{Springel}, V., {Yoshida}, N., \& {White}, S.~D.~M. 2001, \na, 6, 79

\bibitem[{{Stencel} {et~al.}(1988){Stencel}, {Pesce}, \& {Hagen
  Bauer}}]{Sten88}
{Stencel}, R.~E., {Pesce}, J.~E., \& {Hagen Bauer}, W. 1988, \aj, 95, 141


\bibitem[{{Sutherland} \& {Dopita}(1993)}]{Sut93}
{Sutherland}, R.~S. \& {Dopita}, M.~A. 2003, \apjs, 88, 253

\bibitem[{{Suttner} {et~al.}(1997){Suttner}, {Smith}, {Yorke}, \&
  {Zinnecker}}]{Sut97}
{Suttner}, G., {Smith}, M.~D., {Yorke}, H.~W., \& {Zinnecker}, H. 1997, \aap,
  318, 595

\bibitem[{{Townes} {et~al.}(2009){Townes}, {Wishnow}, {Hale}, \&
  {Walp}}]{Tow09}
{Townes}, C.~H., {Wishnow}, E.~H., {Hale}, D.~D.~S., \& {Walp}, B. 2009, \apjl,
  697, L127

\bibitem[{{Ueta} {et~al.}(2008){Ueta}, {Izumiura}, {Yamamura}, {Nakada},
  {Matsuura}, {Ita}, {Tanab{\'e}}, {Fukushi}, {Matsunaga}, \& {Mito}}]{Ueta08}
{Ueta}, T., {Izumiura}, H., {Yamamura}, I., {et~al.} 2008, \pasj, 60, 407

\bibitem[{{Ueta} {et~al.}(2009){Ueta}, {Izumiura}, {Yamamura}, {Stencel},
  {Nakada}, {Matsuura}, {Ita}, {Tanab{\'e}}, {Fukushi}, {Matsunaga}, {Mito}, \&
  {Speck}}]{Ueta09}
{Ueta}, T., {Izumiura}, H., {Yamamura}, I., {et~al.} 2009, in Astronomical
  Society of the Pacific Conference Series, Vol. 418, Astronomical Society of
  the Pacific Conference Series, ed. {T.~Onaka, G.~J.~White, T.~Nakagawa, \&
  I.~Yamamura}, 117--+

\bibitem[{{Valcke} {et~al.}(2010){Valcke}, {de Rijcke}, {R{\"o}diger}, \&
  {Dejonghe}}]{Val10}
{Valcke}, S., {de Rijcke}, S., {R{\"o}diger}, E., \& {Dejonghe}, H. 2010,
  \mnras, 408, 71

\bibitem[{{van Marle} {et~al.}(2006){van Marle}, {Langer}, {Achterberg}, \&
  {Garc{\'{\i}}a-Segura}}]{Van06}
{van Marle}, A.~J., {Langer}, N., {Achterberg}, A., \& {Garc{\'{\i}}a-Segura},
  G. 2006, \aap, 460, 105

\bibitem[{{van Marle} {et~al.}(2011){van Marle}, {Meliani}, {Keppens}, \&
  {Decin}}]{Van11}
{van Marle}, A.~J., {Meliani}, Z., {Keppens}, R., \& {Decin}, L. 2011, \apjl,
  734, L26+

\bibitem[{{Villaver} {et~al.}(2003){Villaver}, {Garc{\'{\i}}a-Segura}, \&
  {Manchado}}]{Vil03}
{Villaver}, E., {Garc{\'{\i}}a-Segura}, G., \& {Manchado}, A. 2003, \apjl, 585,
  L49

\bibitem[{{Wadsley} {et~al.}(2008){Wadsley}, {Veeravalli}, \&
  {Couchman}}]{Wad08}
{Wadsley}, J.~W., {Veeravalli}, G., \& {Couchman}, H.~M.~P. 2008, \mnras, 387,
  427

\bibitem[{{Wareing} {et~al.}(2007{\natexlab{a}}){Wareing}, {Zijlstra}, \&
  {O'Brien}}]{War07a}
{Wareing}, C.~J., {Zijlstra}, A.~A., \& {O'Brien}, T.~J. 2007{\natexlab{a}},
  \apjl, 660, L129

\bibitem[{{Wareing} {et~al.}(2007{\natexlab{b}}){Wareing}, {Zijlstra},
  {O'Brien}, \& {Seibert}}]{War07b}
{Wareing}, C.~J., {Zijlstra}, A.~A., {O'Brien}, T.~J., \& {Seibert}, M.
  2007{\natexlab{b}}, \apjl, 670, L125

\bibitem[{{Weaver} {et~al.}(1977){Weaver}, {McCray}, {Castor}, {Shapiro}, \&
  {Moore}}]{Wea77}
{Weaver}, R., {McCray}, R., {Castor}, J., {Shapiro}, P., \& {Moore}, R. 1977,
  \apj, 218, 377

\bibitem[{{Wilkin}(1996)}]{Wil96}
{Wilkin}, F.~P. 1996, \apjl, 459, L31+

\bibitem[{{Wilkin} {et~al.}(1997){Wilkin}, {Canto}, \& {Raga}}]{Wil97}
{Wilkin}, F.~P., {Canto}, J., \& {Raga}, A.~C. 1997, in IAU Symposium, Vol.
  182, Herbig-Haro Flows and the Birth of Stars, ed. {B.~Reipurth \&
  C.~Bertout}, 343--352

\bibitem[{{Wolfire} {et~al.}(1995){Wolfire}, {Hollenbach}, {McKee}, {Tielens},
  \& {Bakes}}]{Wolf95}
{Wolfire}, M.~G., {Hollenbach}, D., {McKee}, C.~F., {Tielens}, A.~G.~G.~M., \&
  {Bakes}, E.~L.~O. 1995, \apj, 443, 152

\bibitem[{{Young} {et~al.}(1993{\natexlab{a}}){Young}, {Phillips}, \&
  {Knapp}}]{You93b}
{Young}, K., {Phillips}, T.~G., \& {Knapp}, G.~R. 1993{\natexlab{a}}, \apj,
  409, 725

\bibitem[{{Young} {et~al.}(1993{\natexlab{b}}){Young}, {Phillips}, \&
  {Knapp}}]{You93a}
{Young}, K., {Phillips}, T.~G., \& {Knapp}, G.~R. 1993{\natexlab{b}}, \apjs,
  86, 517

\end{thebibliography}

\appendix
\section{The effect of resolution}
\label{sec: appA}

\begin{figure}[!h]
\centering
\includegraphics[scale=.363, angle=0,trim= 0   13 143 10, clip=true]{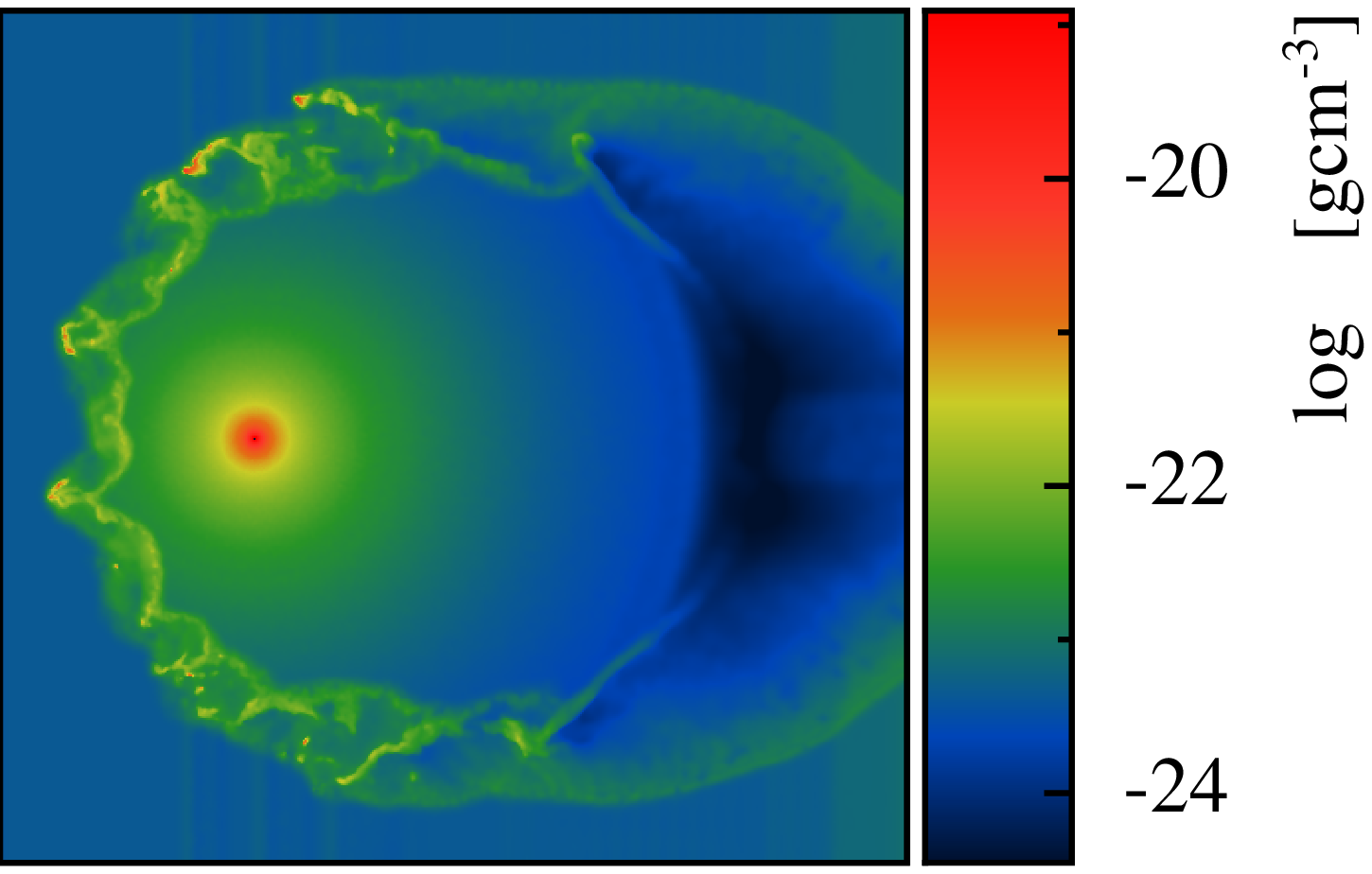}
\includegraphics[scale=.33,   angle=0,trim= 10 0    0     110, clip=false]{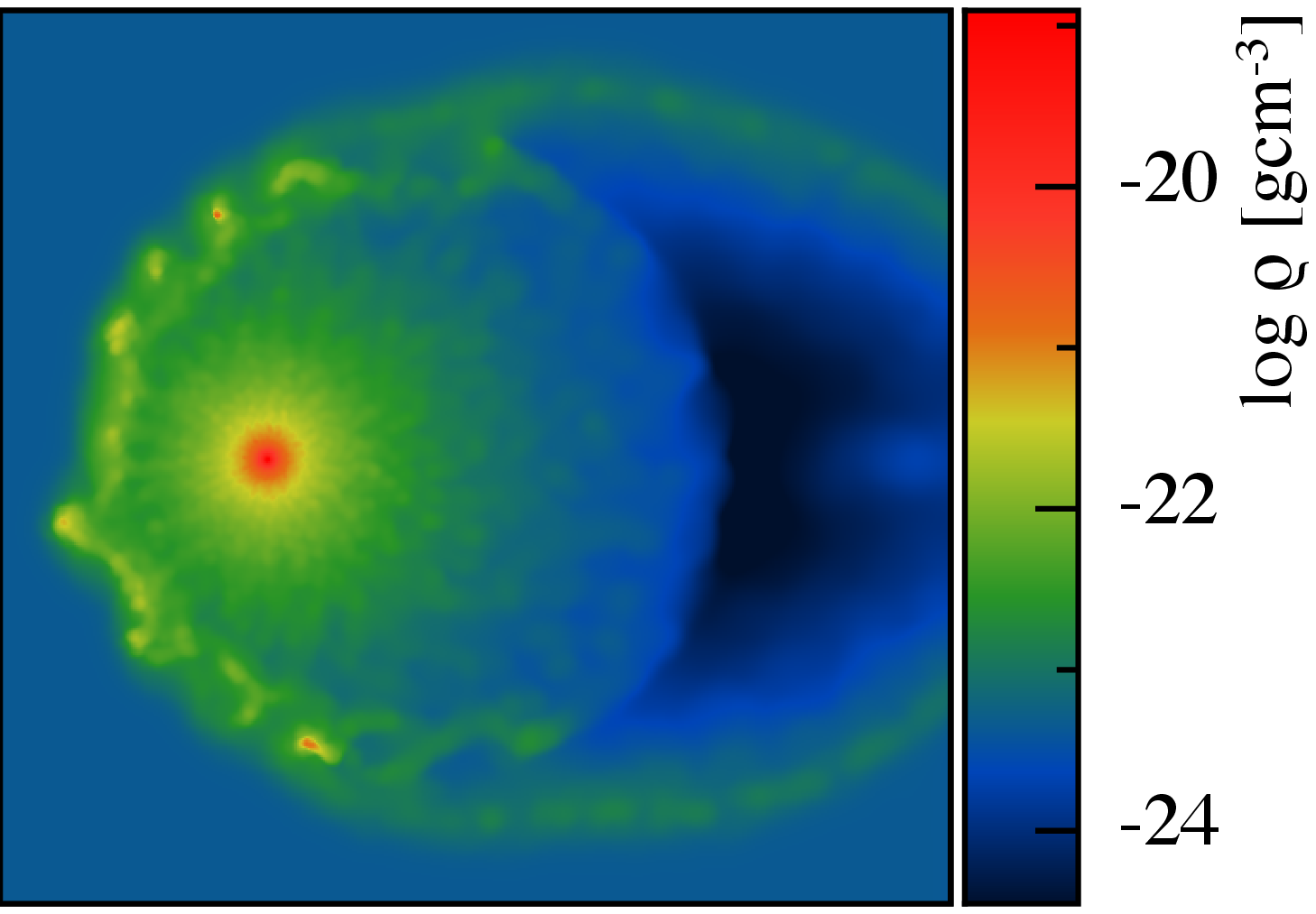}\\
\includegraphics[scale=.362, angle=0,trim= 0   13 143 20, clip=true]{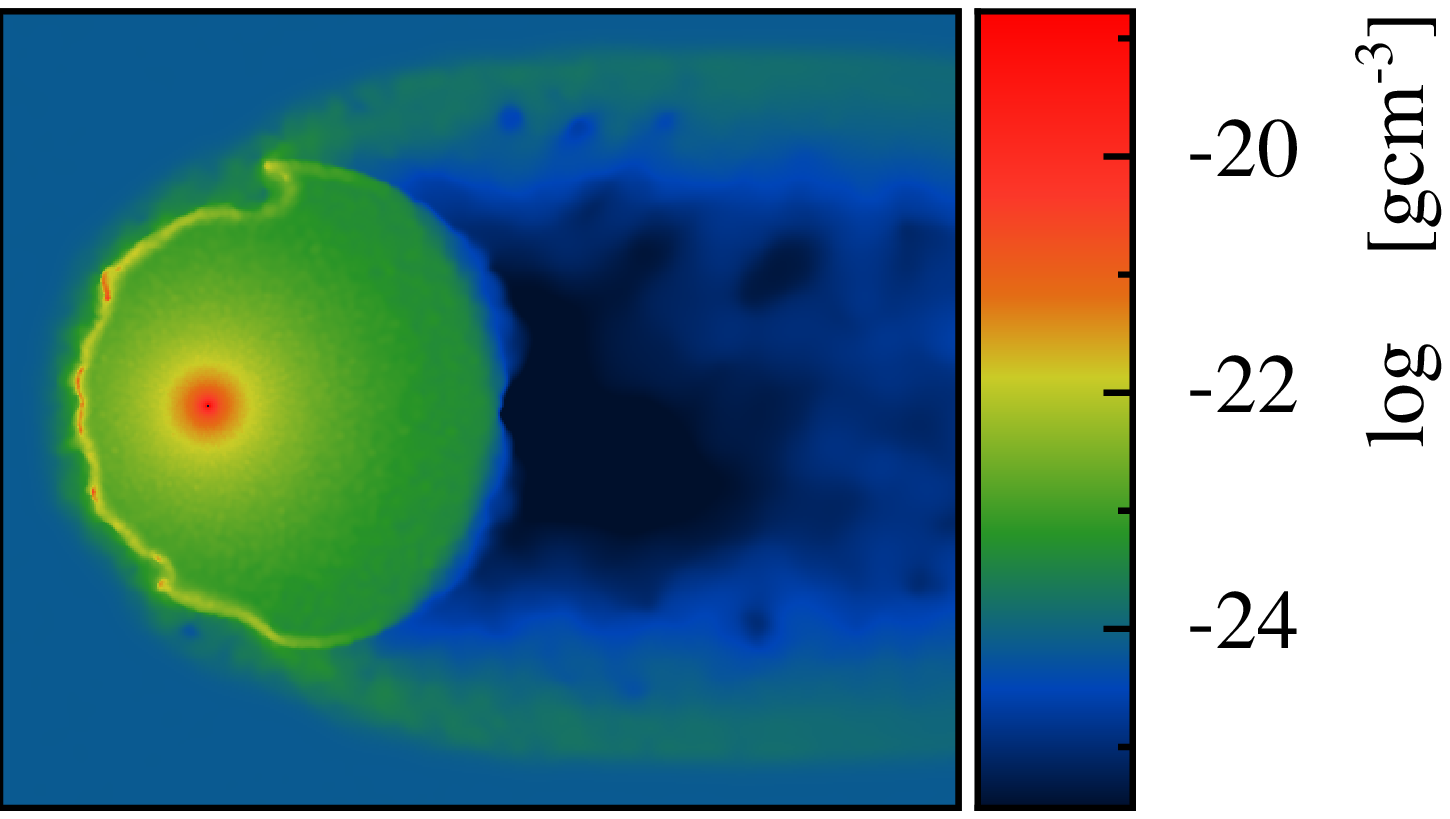}
\includegraphics[scale=.33,   angle=0,trim= 10   0   0    120, clip=false]{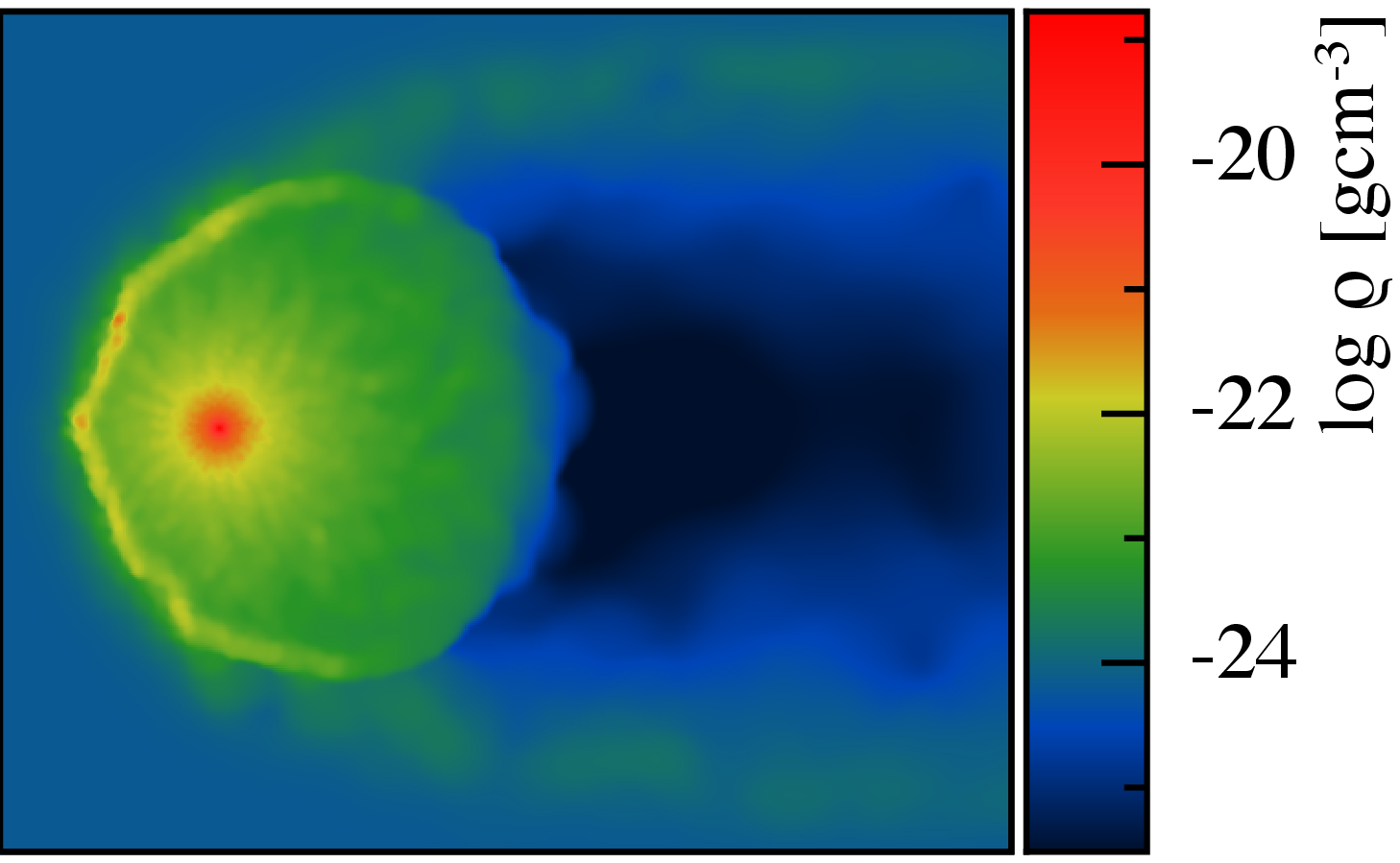}
\caption{Cross-sectional density profiles for models A$_{\rm H}$ [top, left], A 
[top, right], D [bottom, left], and D$_{\rm L}$ [bottom, right] in the symmetry
plane. 
\label{fig: resol}}
\end{figure}

As discussed in Sect.~\ref{sec: res}, it is important to use a
large number of particles to reduce shock broadening. High resolution
is also necessary in order to resolve instabilities, particularly for the 
slow models. To test this, the slowest model (A) was rerun as model 
A$_{\rm H}$ with an average six times higher 
mass resolution, and the results are compared in cross-section in 
 Fig.~\ref{fig: resol} [top].  A higher resolution version of model D was 
 not computationally feasible, so a lower resolution 
 version (model D$_{\rm L}$) was run instead with an average mass 
 resolution that is  two times lower, 
 and the results are shown in the lower two panels of Fig.~\ref{fig: resol}.
In both cases, as expected, the higher resolution model shows a 
smoother RSG wind and stronger instabilities developing in the 
bow shock.  Model D$_{\rm L}$ appears to have insufficient resolution 
to fully develop the instabilities seen in model D (e.g.~the vortex in the 
upper right of the bow shock), whereas the difference between 
models A and A$_{\rm H}$ is less dramatic because both models show the 
development of R-T instabilities. Despite these differences, the overall agreement 
between the larger scale fluid properties of the high and low resolution simulations 
is very good,  and they do not differ by more than a few percent. Thus we conclude 
that our calculations are sufficiently numerically converged in terms of 
the quantities we are interested in measuring.

\section{The effect of varying $T_{\rm ISM}$ and $f$}
\label{sec: appB}
\begin{figure}
\centering
\includegraphics[scale=.38, angle=0,trim= 0 80 0 0, clip=true]{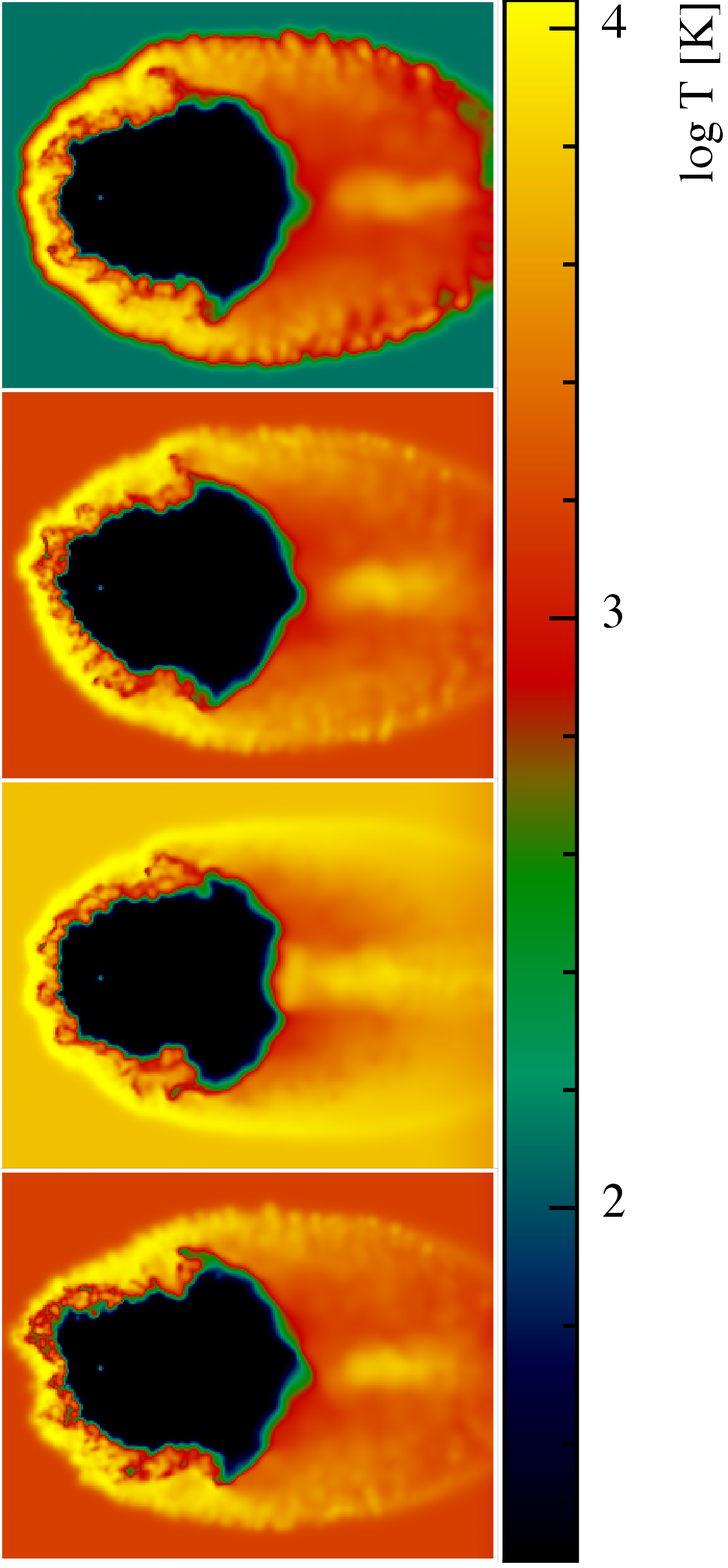}
\includegraphics[scale=.38, angle=0,trim= 0 80 0 0, clip=true]{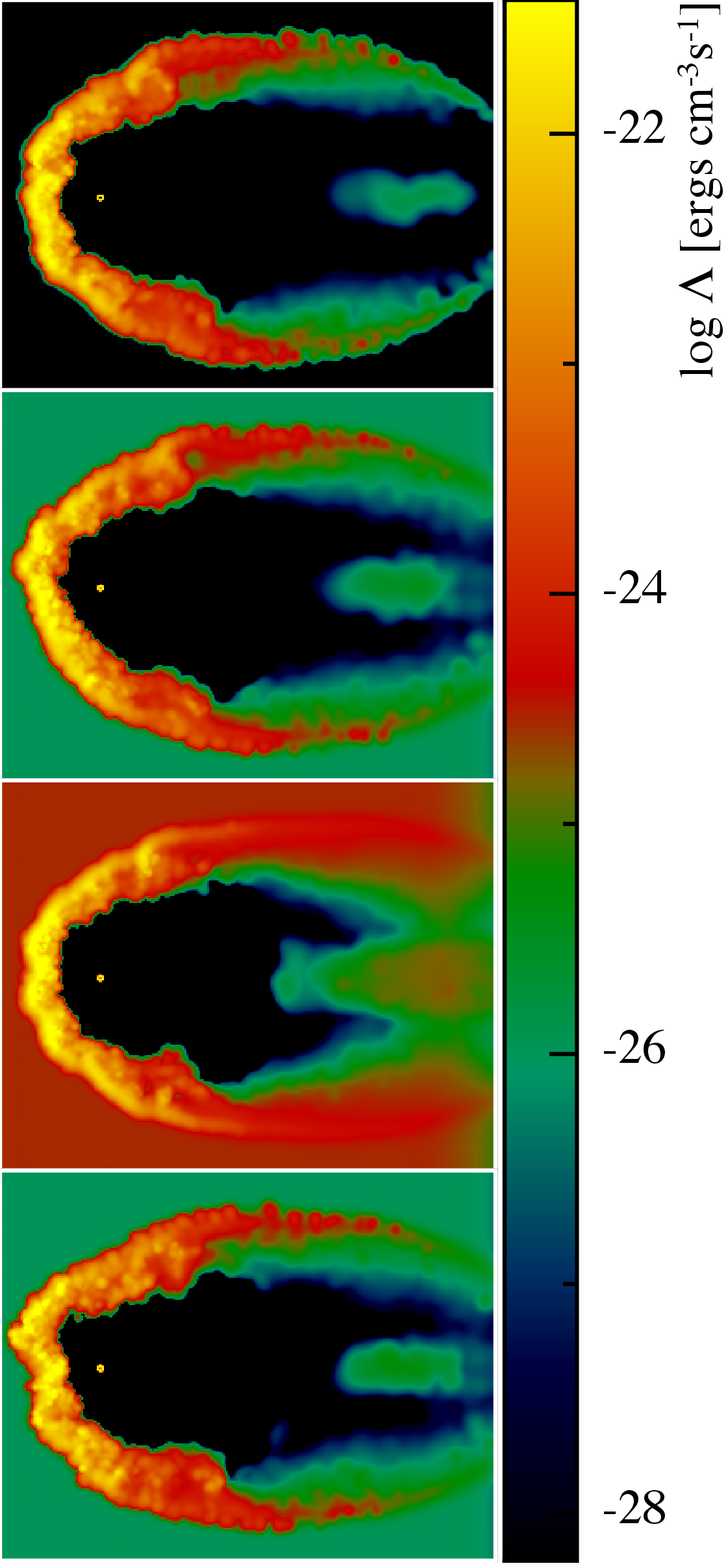}
\caption{Temperature [left] and total emissivity [right] in the symmetry plane
for models C$_{\rm L650}$ [top row], C$_{\rm L1600}$ [second row], C$_{\rm L8000}$ [third row],  and C$_{\rm Lf}$ [bottom row].
\label{fig: Ttest}}
\end{figure}

\begin{figure*}
\centering
\includegraphics[scale=.53, angle=0,trim= 0 125 0 0, clip=false]{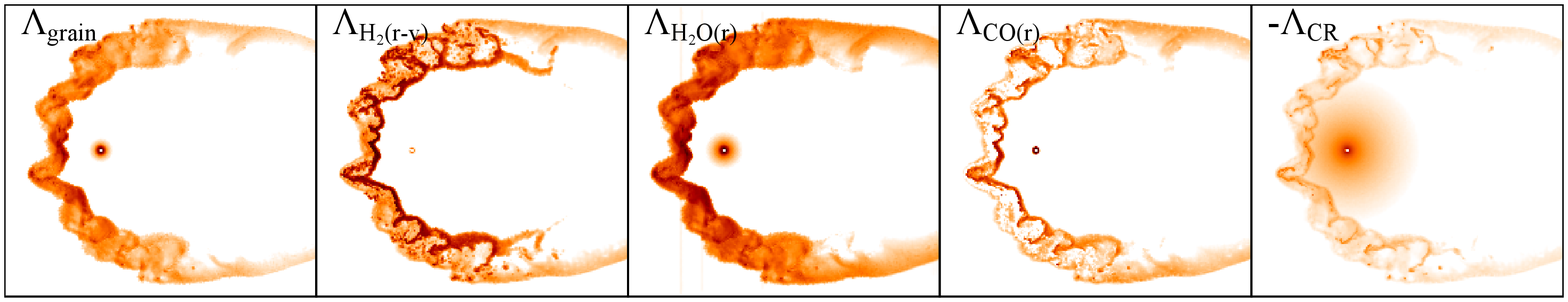}
\includegraphics[scale=.525, angle=0,trim= 150 122 600 0, clip=true]{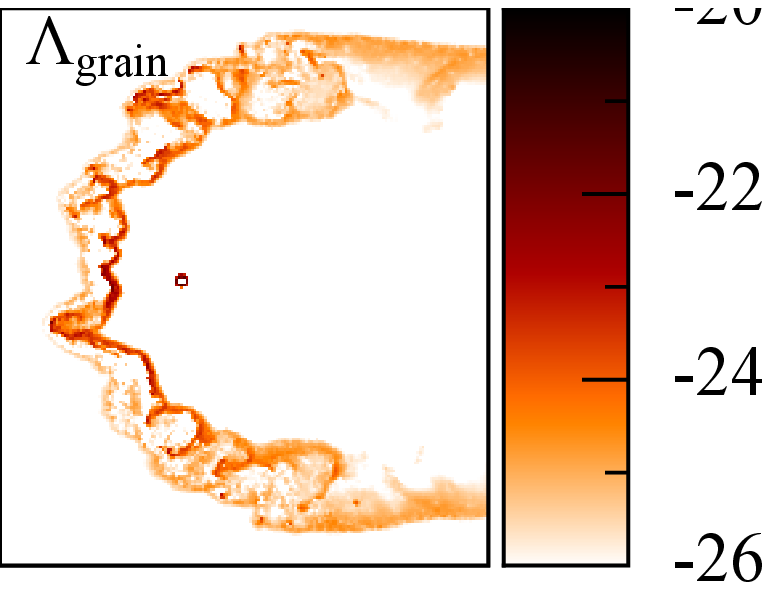}
\includegraphics[scale=.53, angle=0,trim= 0 125 0 0, clip=false]{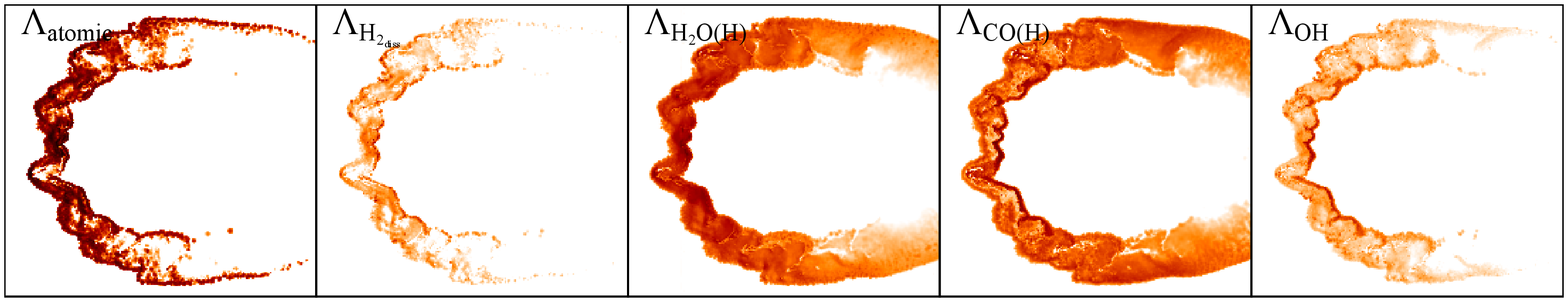}
\includegraphics[scale=.525, angle=0,trim= 150 125 600 0, clip=true]{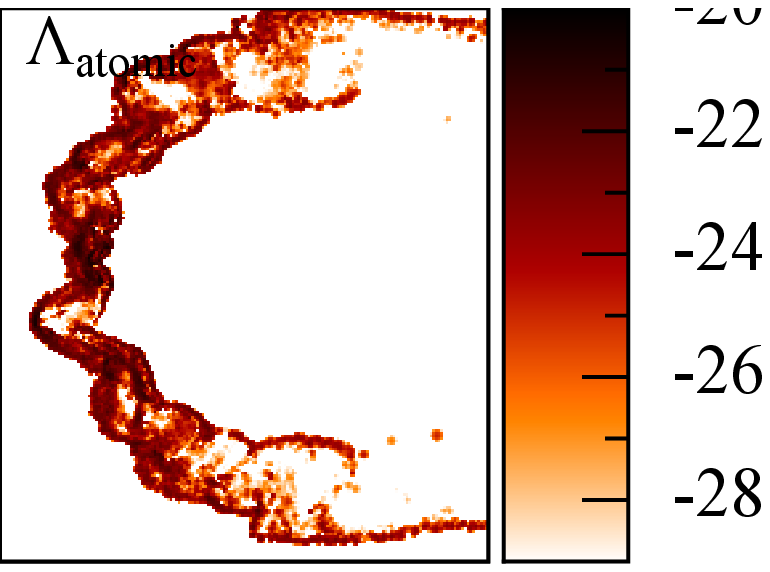}
\includegraphics[scale=.53, angle=0,trim= 0 125 0 0, clip=false]{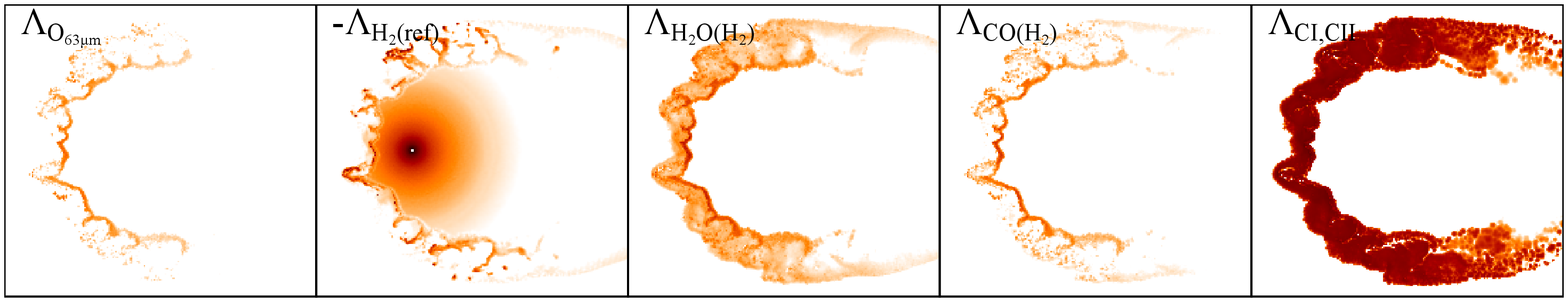}
\includegraphics[scale=.525, angle=0,trim= 150 122 600 0, clip=true]{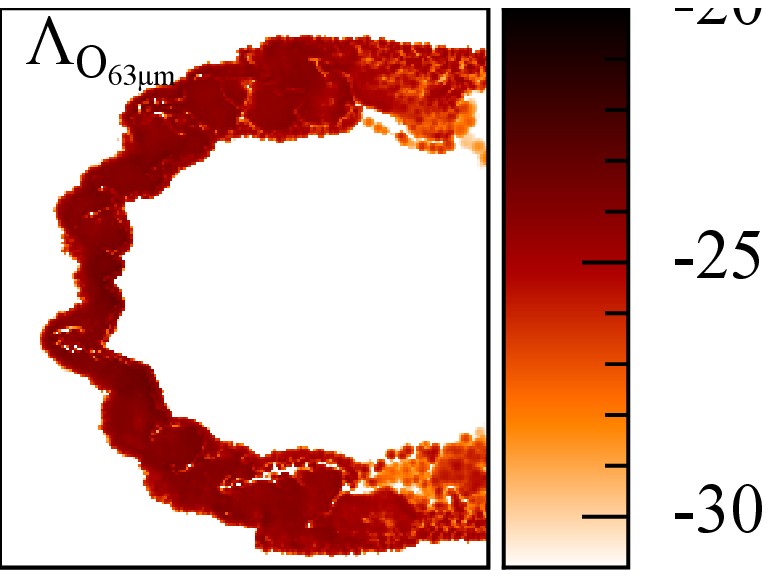}
\includegraphics[scale=.53, angle=0,trim= 0 125 0 -30, clip=false]{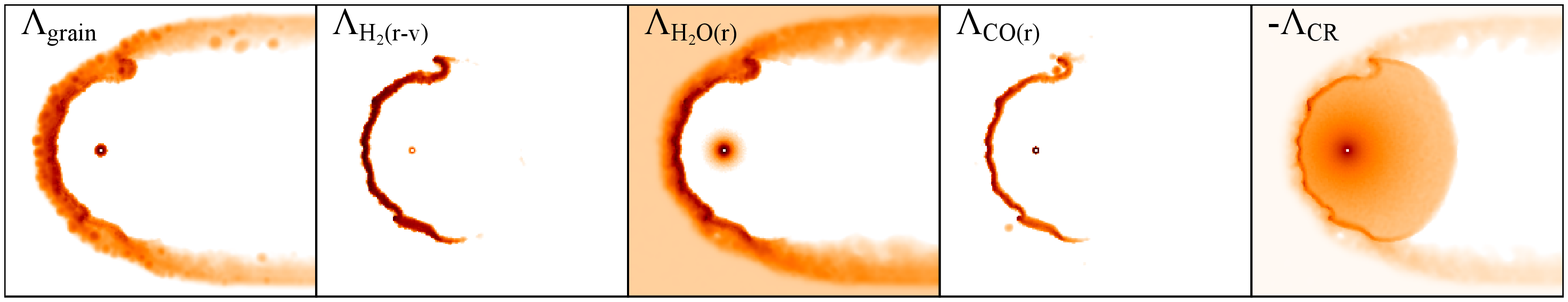}
\includegraphics[scale=.53, angle=0,trim= 150 124 600 0, clip=true]{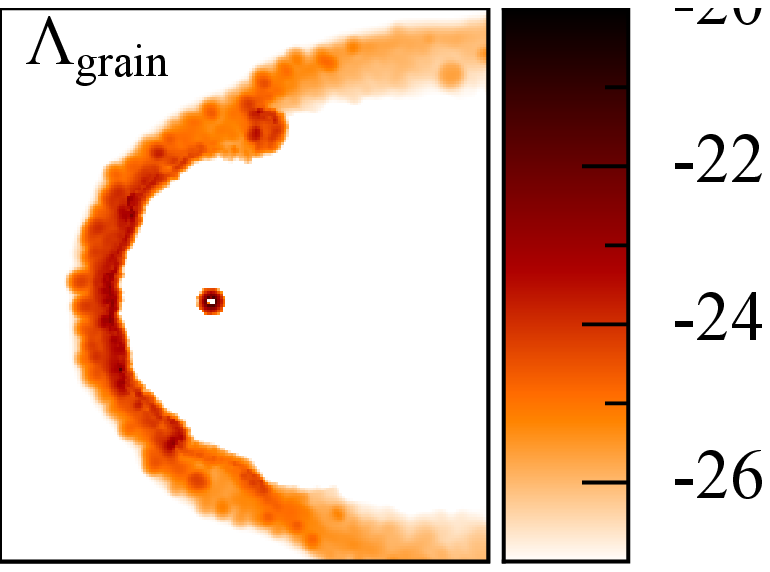}
\includegraphics[scale=.53, angle=0,trim= 0 125 0 0, clip=false]{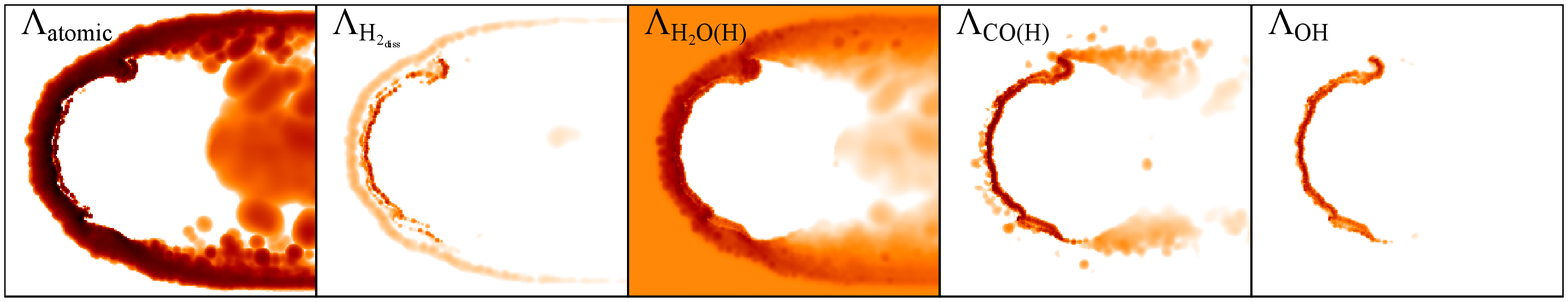}
\includegraphics[scale=.53, angle=0,trim= 150 124 600 0, clip=true]{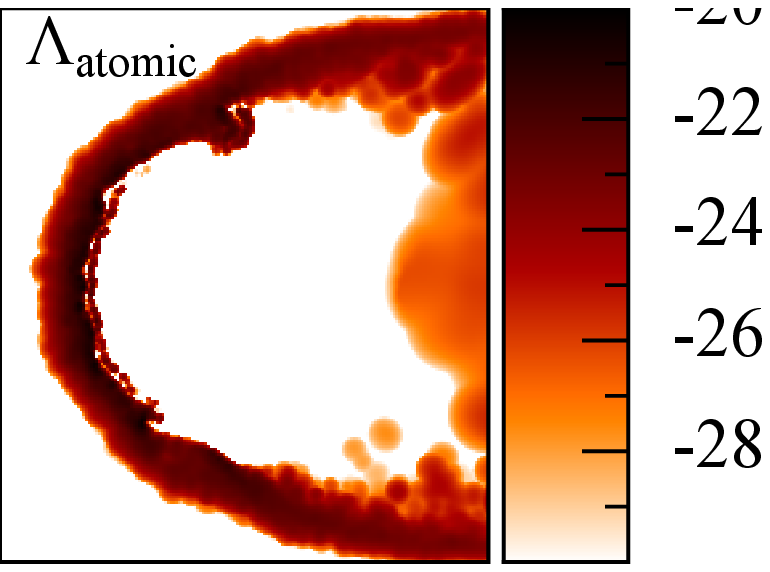}
\includegraphics[scale=.53, angle=0,trim= 0 125 0 0, clip=false]{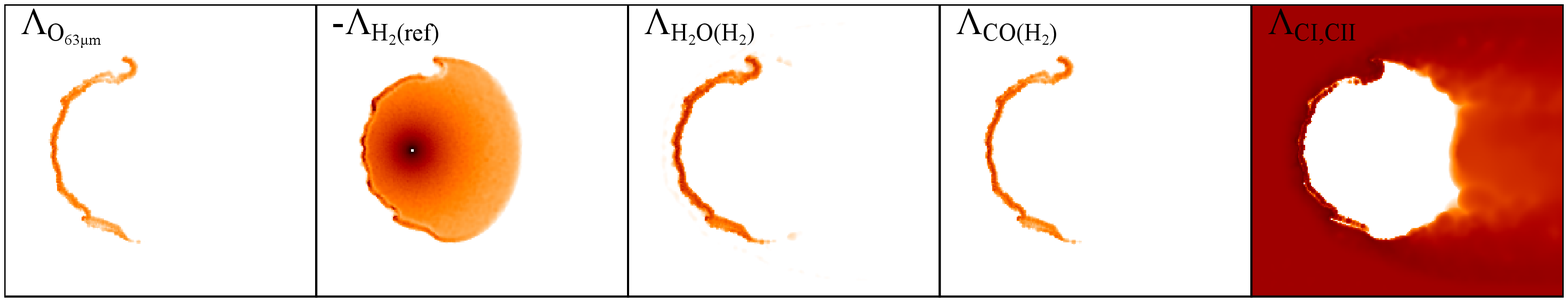}
\includegraphics[scale=.53, angle=0,trim= 150 124 600 0, clip=true]{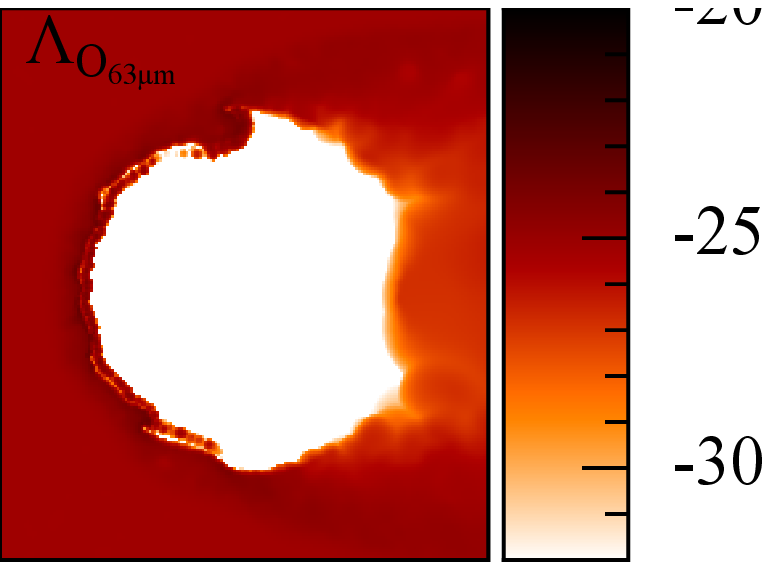}
\caption{Cross-sectional profile of the emissivity (ergs\,cm$^{-3}$\,s$^{-1}$) for the various heating
and cooling species in model A$_{\rm H}$ [top three rows] and model D [bottom three rows].
 \label{fig: Lcross}}
\end{figure*}

While the thermal pressure of the ISM is insignificant compared to its ram pressure 
in all models, it could affect the bow shock thermal and emission properties. The wind 
thermal pressure is insignificant owing to adiabatic expansion, but the hydrogen molecular 
fraction, $f$, entering the reverse shock may also affect the emissivity and cooling of shocked 
gas. The effects of the ISM temperature were tested by using three ISM temperatures for 
different runs of model C: $T_{\rm ISM}$ = 650 K for model C$_{\rm L650}$, 1\,600 K for 
model C$_{\rm L1600}$ and 8\,000 K for model C$_{\rm L8000}$. All models use a 
molecular fraction $f=0.001$ (on a scale where 0 is atomic and 0.5 is fully molecular) 
for the wind boundary condition, so a comparison model C$_{\rm Lf}$ was run with $f=0.18$.
Cross-sections of the gas temperature and total emissivity for these four models are shown 
in Fig.~\ref{fig: Ttest}, and can be compared to the default model C in Fig.~\ref{fig: series}, 
although it should be noted that the logarithmic scales are different in the two figures.
 Higher ISM temperatures produce a smoother forward shock and result in a 
 much fainter tail. The hotter gas that accumulates along 
the symmetry axis in the tail is also closer to the reverse shock, filling the void behind 
the RSG wind more efficiently. This is a relatively minor difference between the models, 
however, so the strongest impression we get from comparing the top three panels in 
Fig.~\ref{fig: Ttest} is how similar they are in both the structure and emissivity of the bow 
shock.
 
The molecular-to-atomic hydrogen fraction in the RSG wind increases as the wind 
cools and expands away from the star. The initial value depends on the temperature 
structure and chemistry near the stellar surface. We assumed a low value since  
Betelgeuse is thought to have a hot, $\sim$8\,000 K chromosphere that would 
dissociate molecules \citep{Hart84}. However, more recent studies suggest that the 
temperature in this region could be much lower at approximately 4\,000 K \citep[and 
references therein]{Harp10}, and thus $f$ could be higher. This was tested in the $f=0.18$ 
model C$_{\rm Lf}$ whose results are shown in the bottom panel of Fig.~\ref{fig: Ttest}. 
As expected, the cooling is much stronger for larger $f$, which leads to a more unstable 
bow shock -- the apex of the bow shock has moved inwards in model C$_{\rm Lf}$ and 
its shape is far from a smooth curve for both the forward and reverse shock. The regular 
appearance and relative stability of the observed bow shock on similar scales \citep{Ueta08}  
 suggests that the hydrogen is indeed mostly in atomic form. 

\section{Emission from the bow shock shell}
\label{sec: appC}

The emissivities for each of the cooling and heating species included 
in the simulations are shown in cross-section in Fig.~\ref{fig: Lcross} 
for models A$_{\rm H}$ [top] and D [bottom]. These cross-sectional profiles show 
the location and morphology of the emitting regions for each species more clearly 
than the projections shown in Figs.~\ref{fig: Lproj15} and \ref{fig: Lproj03}.  
Here we also show results for each coolant individually.
As expected, the processes that involve molecular hydrogen, 
e.g.~$\Lambda_{\rm H_2 (r-v)}$, tend to emit strongly from the gas in the reverse shock. 
The OI fine structure and CO rotational cooling are also more concentrated in 
the denser, cooler regions of the reverse shock. The other species tend to emit 
more evenly from the entire bow shock, with the atomic cooling greatest in the 
forward shock. In the slow models, the atomic cooling is largely absent from the 
cool, dense R-T `fingers'.  In the fast model, several species that do emit from 
both the forward and reverse shocks tend to show a strong ridge of emission just 
ahead of the contact discontinuity, e.g.~the grain, water, atomic, and carbon 
fine-structure cooling. The lower ISM density in the fast model results in much weaker  
emission from the bow shock tail than in the slow models, e.g.~$\Lambda_{\rm H_2O (H_2)}$
 and $\Lambda_{\rm CO (H_2)}$.

\Online
\begin{figure*}
\centering
\includegraphics[scale=.25]{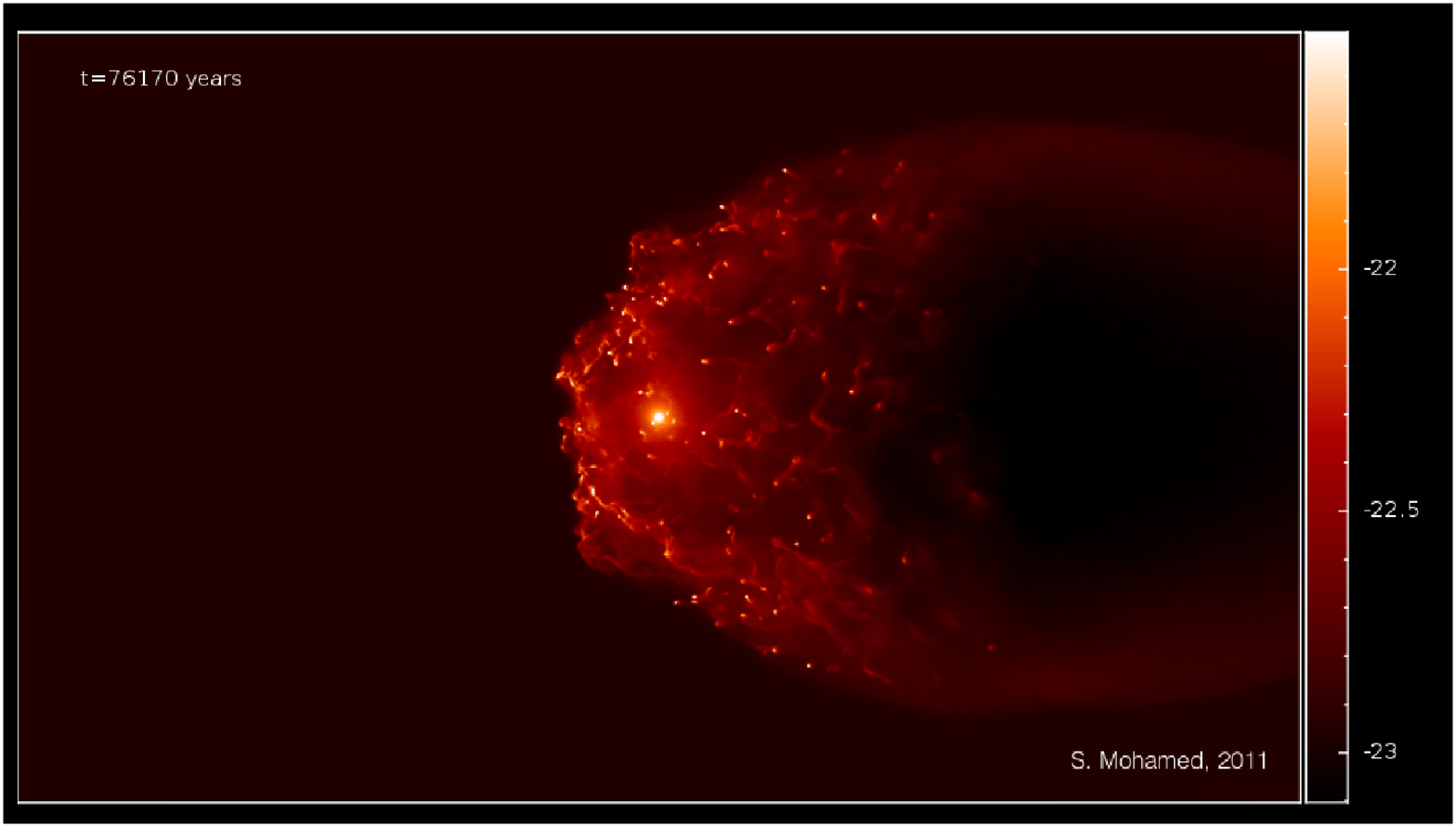}
\includegraphics[scale=.25]{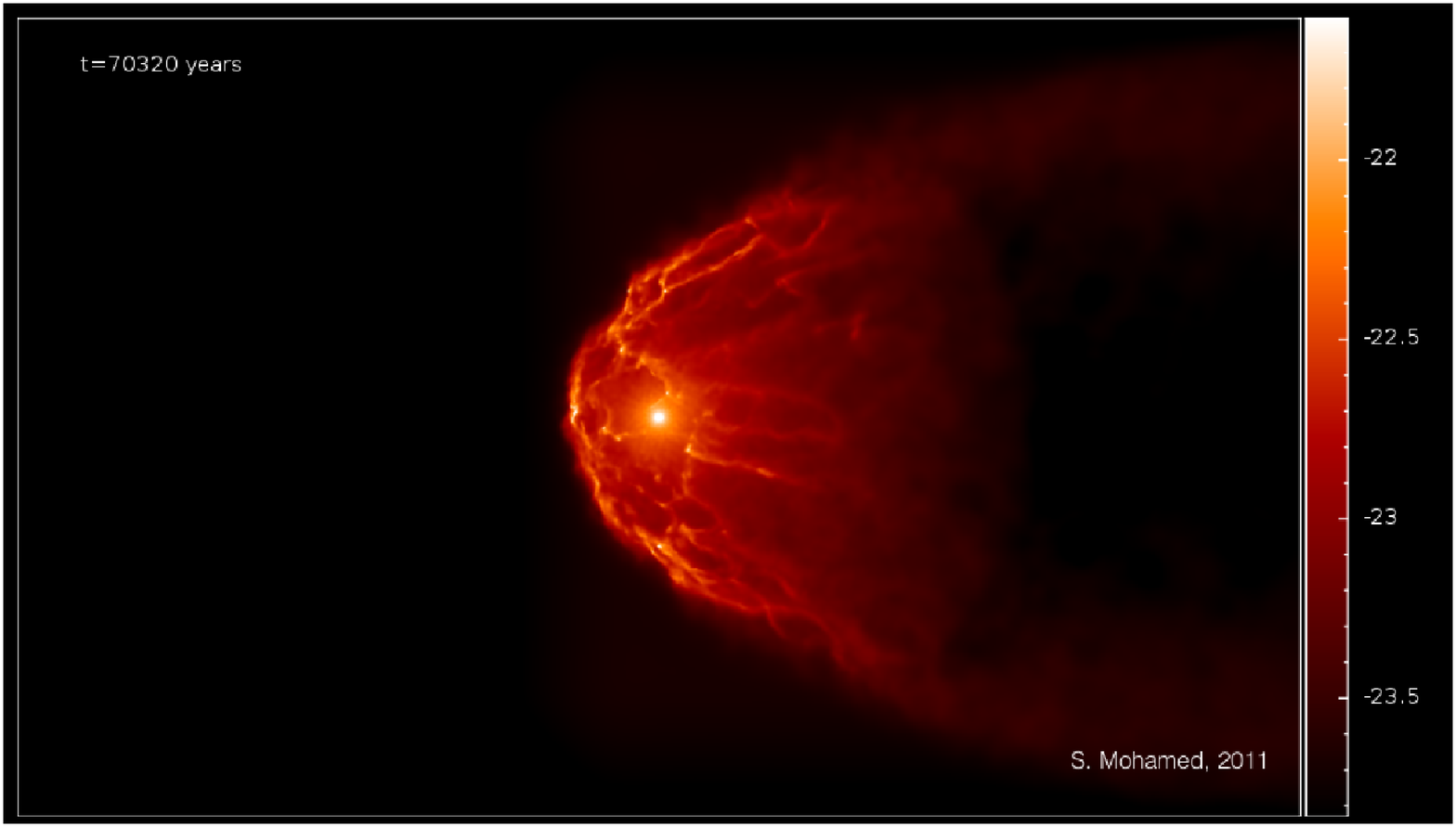}
\caption{Snapshots of the movies for model B [left] and D [right] simulations. First, the 
evolution of the gas column density is followed, then the final structure is 
rotated 360$^\circ$, demonstrating the changing morphology of the bow shock 
with different inclination angles.
 \label{mov: movie}}
\end{figure*}
\end{document}